\documentclass[a4paper,11pt]{article}

\usepackage{jheppub} 

\usepackage{graphicx}
\usepackage{amsmath}
\usepackage{enumitem}

\usepackage{epsfig}

\usepackage{subfigure}
\usepackage{longtable}
\usepackage{times}

\title{Estimating the possible QPOs of $M87^*$ from the parameters of a hairy Kerr black hole}

\author{O. Donmez and F. Dogan}
\affiliation{College of Engineering and Technology, American
  University of the Middle East, Egaila 54200, Kuwait}

\emailAdd{orhan.donmez@aum.edu.kw}
\emailAdd{fatih.dogan@aum.edu.kw}

\abstract{In this paper, we study the dynamics of the shock cone formed around a hairy Kerr black hole due to Bondi-Hoyle-Lyttleton (BHL) accretion and investigate the quasi-periodic oscillations (QPO) behaviors resulting from the black hole-cone interaction, aiming to predict the QPO frequencies that may occur around the $M87^*$ black hole. To achieve this, we use the hairy Kerr black hole parameters consistent with the observed shadow parameters of $M87^*$ as initial conditions in numerical simulations, revealing the structure of the resulting shock cone and QPOs in a strong gravitational field. Numerical calculations show that the deviation of the hairy Kerr black hole from the Kerr metric and the hair parameters significantly influence the complex behavior of the resulting QPOs. It is found that the Lense-Thirring effect and the pressure modes trapped within the cone lead to the excitation of QPOs. The hair parameter has been observed to suppress the resulting QPO frequencies. The Lense-Thirring effect, in a strong gravitational field with a black hole spin parameter of $a/M > 0.7$, also suppresses other modes and generates high-frequency QPOs. It is predicted that the QPO frequencies observed around the $M87^*$ black hole could span a wide range from nHz to mHz. Using both Kerr and hairy Kerr gravities, the QPO frequencies formed around the $M87^*$ black hole can be explained. Especially in cases where the black hole is spinning rapidly, power spectrum density (PSD) analyses have shown very distinct low-frequency QPOs and resonance conditions in both gravities. By comparing the results obtained from numerical calculations with observational and analytical results, we discuss the observability of the QPO frequencies that may occur around the $M87^*$ black hole.}

\keywords{
numerical relativity, $M87^*$, hairy Kerr black hole, Bondi-Hoyle-Lyttleton, QPOs}

\begin{document} 
\maketitle
\flushbottom


\section{Introduction}
\label{Introduction}

Black holes are among the most mysterious objects in the universe. Understanding the events occurring inside and around these objects would contribute to a more detailed understanding of the universe and the nature of dark matter. In the last decade, observations of the black holes using the ground-based and space-based telescopes have enabled the understanding of their physical structure and their interactions with surrounding matter. One such observation is the shadow of the $M87^*$ black hole observed by the EHT collaboration \citep{Akiyama1, Akiyama2, Akiyama3, Akiyama4, Akiyama5}. This observation received great attention, and many scientists have conducted numerous studies to theoretically and numerically support the observed results. EHT collaboration measured the angular gravitational radius of the $M87^*$ black hole as $3.8\pm0.4\mu as$, the asymmetric diameter as $42\pm 3 \mu as$, and the deviation from a perfect circle as $\Delta C\leq 0.1$. Theoretical studies have shown that the shadow of the $M87^*$  black hole is consistent with a Kerr black hole. Although the observed shadow was not initially tested with alternative theories, recent efforts using modified gravity have aimed to theoretically validate the observed shadow. Thus, by testing modified gravity in the region close to the black hole's shadow, i.e., in the strong gravitational field, some conclusions about the nature of dark matter can be drawn \citep{Johannsen2010ApJ}. Additionally, the frequency range of QPOs, which currently face significant technical challenges in being observed, can be identified if they form in this area.

  It is thought that a hairy Kerr black hole could arise due to the interaction between the black hole and additional matter, such as surrounding dark matter. Possible dark matter or scalar fields around the black hole could interact with it, resulting in a modified black hole and consequently a modified gravity \citep{Hod2012PhLB, Luo2024arXiv240410742L, Errehymy2024EPJC, Maurya2024ptep}. It is also considered that scalar fields could form halos or clouds around the black hole \citep{Hertog2004JHEP}. Thus, the hairy parameter of the black hole could affect QPO  that might occur in the strong gravitational field around it \citep{Konoplya2006PhRvD}. On the other hand, axion-type hypothetical particles, which are strong candidates for dark matter particles, are thought to form a cloud around the black hole under certain conditions, thus creating a hairy black hole \citep{Campbell1990PhLB, Delgado2021PhRvD, Burrage2023CQGra, DonmezMPLA2024}. This hair parameter could change the black hole's shadow and also affect the characteristics of gravitational and electromagnetic waves \citep{Cunha2015PhRvL, Guo2023EPJC}. Dark matter particles in a strong gravitational field can also influence the distribution of normal matter. In this article, we model  the QPOs around a hairy Kerr black hole in a strong gravitational field that may bear the traces of the black hole-dark matter interaction.

The QPOs  are important tools for testing different gravity theories around the black holes and the neutron stars, particularly in strong gravitational fields. These oscillations, observed in X-ray data, provide crucial information about the behavior of matter near the black hole and its interaction with the black hole itself. The frequency, amplitude, stability, and resonance conditions of the resulting QPO oscillations depend on the spacetime geometry. Therefore, studying alternative gravity theories, identifying different QPO frequencies, and explaining observational results are significant. For example, in the Black–Bounce–Reissner–Nordström geometry, QPO behavior has been examined for electrically charged particles orbiting around a compact object. The theoretical results obtained have led to parametric studies determining whether different sources should be classified as black holes or wormholes \citep{2024PDU....4601561M}. Additionally, QPO behavior has been investigated for test particles around the Kiselev black hole with a string cloud. For sources like RO J1655-40, GRS 1915+105, and the Milky Way galaxy, the possible values of the string cloud parameter have been identified \citep{2024ChPhC..48e5104R}. Similarly, various gravity tests have been conducted by examining QPO behavior, such as the study of QPO behavior around Regular-Kiselev Black Holes \citep{2023Galax..11..113R}, the behavior of test particles and the discovery of QPOs around Aether Black Holes \citep{2023Galax..11...95R}, the behavior of test particles and the discovery of QPOs in the presence of an external magnetic field in the Simpson–Visser spacetime \citep{2023EPJC...83..854V}, the investigation of QPOs and test particle behavior around Quasi- and Non-Schwarzschild Black Holes \citep{2023Univ....9..391M}, and the study of QPOs in Horndeski gravity \citep{2023EPJC...83..572R, Donmez2024arXiv240216707D, Donmez2024Submitted}.

In the region close to the black hole's horizon, QPO frequencies arise as a result of the interaction between matter and the black hole. By analyzing these frequencies, it is possible to infer the spin, mass, and physical structure of the black hole. Additionally, QPOs are important for testing the strong gravitational field and understanding the behavior of matter in that region. Therefore, numerically modeling QPOs in the strong gravitational field around the $M87^*$  black hole can help better understand the characteristics of the $M87^*$ black hole observed by the EHT detector. Given the current technology and sensitivity, QPOs formed from the strong gravitational field of the $M87^*$  black hole have not yet been observed. Thus, understanding these QPOs would make a significant contribution to the literature. The main aim of this paper is to numerically identify QPOs occurring in the strong gravitational field around the $M87^*$ black hole. This is achieved by using the shadow parameters of the $M87^*$  black hole, which are identified through the hairy Kerr black hole model, and applying the BHL accretion mechanism previously used for different sources \citep{Donmez2024Univ, Donmez2024arXiv240216707D, Donmez2024Submitted}.

In Ref.\citep{Afrin_2021}, the results obtained from observations of the black hole in $M87^*$ \citep{Akiyama1, Akiyama2, Akiyama3, Akiyama4, Akiyama5} have been theoretically derived using a hairy Kerr black hole. As a result of comparing the observations with theory, the possible values of the spin ($a/M$), deviation from the Kerr black hole ($\eta$), and primary hair ($l_0/M$) parameters have been determined to be consistent with the observational results of $M87^*$. In this paper, utilizing these parameters from the hairy Kerr black hole in Ref.\citep{Afrin_2021}, the dynamic structure of the accreted matter around the hairy Kerr black hole has been revealed through the  BHL accretion, resulting in the formation of the shock cone. Additionally, the trapped QPOs within the shock cone have been identified. It is predicted that these QPOs could be possible for $M87^*$. The discovery of QPOs around the $M87^*$ black hole could have several important implications. These include: (a) general relativity can be tested in a strong gravitational field \citep{Remillard2006ARA&A}, (b) the physical properties of the black hole can be revealed and compared with observational data \citep{Akiyama1}, (c) it can provide us with information about the results of black hole-matter interactions \citep{Belloni2012}, allowing predictions about the formation and evolution of black holes, and predictions can be made about the nature of the dark matter known to exist within the galaxy  with modified gravity \citep{Johannsen2010ApJ}.

Black holes-accreting matter system can produce QPOs through various physical mechanisms. These mechanisms include relativistic precession \citep{Mottamnras2017, Motta2022MNRAS}, magnetohydrodynamic instabilities \citep{McKinney2009MNRAS, Dongmnras2020}, resonance conditions \citep{Molteni1996ApJ, Deligianni2020PhRvD}, Rayleigh-Taylor instability \citep{Toma2017MNRAS}, $g-$, $p-$, and $c-$mode oscillations \citep{Hawley1999ApJ, Rezzolla2003MNRAS}, and instabilities created by shock waves \citep{Nakayamamnras1994, Donmez2024Univ, DonmezMPLA2024}. Here, we model the formation of a shock cone around the black hole using the BHL accretion mechanism. Once this shock cone reaches the steady-state and creates instability, particularly with $p-$mode oscillations trapped inside the cone, it generates QPOs. These modes oscillate either angular direction within the cone or between the black hole's horizon and the stagnation point, producing QPO frequencies. Additionally, the resulting QPO frequencies can produce new frequencies through nonlinear coupling. The modes formed in this strong gravitational field are crucial for revealing the physical properties of the black hole, as mentioned earlier.

Our paper is organized as follows. In Section \ref{HairyKBH}, we present the geometrical structure of the black hole and the metric of the hairy Kerr black hole that describes the surrounding spacetime, and we calculate the lapse function and shift vector of this matrix required for numerical calculations. In Section \ref{GRH_Equation}, we briefly introduce the GRH equations in the equatorial plane. The boundary conditions, initial conditions, and all constraints necessary for BHL accretion used in the models are explained in Section \ref{Model_Initial}. In Section \ref{NumRes}, after briefly explaining the physical structure of the shock cone that forms around the black hole due to the BHL mechanism, we detail the QPO modes trapped within this cone for each model situation, as well as the QPO oscillation frequencies resulting from the black hole-cone interaction. We also discuss how the observed results can determine the observability of QPOs around the $M87^*$ black hole. We used the parameters of the hairy Kerr black hole compatible with the observed shadow of the $M87^*$ black hole in these analyses. We then discuss in Section \ref{Theory_prediction} the compatibility of the predicted QPO frequencies for $M87^*$ with theoretically known relativistic precession frequencies. In Section \ref{Expected_frequency}, based on the QPO frequencies observed from other supermassive black holes, we predict the possible frequencies for $M87^*$ and compare them with the frequencies calculated in this paper. Finally, in Section \ref{Concl}, we summarize the findings. Throughout this paper, we use geometrized units where $G=c=1$. Greek indices range from 0 to 3, while Latin indices range from 1 to 3.

\section{Hairy Kerr Black Holes}
\label{HairyKBH}

The framework of the modified gravity could shed light on the interactions of dark matter with other matter in the universe. In this context, the dark matter interacts with black holes. During the interaction of the black holes with the surrounding matter, they also interact with the unobservable but indirectly observed dark matter, which is confirmed to exist in galaxies. Consequently, this interaction could lead to changes in the shadow of the black hole. Therefore, the existence of hairy black holes and their ability to explain situations such as dark matter and energy is considered important. One of the recent studies in this regard is on hairy Kerr black holes. These black holes have been theoretically discovered using the gravitational decoupling (GD) method. Ini ially, the Schwarzschild hairy black hole \citep{OvallePDU2021} was identified using this approach \citep{OvallePRD2017, OvallePLB2019}. The appeal of this solution lies in its development under very few conditions, requiring only a well-defined event horizon and knowledge of energy conditions, making it quite general. This solution was later extended to the rotating case \citep{Contreras2021PhRvD, Mahapatra2023PDU}, leading to the discovery of the hairy Kerr solution. GD approximately describes the deformation occurring in General Relativity (GR) as a result of the interaction between dark matter or energy and Kerr black holes \citep{Zhang_2023}. Through this method, the extended solution of the Kerr black hole, namely the hairy black hole, is defined. Unlike the Kerr black hole in GR, which produce new solutions, these solutions lead to more chaotic situations. These black holes, defined as stationary, depend on either scalar \citep{Guo_2021} or proca \cite{Herdeiro_2016} hair parameters.

The stationary and axisymmetric hairy Kerr black hole, distinct from the GR Kerr black hole, depends on
the deviation ($\eta$) and primary hair ($l_0/M$) parameters. When expressed in terms of the mass ($M$)and
spin parameter ($a/M$) of the black hole, along with these parameters, the hairy Kerr black hole in
Boyer-Lindquist coordinates can be written as \citep{Contreras2021PhRvD},

\begin{eqnarray}
  ds^2 = -\left(\frac{\bigtriangleup - a^2Sin^2\theta}{\Sigma}\right)dt^2 + \frac{\Sigma}{\bigtriangleup}dr^2
  + \Sigma d\theta^2 - 2a Sin^2\theta \left( 1-\frac{\bigtriangleup-a^2 Sin^2\theta}{\Sigma}\right)dt d\phi \nonumber \\
  + Sin^2\theta \left(\Sigma + a^2Sin^2\theta \left(2- \frac{\bigtriangleup-a^2 Sin^2\theta}{\Sigma} \right) \right)
   d\phi^2, \;\;\;\;\;\;\;\;\;\;\;\;\;\;\;
\label{Eq1}
\end{eqnarray}

\noindent
where $\bigtriangleup = r^2+a^2-2Mr + \eta r^2 e^{-r/(M-\frac{l_0}{2})}$  and
$\Sigma=r^2+a^2 Cos^2\theta$.  
$\eta$ determines the potential  deviation of the hairy Kerr black hole from the standard Kerr. When  $\eta=0$,
the hairy Kerr black hole reduces to the standard Kerr. In the case where $a = 0$, the hairy Schwarzschild black
hole spacetime matrix is obtained \citep{Contreras2021PhRvD}. $l_0 \leq 2M$ is necessary to ensure asymptotic
smoothness.

For the numerical solution of the GRH equations, along with the spacetime matrix given in Eq.\ref{Eq1} which represents the
the curvature of the hairy Kerr black hole and the black hole itself, it is necessary to define the
lapse function ($\alpha$) and shift vector $\beta_i$ around the  hairy Kerr black hole.
For this purpose, the relationship below with the 4-metric, 3-metric, and these parameters are
\citep{Misner1977},

\begin{eqnarray}
  \left( {\begin{array}{cc}
   g_{tt} & g_{ti} \\
   g_{it} & \gamma_{ij} \\
  \end{array} } \right)
=  
  \left( {\begin{array}{cc}
   (\beta_k\beta^k - \alpha^2) & \beta_k \\
   \beta_i & \gamma_{ij} \\
  \end{array} } \right),
\label{Eq2}
\end{eqnarray}

\noindent
where i, j, and k represent only the spatial elements of the spacetime matrix. For example, i corresponds to r, $\theta$, and $\phi$.
The lapse function for hairy Kerr black hole is

\begin{eqnarray}
  \alpha = \sqrt{-\frac{a^2-\bigtriangleup}{\Sigma}+\frac{a^2 \left(1-\frac{-a^2+\bigtriangleup}{\Sigma}\right)^2
      \left(-a^2+\frac{a^4}{\bigtriangleup}+\frac{2a^2 \Sigma}{\bigtriangleup}+\frac{\Sigma^2}{\bigtriangleup}\right)}
    {-a^2 \Sigma+\frac{a^4 \Sigma}{\bigtriangleup}+\frac{2 a^2 \Sigma^2}{\bigtriangleup}+\frac{\Sigma^3}{\bigtriangleup}}},
\label{Eq3}
\end{eqnarray}

\noindent
and the shift vectors can be represented as  

\begin{eqnarray}
  \beta_r = 0, \;\;\;  \beta_{\theta}= 0, \;\;\;  \beta_{\phi}= -a \left(1-\frac{-a^2+\bigtriangleup }{\Sigma}\right).
\label{Eq4}
\end{eqnarray}


\section{Formulations of General Relativistic Hydrodynamic Equations}
\label{GRH_Equation}

The covariant form of the GRH equations, depending on the stress-energy tensor $T^{\mu\nu}$ and the current density of matter $J^{\mu}$, is as follows \citep{Font2000LRR, Donmez1},

\begin{eqnarray}
  \triangledown_{\mu}T^{\mu\nu}=0, \;\;\; \triangledown_{\mu}J^{\mu} = 0.
\label{GRHEq1}
\end{eqnarray}

\noindent To solve these equations numerically, they must first be written in conservation form. For this, it is necessary to define the stress-energy tensor and current density first. In non-ideal conditions, solving these equations is very difficult; therefore, when viscosity and heat conduction effects are neglected, the stress-energy tensor for an ideal fluid is as follows,

\begin{eqnarray}
  T^{\mu\nu}= \rho h u^{\mu}u^{\nu}+Pg^{\mu\nu}.
\label{GRHEq2}
\end{eqnarray}

\noindent Here, $\rho$ represents the rest-mass density of matter, $h=1+\epsilon + \frac{P}{\rho}$ is the specific enthalpy, $u^{\mu}$ is the flow 4-velocity of the ideal gas in spacetime, $P = (\Gamma-1)\rho \epsilon$ is the pressure from the ideal gas equation, and $g^{\mu\nu}$ denotes the metric defining the geometry of the spacetime. This metric, used in this paper for the hairy Kerr black hole, is provided in Section \ref{HairyKBH}. Here, $\epsilon$ is the specific internal energy and $\Gamma$ is the adiabatic index.

The covariant form of the GRH equations given in Eq.\ref{GRHEq1} is not suitable for high-resolution shock capturing methods used in fluid dynamics. Therefore, they are written in conservation form using the 3+1 formalism \citep{Font2000LRR, Donmez1}. In conservation form, the GRH equations  on the equatorial plane are

\begin{eqnarray}
  \frac{\partial U}{\partial t} + \frac{\partial F^r}{\partial r} + \frac{\partial F^{\phi}}{\partial \phi}
  = S,
\label{GRHEq3}
\end{eqnarray}

\noindent where  $U$ is the conserved variables, $F^r$ is the fluxe in  the $r$ direction , $F^{\phi}$ is the fluxe in  the $\phi$ direction , and $S$ is  the source. The conserved variables, fluxes, and sources are expressed in terms of primitive variables, the 3-metric, and other variables defined later, as follows.

\begin{eqnarray}
  U =
  \begin{pmatrix}
    D \\
    S_j \\
    \tau
  \end{pmatrix}
  =
  \begin{pmatrix}
    \sqrt{\gamma}W\rho \\
    \sqrt{\gamma}h\rho W^2 v_j\\
    \sqrt{\gamma}(h\rho W^2 - P - W \rho)
    \end{pmatrix},
\label{GRHEq4}
\end{eqnarray}

\begin{eqnarray}
  \vec{F}^i =
  \begin{pmatrix}
    \alpha\left(v^i - \frac{1}{\alpha\beta^i}\right)D \\
    \alpha\left(\left(v^i - \frac{1}{\alpha\beta^i}\right)S_j + \sqrt{\gamma}P\delta^i_j\right)\\
    \alpha\left(\left(v^i - \frac{1}{\alpha\beta^i}\right)\tau  + \sqrt{\gamma}P v^i\right)
    \end{pmatrix},
\label{GRHEq5}
\end{eqnarray}

\noindent and source is

\begin{eqnarray}
  \vec{S} =
  \begin{pmatrix}
    0 \\
    \alpha\sqrt{\gamma}T^{ab}g_{bc}\Gamma^c_{aj} \\
    \alpha\sqrt{\gamma}\left(T^{a0}\partial_{a}\alpha - \alpha T^{ab}\Gamma^0_{ab}\right)
   \end{pmatrix}.
\label{GRHEq6}
\end{eqnarray}

\noindent Here, $\gamma$ is the determinant of the 3-metric,  $\gamma_{i,j}$ is the 3-metric,  $v_j$ is the 3-velocity of the matter, $W = (1 - \gamma_{i,j}v^i v^j)^{1/2}$ is the Lorentz factor, and $v^i = u^i/W + \beta^i$ is the 3-velocity of the fluid. $\alpha$ and $\beta^i$ are the lapse function and shift vector, respectively.
The three-velocity, $v_j$,  is the velocity of matter as defined with respect to the observer's reference frame, representing only the spatial components of the timelike four-velocity. This velocity is measured as the speed of the matter relative to the observer's rest frame.
The lapse function, shift vector, and 3-metric of the system  are calculated using the spacetime matrix of the hairy Kerr black hole, as provided in Section \ref{HairyKBH}.

The numerical solution of the GRH equations was carried out using the High-Resolution Shock-Capturing (HRSC) method, with fluxes computed using the Marquina method. In each time step, the conservative variables are updated first, followed by the primitive variables. The ideal gas equation of state was used to model the system's behavior by solving the GRH equations. Detailed explanations of the numerical code used and the results of the test problems can be found in Refs.\citep{Donmez1, Donmez2}.


\section{Model Setups}
\label{Model_Initial}

The BHL mechanism is one of the most important accretion mechanisms around the black holes. Because, as a result of various astrophysical events like supernova explosions, interstellar matter is created and it falls into the gravitational field of the black hole. This infall results in the formation of an accretion disk and also generates a shock wave around the black hole \citep{Bondi1, Bondi1952MNRAS, Edgar1}. Numerical analyses over many years have shown that the shock waves around the black hole form a shock cone \citep{Donmez6, Penner2, Koyuncu1, Donmez5, Wenrui1, Donmez3, Donmez2024arXiv240216707D, Donmez2024Univ}. Since oscillation modes can be trapped within these cones, this physical mechanism is proposed as a mechanism for QPOs \citep{CruzOsorio2023JCAP, Donmez2024arXiv240216707D, Donmez2024Submitted}. In this section, the initial conditions for matter falling toward the black hole to form a shock cone are given. In this article, as we examine the formation of QPOs around the black hole in two dimensions on the equatorial plane, matter moving with asymptotic velocity in interstellar space falls toward the black hole with the radial and angular velocities given below after entering the gravitational field,

\begin{eqnarray}
  V^r = \sqrt{\gamma^{rr}}V_{\infty}cos(\phi) \;\;\; and  \;\;\;  V^{\phi} = -\sqrt{\gamma^{\phi \phi}}V_{\infty}sin(\phi).
\label{Asymptotic_veloc}
\end{eqnarray}

\noindent
Here, the $V_{\infty}$ is the asymptotic velocity of the matter, as given in Tables \ref{Inital_Con_Kerr}, \ref{Inital_Con_1}, and \ref{Inital_Con_2}.
The velocities,  $V^r$ and $V^{\phi}$,  ensure that the velocity of the gas falling towards the black hole at infinity is equal to the asymptotic speed, $V^2=V_iV^i=V_{\infty}^2$, while also causing the gas moving toward the black hole in the interstellar medium to fall toward the black hole's horizon parallel to the x-axis.
This velocity significantly affects the physical structure of the resulting shock cone \citep{Donmezetal2022,Donmez6}. As a requirement of the BHL mechanism, matter is made to fall toward the black hole from the upstream side. Other initial conditions used here are density $\rho = 1$ and sound speed $C_s/c = 0.1$.
The reason for choosing $C_s/c$  as $0.1$  is to ensure that the gas falls supersonically into the black hole. This way, accretion towards the black hole is facilitated from a point far away from the black hole.
On the other hand, the pressure of the falling matter is calculated using the ideal gas equation $P = (\Gamma-1)\rho\epsilon$ with an adiabatic index $\Gamma = 4/3$.

  Before the onset of accretion, there is no matter in the computational domain. However, when modeling with Euler-type methods, the density cannot be zero. Therefore, we fill the computational domain with values that are negligible compared to the matter sent from the outer boundary. 
This computational domain is filled with matter, with density and pressure values corresponding to the same speed of sound value given above, which are negligible. For this matter, radial and angular velocities are zero. However, as a requirement of the BHL accretion mechanism, the matter is sent from the upstream region towards the black hole with the initial conditions given above. Since the main purpose of this article is to propose possible QPOs around a hairy Kerr black hole and the $M87^*$ black hole, other initial values are defined by comparing them with EHT observations data given in Ref.\citep{Afrin_2021}, and using the values given in Tables \ref{Inital_Con_1} and \ref{Inital_Con_2}. Thus, using the parameters of the hairy Kerr black hole that are compatible with the observed shadow parameters of the $M87^*$ black hole, we attempt to identify the QPOs that may be observed around the $M87^*$ black hole.
In numerical computations, the boundary points are defined using first-order extrapolation at the inner and outer boundaries along the radial direction. Since this is not an analytical calculation, there is a high likelihood of erroneous solutions forming at the boundary. For this reason, it is undesirable for matter that reaches the inner and outer boundary points to re-enter the computational domain. If this matter re-enters the computational domain, it could generate non-physical oscillations, which could not only affect the QPOs but also potentially cause the code to crash. 
To prevent  non-physical oscillations in numerical simulations, matter reaching the black hole's horizon is made to fall into the black hole using the outflow boundary condition. Matter reaching the outside of the computational domain on the downstream side is also ejected using the outflow boundary condition to prevent non-physical  oscillations.
Periodic boundary conditions are used in the angular direction to ensure continuity. Since the aim of this article is to investigate the formation of QPOs in the strong gravitational field, i.e., in the region $r<10M$, the inner boundary is set at $r_{min}=2.3M$ and the outer boundary at  $r_{max}=100M$. We use $1024$ points in the radial direction and $512$ points in the angular direction. As seen in Tables \ref{Inital_Con_1} and \ref{Inital_Con_2}, the shock cone generally reaches a steady state around or less than $t=3000M$, so the code is run up to $t=35000M$. This allows the matter at the last stable orbit around the Schwarzschild black hole at $r=6M$ to oscillate at least $360$ times, showing that the oscillations and thus the QPOs are persistent. The proof of the persistence of QPOs is explained in detail while analyzing the numerical results in Section \ref{NumRes} for the Kerr black hole models.

\begin{table}
\footnotesize
\caption{ Kerr black hole parameters and the numerical results are given. The parameters, $Model$, $a/M$, $R_{BH}/M$,  $V_{\infty}/c$, $r_{stag}/M$, $\theta_{sh}/rad$, $\tau_{ss}/M$, represent the model name, the black hole rotation parameter, black hole event horizon, the asymptotic velocity of the matter,  the position of the stagnation point, the shock cone opening angle, and the time required to reach the steady-state, respectively.}
 \label{Inital_Con_Kerr}
\begin{center}
  \begin{tabular}{ccccccc}
    \hline
    \hline
    
    $Model$  & $a/M$  & $R_{BH}/M$ & $V_{\infty}/c$ & $r_{stag}/M$ & $\theta_{sh}/rad$ &$\tau_{ss}/M$\\ 
    \hline
    $Kerr04$ & $0.4$& $1.6$   & $0.2$      & $27.10$ & $1.052$  & $2700$ \\
    $Kerr09$ & $0.9$& $1.436$ & $0.2$      & $26.72$ & $1.077$  & $2400$ \\ 
    \hline
    \hline
  \end{tabular}
\end{center}
\end{table}

\begin{table}
\footnotesize
\caption{
  The parameters of the hairy Kerr black hole, obtained from the data gathered by the Event Horizon Telescope
  (EHT) \citep{Akiyama1, Akiyama2, Akiyama3, Akiyama4, Akiyama5} observations of $M87^*$ \citep{Afrin_2021}
  in the equatorial plane, which serve as initial conditions in the numerical calculations, are presented.
  Additionally, it includes numerical results computed from the simulations.
  The parameters, including $Model$, $R_s/M$, $\delta_s$, $\eta$, $a/M$, $l_0/M$, $R_{BH}/M$, $V_{\infty}/c$,
  $r_{stag}/M$, $\theta_{sh}/rad$, and $\tau_{ss}$, represent the model name, the radius of the
  observable circle (photon sphere),  the deviation of the left edge of the shadow from the circle boundary,
  the potential deviation of the metric from the Kerr black hole, the black hole rotation parameter,
  the primary hair parameter in the Kerr black hole, black hole event horizon, the asymptotic velocity of gas injected from the
  outer boundary, the position of the stagnation point, the shock cone opening angle, and the time
  required to reach the steady-state, respectively.
}
 \label{Inital_Con_1}
\begin{center}
  \begin{tabular}{ccccccccccc}
    \hline
    \hline

    $Model$ & $R_s/M$  & $\delta_s$ & $\eta$& $a/M$   & $l_0/M$ & $R_{BH}/M$ &$V_{\infty}/c$ & $r_{stag}/M$ & $\theta_{sh}/rad$ &$\tau_{ss}/M$\\ 
    \hline
    $HK05A$      & $5.000$&  $0.001$ & $0.5$ & $0.0682$& $0.0608$& $1.861$ & $0.2$      & $27.29$   & $1.077$          & $2300$ \\
    $HK05B$      & $5.075$&  $0.010$ & $0.5$ & $0.2225$& $0.3119$& $1.872$& $0.2$       & $27.10$   & $1.055$          & $2670$ \\
    $HK05C$      & $5.150$&  $0.100$ & $0.5$ & $0.6204$& $0.6557$& $1.707$& $0.2$       & $27.10$   & $1.076$          & $2900$ \\
    $HK05D$      & $5.182$&  $0.210$ & $0.5$ & $0.7923$& $0.9113$& $1.550$& $0.2$       & $27.29$   & $1.076$          & $2100$ \\
    \hline
    $HK1A$      & $4.750$&  $0.001$ & $1.0$ & $0.0657$& $0.0583$& $1.702$& $0.2$        & $27.10$    & $1.052$          & $2360$ \\
    $HK1B$      & $4.900$&  $0.010$ & $1.0$ & $0.2181$& $0.2669$& $1.739$& $0.2$        & $27.10$    & $1.052$          & $2200$  \\
    $HK1C$      & $5.045$&  $0.035$ & $1.0$ & $0.3986$& $0.5289$& $1.746$& $0.2$        & $27.29$    & $1.052$          & $2500$ \\
    $HK1D$      & $5.125$&  $0.100$ & $1.0$ & $0.6147$& $0.7405$& $1.651$& $0.2$        & $26.72$    & $1.077$          & $2850$  \\
    $HK1E$      & $5.170$&  $0.250$ & $1.0$ & $0.8056$& $0.9387$& $1.464$& $0.2$        & $26.72$    & $1.052$          & $2160$  \\    
    \hline
    \hline
  \end{tabular}
\end{center}
\end{table}

\begin{table}
\footnotesize
\caption{
  Same as Table \ref{Inital_Con_1} but different estimated values of hairy Kerr black hole
  parameters from the the known shadow observable area ($A/M^2$) and
  oblateness ($D$) \citep{Afrin_2021}. 
}
 \label{Inital_Con_2}
\begin{center}
  \begin{tabular}{ccccccccccc}
    \hline
    \hline
    
    $Model$ & $A/M^2$  & $D$ & $\eta$& $a/M$   & $l_0/M$ & $R_{BH}/M$ & $V_{\infty}/c$ & $r_{stag}/M$ & $\theta_{sh}/rad$ &$\tau_{ss}/M$\\ 
    \hline
    $HKA05A$      & $84$&  $0.999$ & $0.5$ & $0.1250$& $0.8815$& $1.962$& $0.2$      & $26.72$    & $1.052$          & $2400$ \\
    $HKA05B$      & $83$&  $0.995$ & $0.5$ & $0.2714$& $0.7172$& $1.913$& $0.2$      & $27.10$    & $1.052$          & $2400$ \\
    $HKA05C$      & $82$&  $0.980$ & $0.5$ & $0.5233$& $0.7616$& $1.798$& $0.2$      & $27.10$    & $1.052$          & $3000$ \\ 
    $HKA05D$      & $80$&  $0.950$ & $0.5$ & $0.7644$& $0.8038$& $1.571$& $0.2$      & $27.29$    & $1.052$          & $2400$ \\
    $HKA05E$      & $78$&  $0.910$ & $0.5$ & $0.9108$& $0.9193$& $1.307$& $0.2$      & $27.10$    & $1.052$          & $2450$ \\
    \hline
    $HKA1A$      & $84$&  $0.999$ & $1.0$ & $0.1211$& $0.9790$& $1.949$& $0.2$       & $27.29$    & $1.052$          & $2500$ \\
    $HKA1B$      & $83$&  $0.990$ & $1.0$ & $0.3706$& $0.9282$& $1.869$& $0.2$       & $27.10$    & $1.076$          & $2400$\\
    $HKA1C$      & $80$&  $0.960$ & $1.0$ & $0.6587$& $0.8244$& $1.633$& $0.2$       & $27.29$    & $1.052$          & $2700$ \\
    $HKA1D$      & $78$&  $0.940$ & $1.0$ & $0.7439$& $0.7515$& $1.493$& $0.2$       & $27.10$    & $1.076$          & $2700$ \\
    $HKA1E$      & $67$&  $0.870$ & $1.0$ & $0.8291$& $0.2678$& $0.990$& $0.2$       & $27.10$    & $1.052$          & $2200$ \\    
    \hline
    \hline
  \end{tabular}
\end{center}
\end{table}
%


\section{Numerical Results}
\label{NumRes}

Independent of different gravitational scenarios, that is, regardless of different modified gravity theories, modeling disk formation through BHL accretion around black holes has led to the formation of shock cones around these black holes. The formation and structure of these shock cones have been demonstrated through numerical models we previously conducted. In Ref.\citep{Donmez6,Donmez5}, the shock cone formed around Schwarzschild and Kerr black holes and the QPO frequencies resulting from the excitation of pressure modes trapped within the cone have been numerically determined and compared with observational results. Using the same method, the dependence of these cones and the resulting QPO frequencies on the EGB coupling constant has been studied for both non-rotating and rotating EGB black holes \citep{Donmez3,Donmez_EGB_Rot}. The compatibility of these frequencies formed around the EGB black hole with observations has been demonstrated by selecting specific sources. Subsequently, applying the same method to slowly rotating neutron stars, and considering that it can explain the events around slowly rotating black holes, the shock cone formed around the Hartle-Thorne black hole and the QPOs trapped within this cone have been numerically calculated and compared with observations \citep{Donmez2024Univ}. Finally, the formation of the shock cone around the Horndeski black hole, which contains a scalar field parameter, and the excitation of QPOs and how they change with respect to the scalar field parameter have been studied separately for non-rotating \citep{Donmez2024Submitted} and rotating \citep{Donmez2024arXiv240216707D} Horndeski black holes. Since the scalar field parameter defining the spacetime around the Horndeski black hole is thought to define the dark matter within the galaxy, it is predicted that revealing the effect of this field contributes to understanding the nature of dark matter \citep{Magana2012JPhCS, TellezTovar2022PhRvD, CruzOsorio2023JCAP, Gomez2024PhRvD}. In fact, in this study, it has been observed that shock cones and thus QPO frequencies disappear depending on the intensity of the scalar field parameter. The numerical results obtained from this study have been compared with a few observed sources, proposing physical mechanisms that could explain the observed frequencies of these sources.

We use the parameters consistent with the $M87^*$ black hole shadow observed with the EHT telescope calculated in Ref.\citep{Afrin_2021}, and using the initial values given in Tables \ref{Inital_Con_1}, and \ref{Inital_Con_2}, we  attempt to determine the possible QPO frequencies around the $M87^*$ black hole with Hairy Kerr gravity, alongside proposing physical reasons for these frequencies and establishing possible lower and upper limits for the QPO frequencies. However, before this, we  numerically discuss the QPOs that can be excited around the Kerr black hole and attempt to compare them with the Hairy Kerr black hole where necessary.

The emergence of QPO frequencies caused by different physical mechanisms such as Lense-Thirring Precession and Diskoseismic Modes in the strong gravitational field around the black hole can be realized by performing a PSD analysis of the mass accretion rate closest to the black hole's horizon. The analyses of power spectra of physical mechanisms formed around black holes are important tools in understanding events occurring in the vicinity of the horizon where strong gravitational forces exist. Since the hair parameter in modified gravities alters the spacetime around black holes, studying power spectrum analyses in strong fields is an important endeavor in understanding the effects of the hair parameter on electromagnetic spectra.

The fundamental QPO frequencies that may be possible due to the effect of the shock cone in the equatorial plane around a Hairy Kerr black hole is  numerically revealed. According to the theory, we believe that fundamental QPO frequencies can be excited for two reasons. One of these reasons is Lense-Thirring Precession \citep{Wu2023PhRvD}. This precession entirely forms a frequency if the black hole's spin parameter is moderate to high. Lense-Thirring Precession results in high-frequency QPOs due to the warping of spacetime where the gravitational field is strong. The other reason is the oscillation of the diskoseismic mode trapped within the shock cone, either in the angular direction or in the radial direction between the stagnation point and the black hole's horizon. Due to the Diskoseismic Mode, two fundamental mode frequencies occur. One of them, referred to as $f_{sh}$, occurs due to the angular oscillation of matter trapped inside the cone. This frequency is proportional to the $\theta_{sh}$ value given in Tables \ref{Inital_Con_Kerr}, \ref{Inital_Con_1}, and \ref{Inital_Con_2}. The other, referred to as $f_{EH}$, occurs due to the oscillation of matter between the stagnation point and the inner radius of the disk located at $r=2.3M$, as given in these tables.

In this paper, the data used for the PSD analysis is obtained by dumping the mass accretion rate at the different points closest to the black hole's horizon, at $r=2.3M$, $r=3.8M$ and $r=6.08M$. In PSD analyses, the units of the frequencies are in terms of black hole mass. This is because, as we mention in the introduction, geometrized units are used in this article. When the QPOs around the $M87^*$ black hole are observed, the possibility of comparing the numerical results obtained here with the observational results is  provided.
Therefore,  the QPO frequency calculated in geometrized units can be expressed in Hz using the equation $f(Hz)=(c^3/GM)f(M)$. Here, c is the speed of light, G is the gravitational constant, and M is the observed mass of the $M87^*$ black hole, $M=6.5\times10^9 M_{\odot}$. When these values are substituted into the equation, the following relation is obtained:

\begin{eqnarray}
  f(\rm{Hz}) = f(M) \times 3.123 \times 10^{-5},
\label{conMtoHz}
\end{eqnarray}

\noindent where $f(M)$ is the frequency obtained from numerical calculations.

The innermost stable orbit, where the matter forming the accretion disk is closest to the black hole, changes depending on the spin parameter of the black hole in the case of spherical accretion. For a non-rotating black hole, this point is at $r=6M$, whereas for a spin parameter of $a=0.9$, it is at $r=2.32M$. However, the innermost stable orbit of the matter trapped within the shock cone formed as a result of BHL accretion changes due to factors such as the pressure forces that arise within the cone. Consequently, the innermost stable orbit moves closer to the black hole's horizon. This phenomenon has been numerically confirmed both in this study and in existing literature \citep{Foglizzo1999A&A, Penner2,  Donmez5, Donmez4, CruzOsorio2023JCAP, Donmez2024arXiv240216707D}. This also allows us to conduct QPO analysis in the strong gravitational field very close to the black hole's horizon.

\subsection{Formation of the Shock Cone around the Hairy Kerr Black Hole}
\label{color_shock_cone}

\begin{figure*}
  \vspace{1cm}
  \center
     \psfig{file=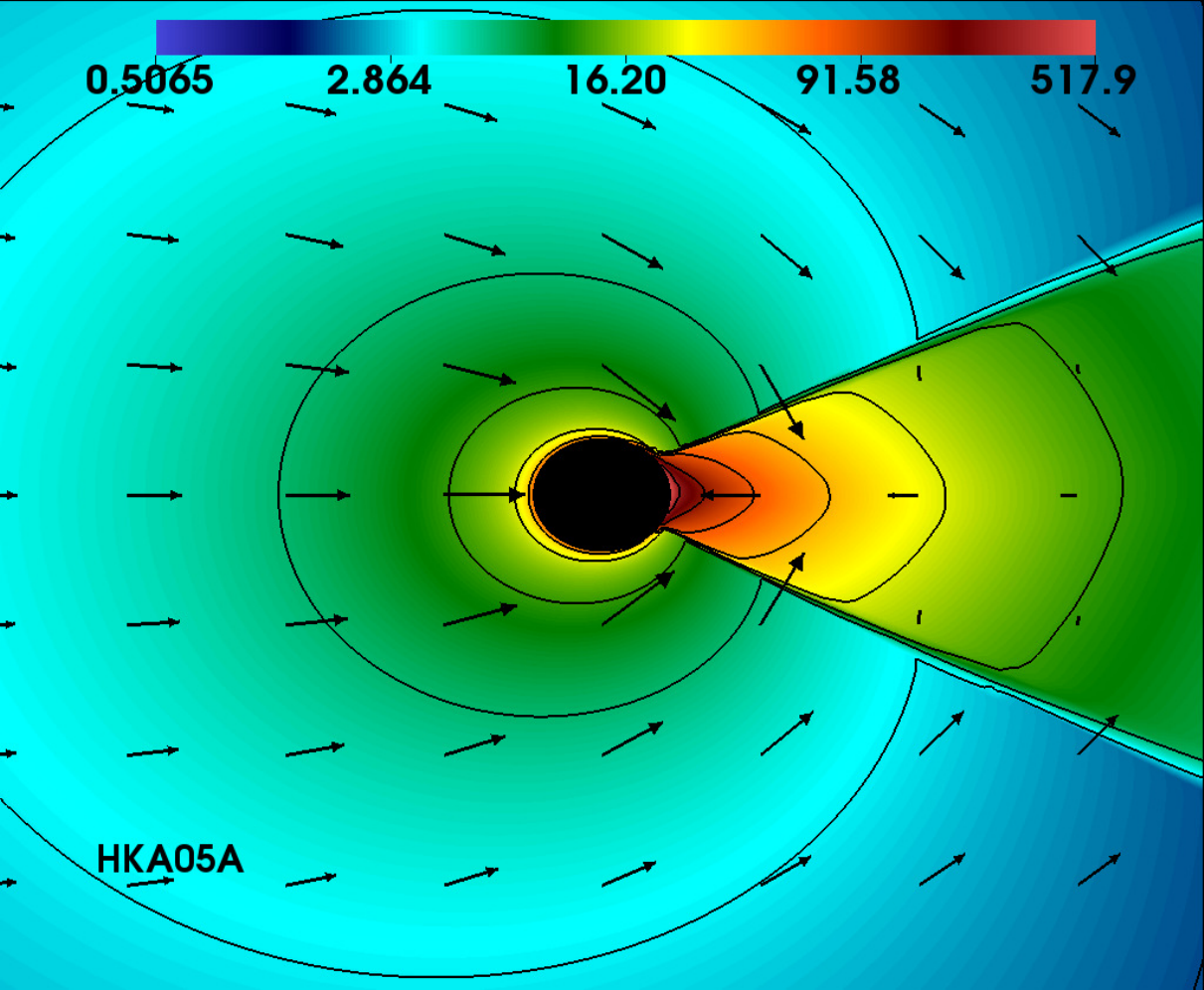,width=7.0cm}
     \psfig{file=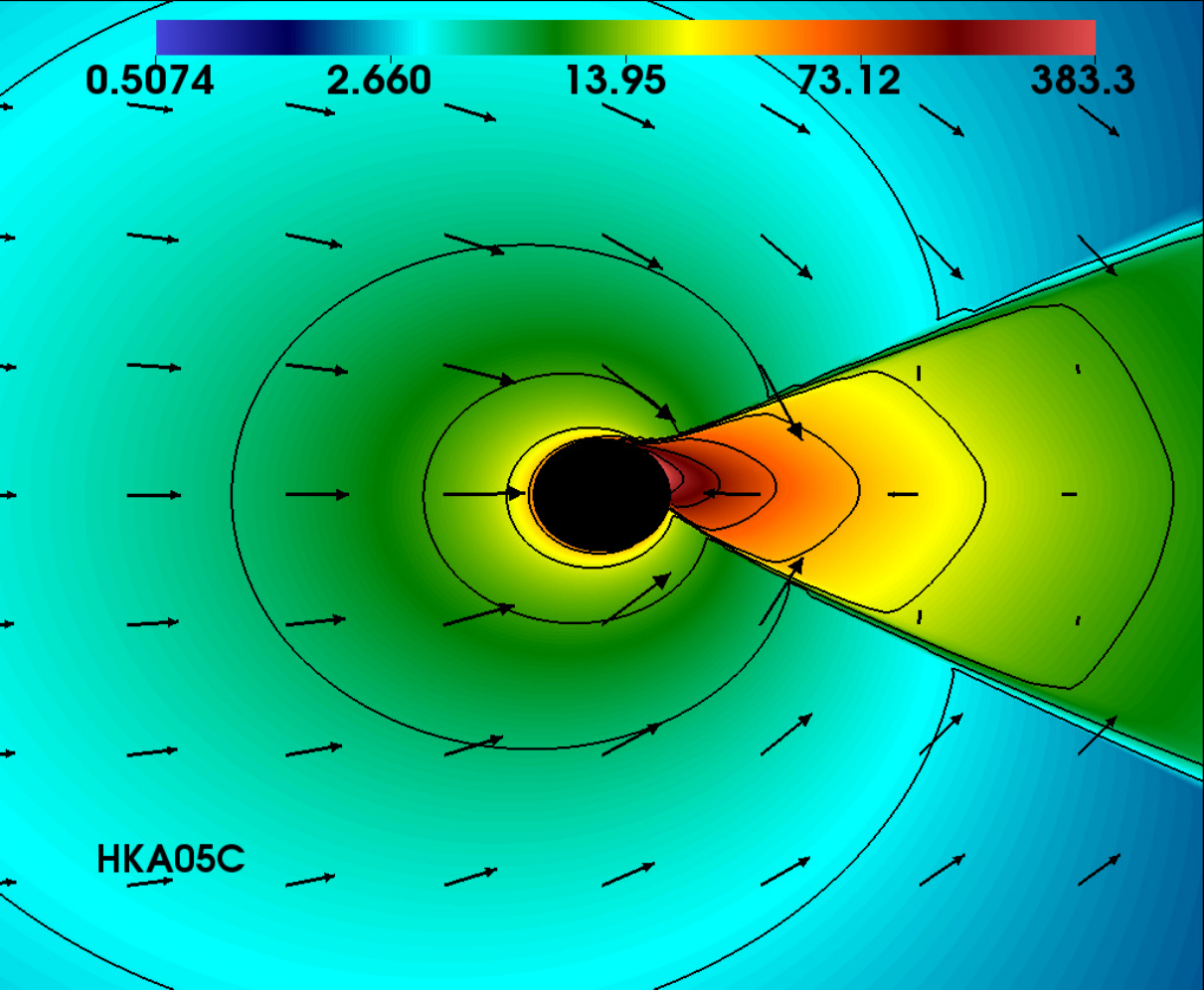,width=7.0cm}
     \psfig{file=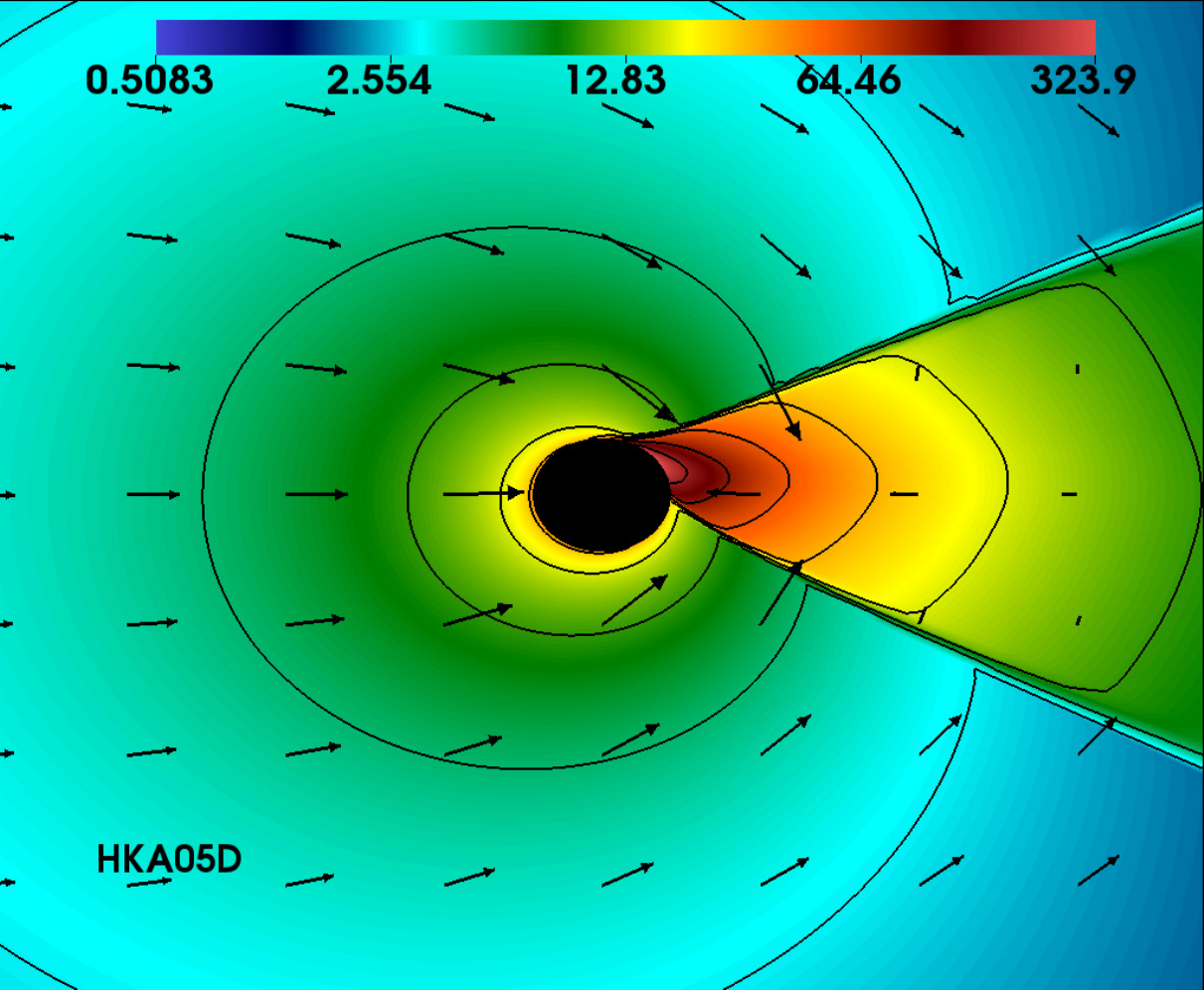,width=7.0cm}
     \psfig{file=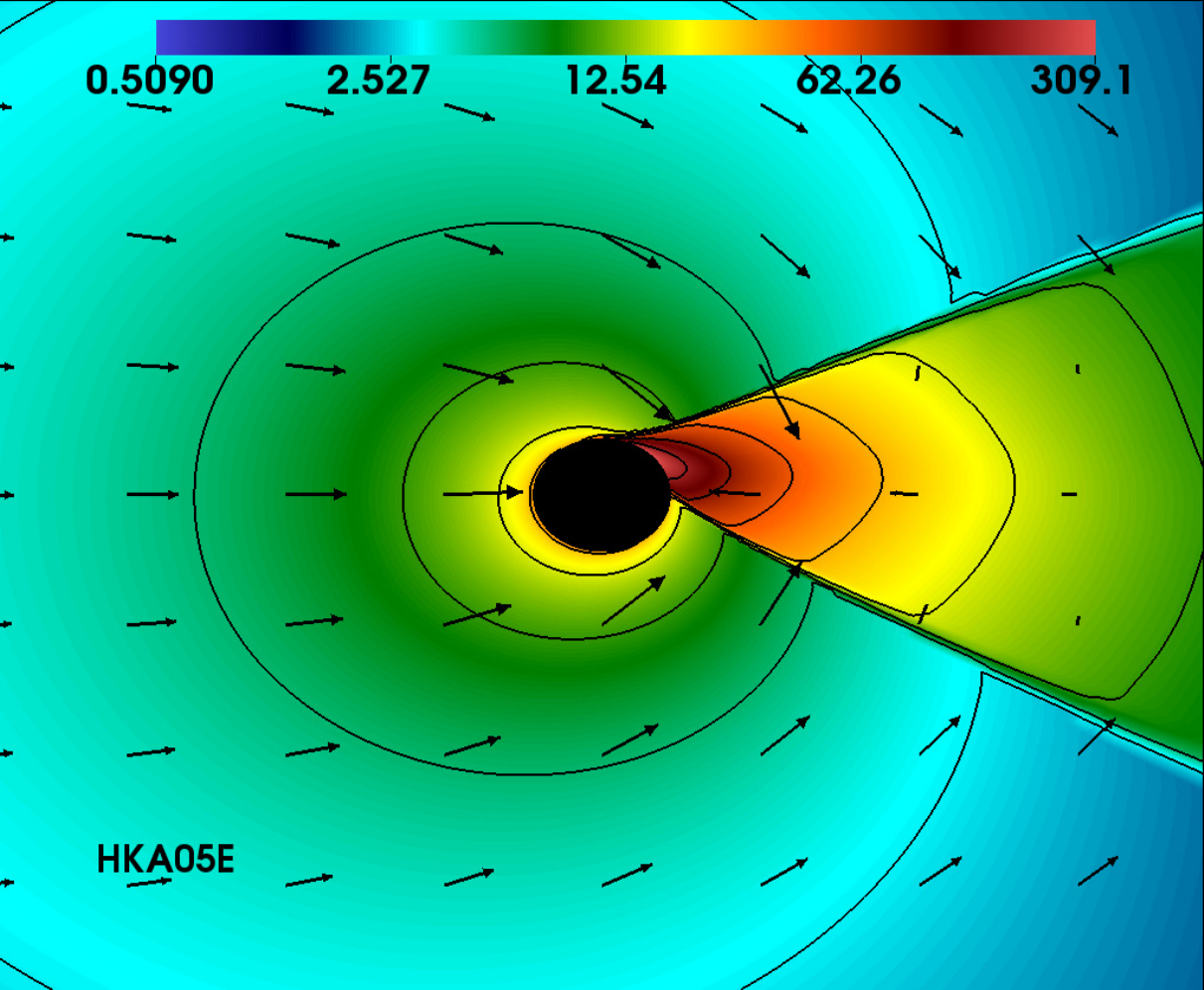,width=7.0cm}  
     \caption{The rest-mass density of the shock cone formed around the black hole is depicted using both colored and contour plots, along with vector plots, for different models. In Table \ref{Inital_Con_2}, for the case of $\eta = 0.5$, the changes in the structure of the shock cone near the black hole's horizon are shown for various spin and hair parameters. To illustrate the effects of the strong gravitational field, the screenshots are displayed within the range of $[x_{min}, y_{min}] = [-20M, -20M]$ and $[x_{max}, y_{max}] = [20M, 20M]$.
    }
\vspace{1cm}
\label{Color_HKA05}
\end{figure*}

Before modeling the QPO frequencies formed around the Hairy black hole, we first investigate the formation and excitation mechanisms of the shock cone and how the black hole's parameters affect this formation. In Figure \ref{Color_HKA05}, to illustrate the causes and formation of the shock cone, the initial conditions for $\eta = 0.5$ from Table \ref{Inital_Con_2} are used, explaining the formation and dynamic structure of the shock cone. Supersonic matter falling towards the black hole from the interstellar medium due to BHL accretion interacts with the black hole's strong gravitational field, creating a shock cone on the opposite side of where the matter is falling. Simulations have shown that this cone forms entirely due to the sudden deceleration of the infalling matter on the other side. Consequently, a discontinuity in velocity, density, and pressure is created on the opposite side of the infall, leading to the formation of the shock cone. As calculated for each model in the tables, a stagnation point forms inside the cone. Matter moves towards the black hole if it is closer to the black hole's horizon than this point, while matter farther away is pushed radially outward. This creates a shock cone on the downstream or on the opposite side of the inflow, and the shock cone eventually reaches a steady state.  A vector plot is drawn to illustrate the direction and speed of the matter's flow. As seen from the vectors, supersonic matter falls towards the black hole, forming the cone, and within the cone, the matter either falls towards the black hole at slower speeds or is ejected outward. Matter falling towards the black hole accelerates further under the influence of the strong gravitational field, moving at relativistic speeds. Therefore, as the intensity of the relativistic effect increases, the dynamic structure of the shock cone changes, and the amount of matter accreted on the black hole's horizon also increases. Additionally, the frame-dragging effect, dependent on the black hole's spin parameter $a/M$, becomes significant. As seen in Figure \ref{Color_HKA05}, as the value of $a/M$ approaches 1, the shock cone is increasingly stretched and begins to envelop the black hole's horizon. This effect, also known as the Lens-Thirring effect, alters and excites physical phenomena near the black hole's horizon \cite{Wu2023PhRvD, Pasham2024arXiv240209689P}. As the amount of matter falling into the black hole increases, pressure also rises, leading to nonlinear and even chaotic behaviors. Consequently, high-energy electromagnetic emissions such as X-rays and gamma rays are triggered in this region. These emissions occur due to quantum  processes such as inverse Compton scattering and pair annihilation. Additionally, the deceleration of particles falling from the upstream side due to strong interactions on the opposite side of the black hole increases the temperature of the matter. These interacting materials within the cone lead to the formation of Rayleigh-Taylor instabilities \citep{Foglizzo1999A&A,Park2014MNRAS}. Such instabilities around the black hole cause continuous X-ray emissions and the formation of QPOs. Rayleigh-Taylor instabilities within the cone lead to turbulence and chaotic behaviors.

\subsection{QPO frequencies of $M87^*$ for the given parameter of the Kerr black hole in Table \ref{Inital_Con_Kerr}}
\label{Num_QPO_Kerr}
As mentioned in the introduction of Section \ref{NumRes} and further detailed here, this paper shows that when the accretion of matter toward the black hole is modeled only in the equatorial plane, QPO frequencies can arise from two different mechanisms around both Kerr and Hairy Kerr black holes. The first mechanism is the Lense-Thirring effect, resulting from the warping of spacetime around the black hole. The second mechanism involves the excitation of pressure modes trapped within the shock cone.

\begin{figure*}
  \vspace{1cm}
  \center
     \psfig{file=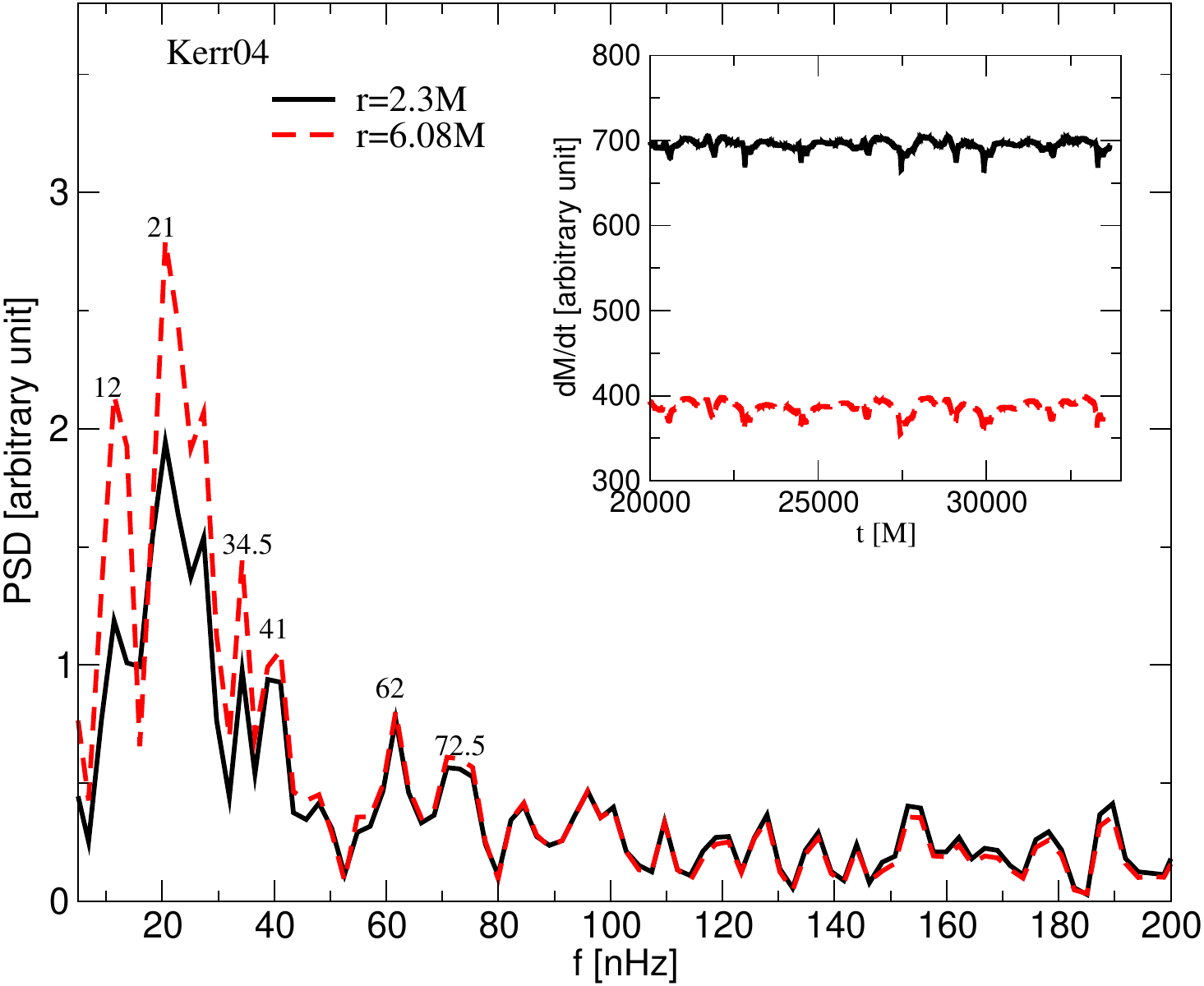,width=7.5cm}\hspace*{0.15cm}
     \psfig{file=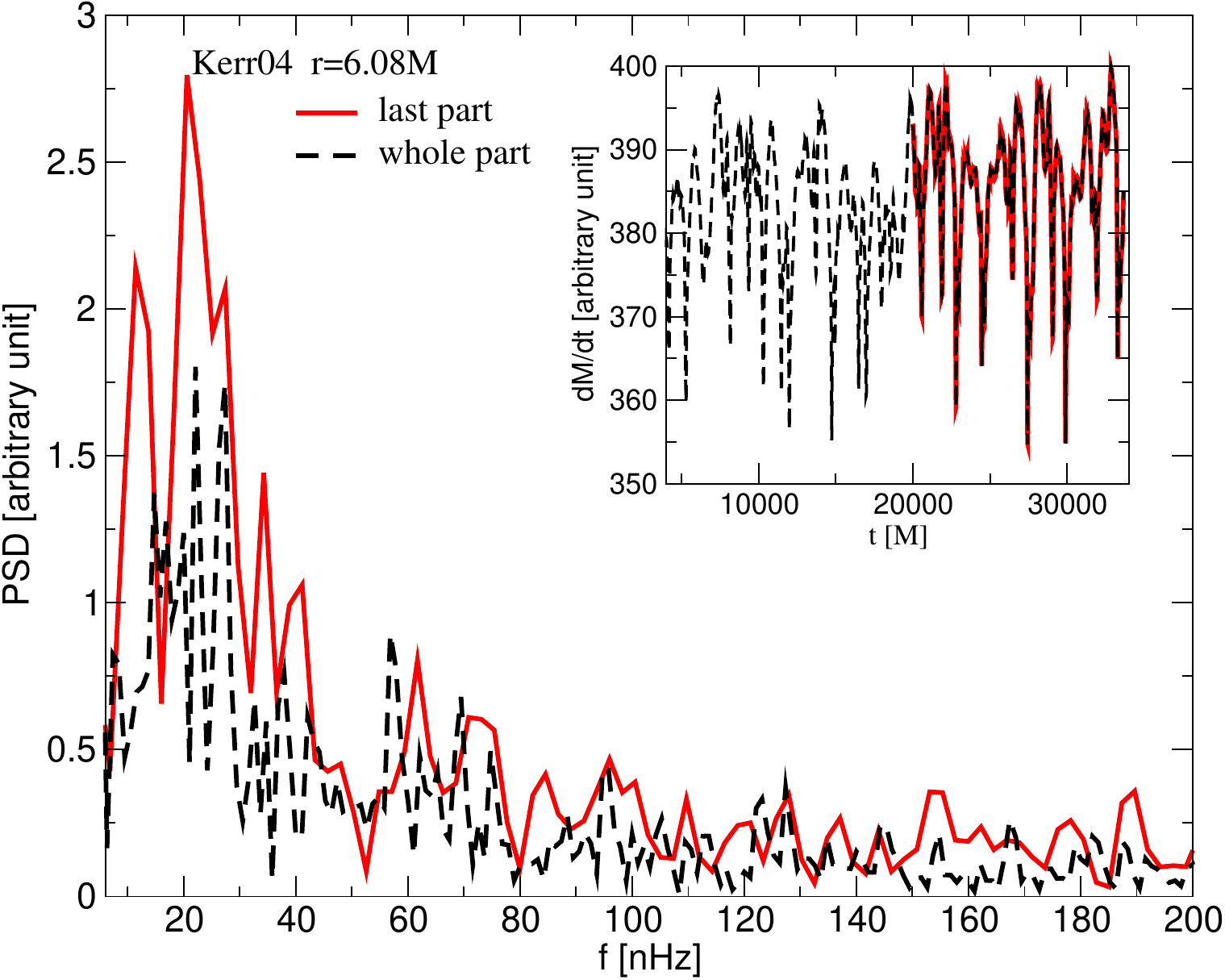,width=7.5cm}\\
     \vspace*{0.25cm}
     \psfig{file=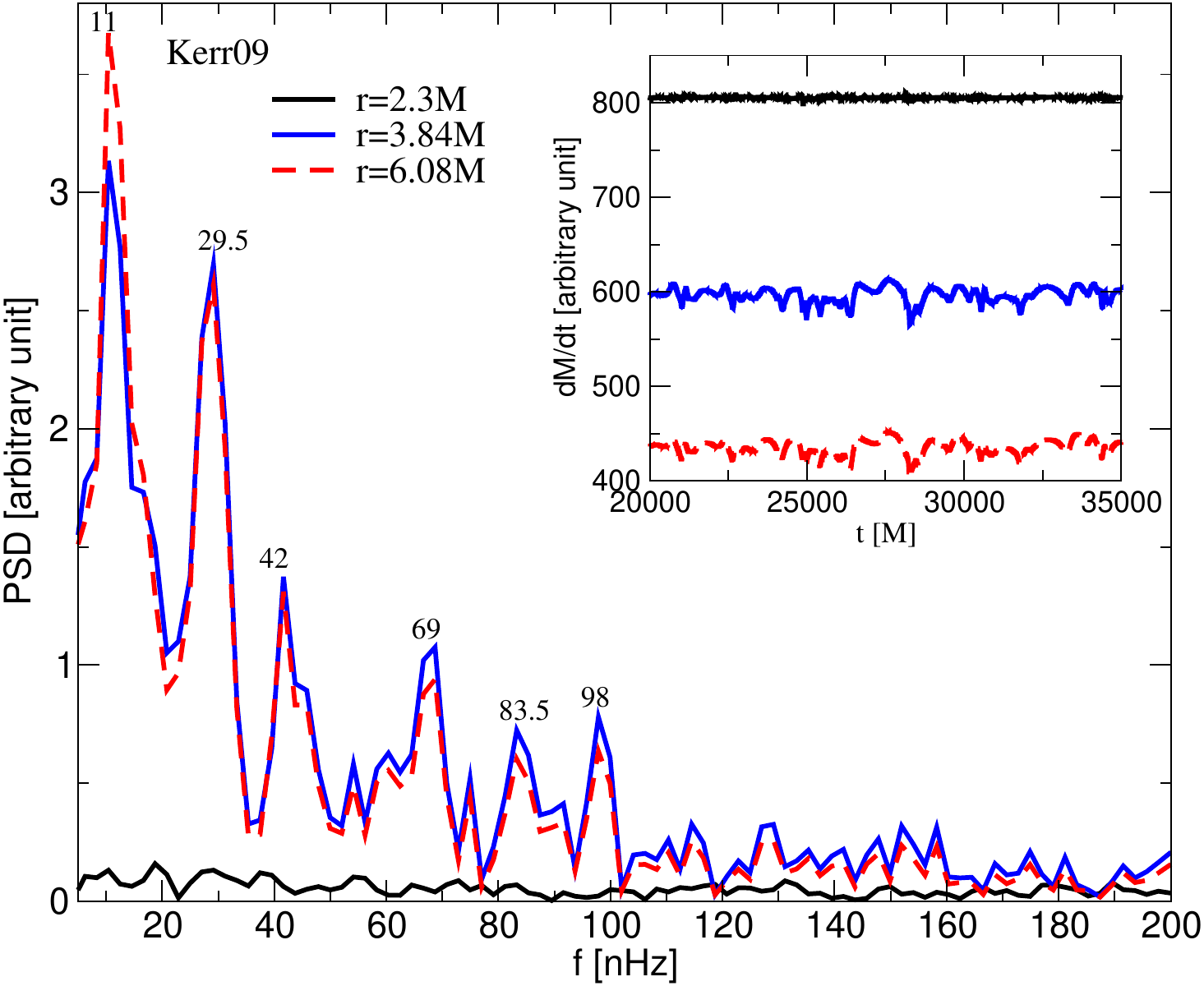,width=7.5cm}\hspace*{0.15cm}
     \psfig{file=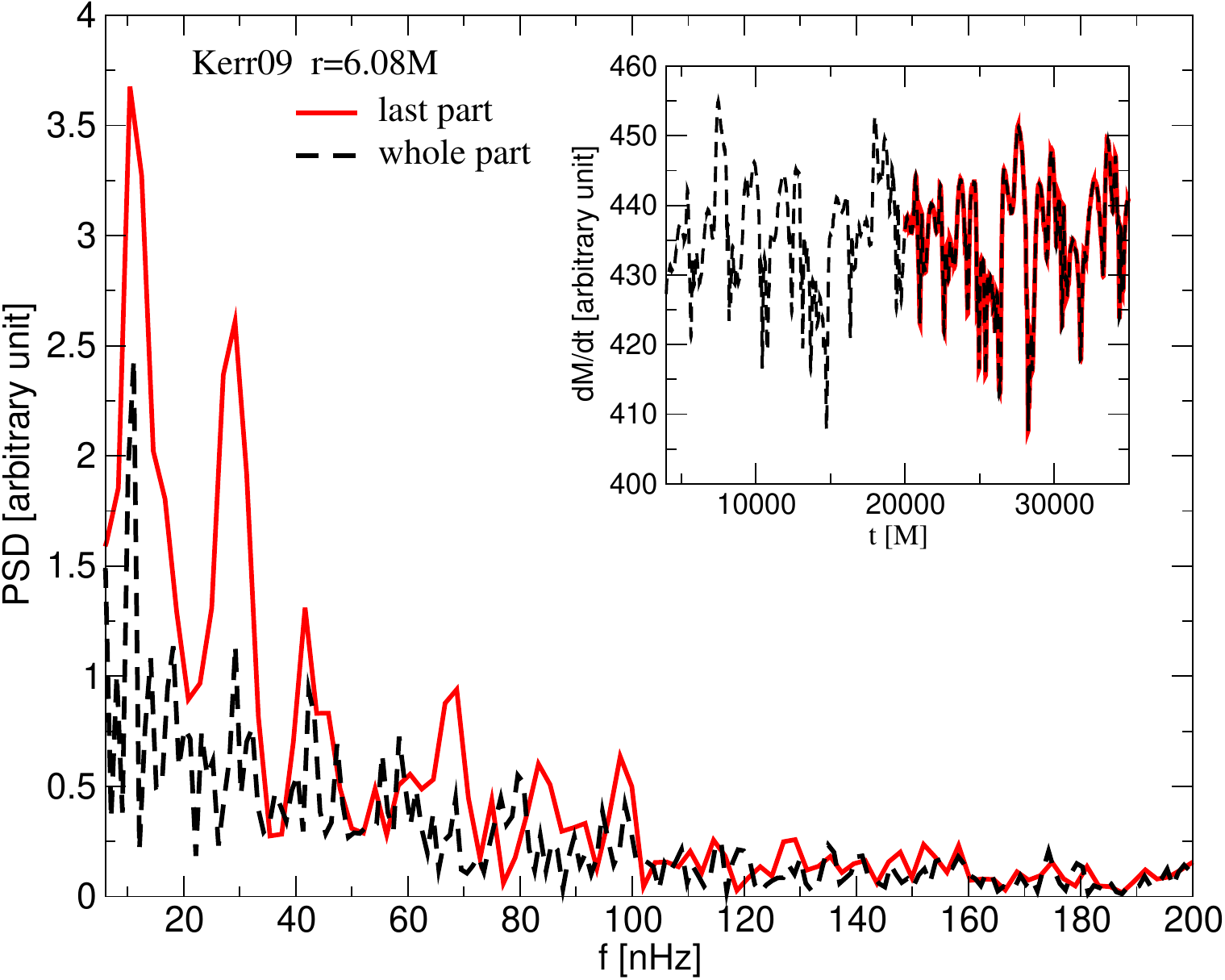,width=7.8cm}\\     
     \caption{For spin parameters $a/M=0.4$ at the top and $a/M=0.9$ at bottom panels, PSD analysis around a Kerr black hole in a strong gravitational field  at $r=2.3M$ and $r=6.08M$. The left column shows PSD analysis results obtained from the mass accretion rates calculated at different radii long after the disk has reached the steady state. The right column compares two PSD analyses from short- and long-duration runs for $r=6.08M$. Thus, it demonstrates the continuity of the resulting QPO frequencies.
    }
\vspace{1cm}
\label{Kerr09}
\end{figure*}

\begin{figure*}
  \vspace{1cm}
  \center
     \psfig{file=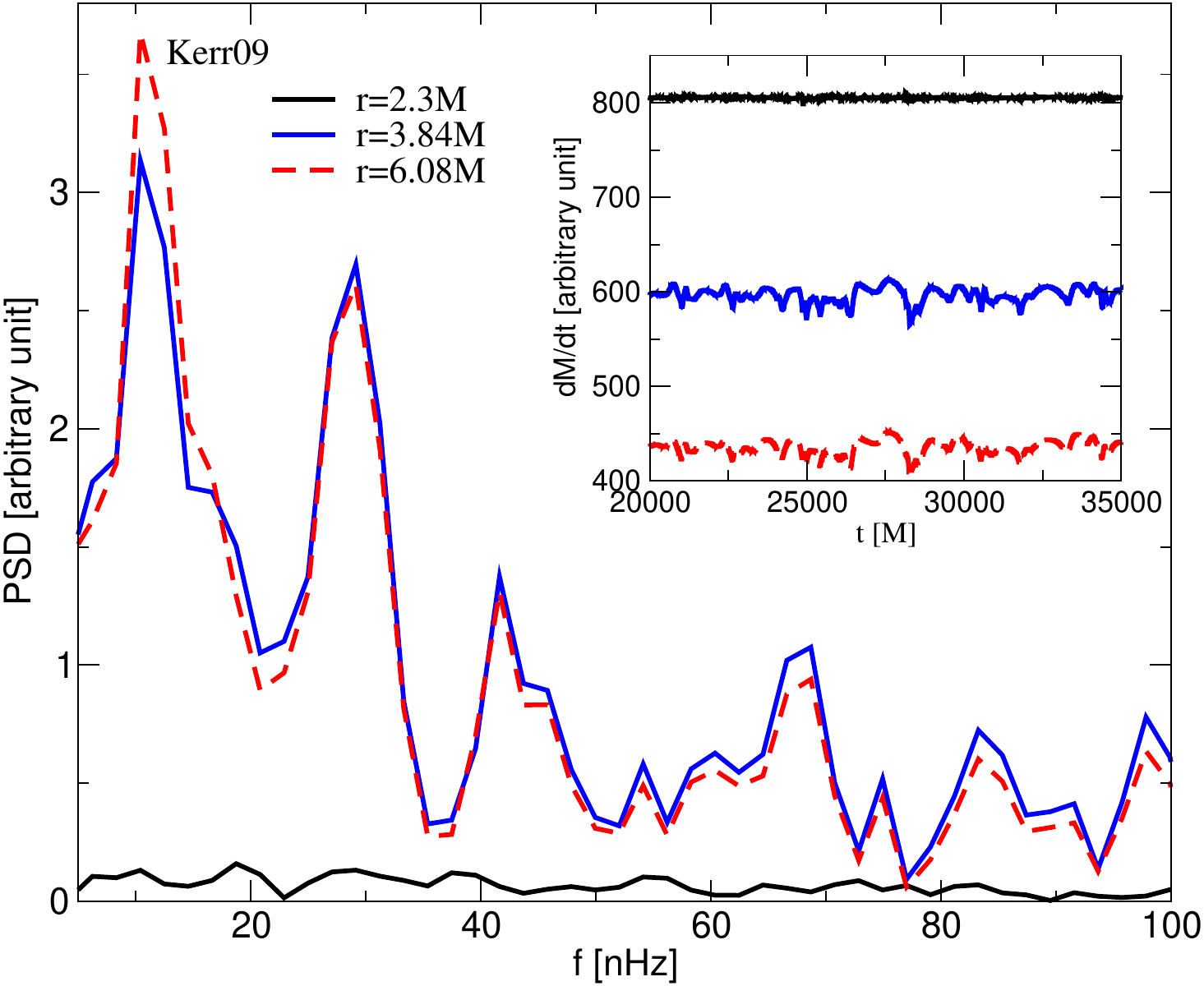,width=3.5cm, height=4.0cm}\hspace*{0.15cm}
     \psfig{file=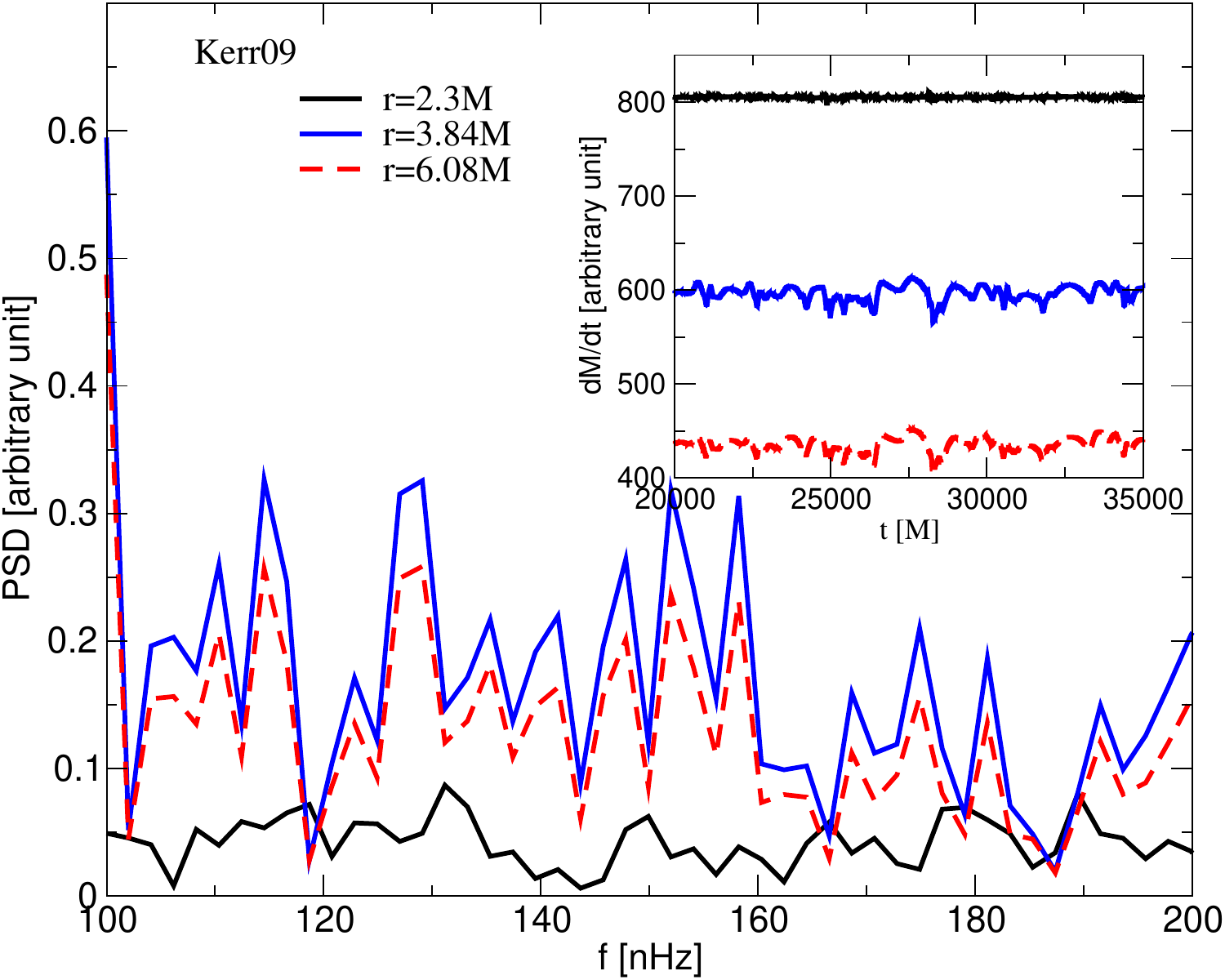,width=3.5cm, height=4.0cm}\hspace*{0.15cm}
     \psfig{file=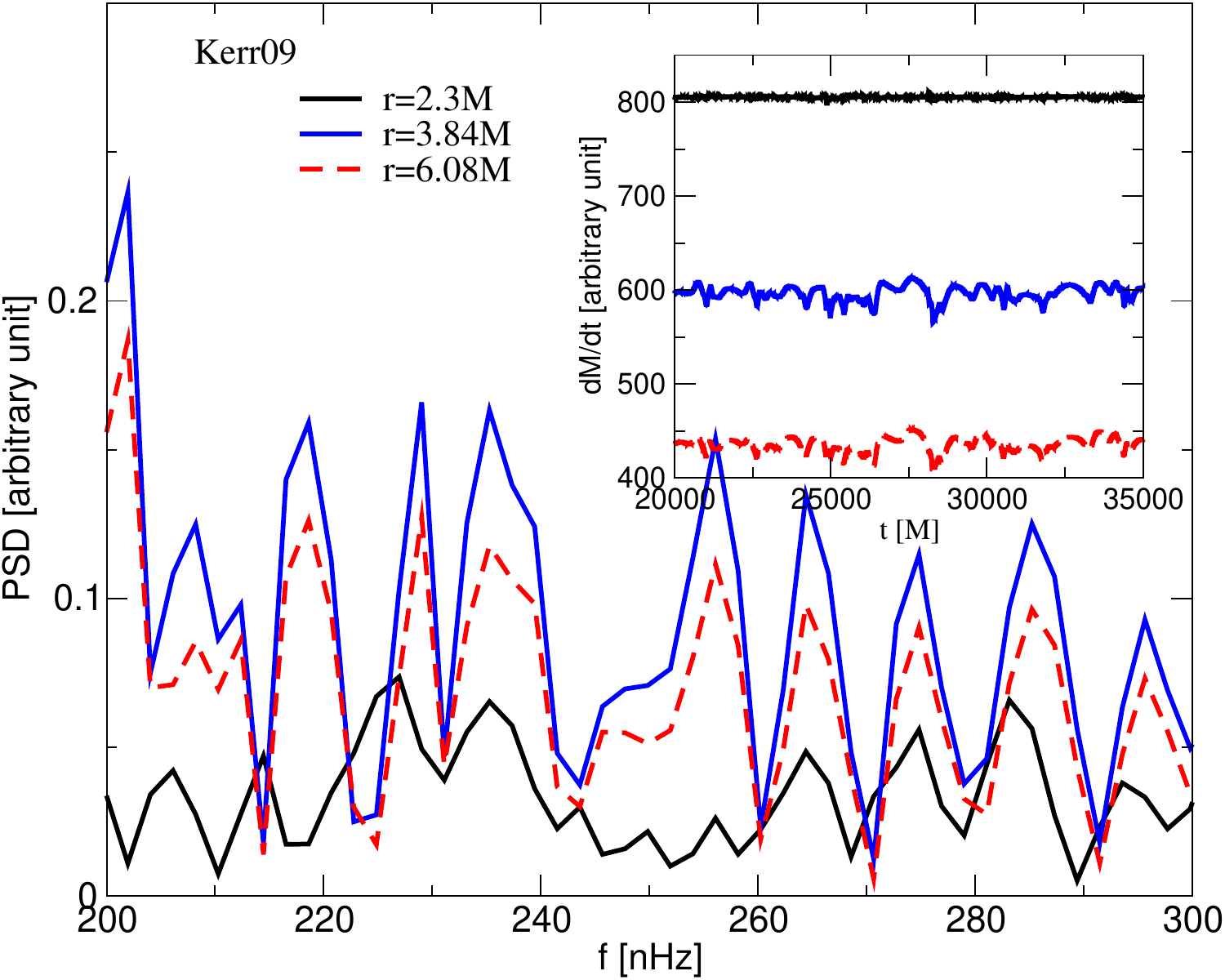,width=3.5cm, height=4.0cm}\hspace*{0.15cm}
     \psfig{file=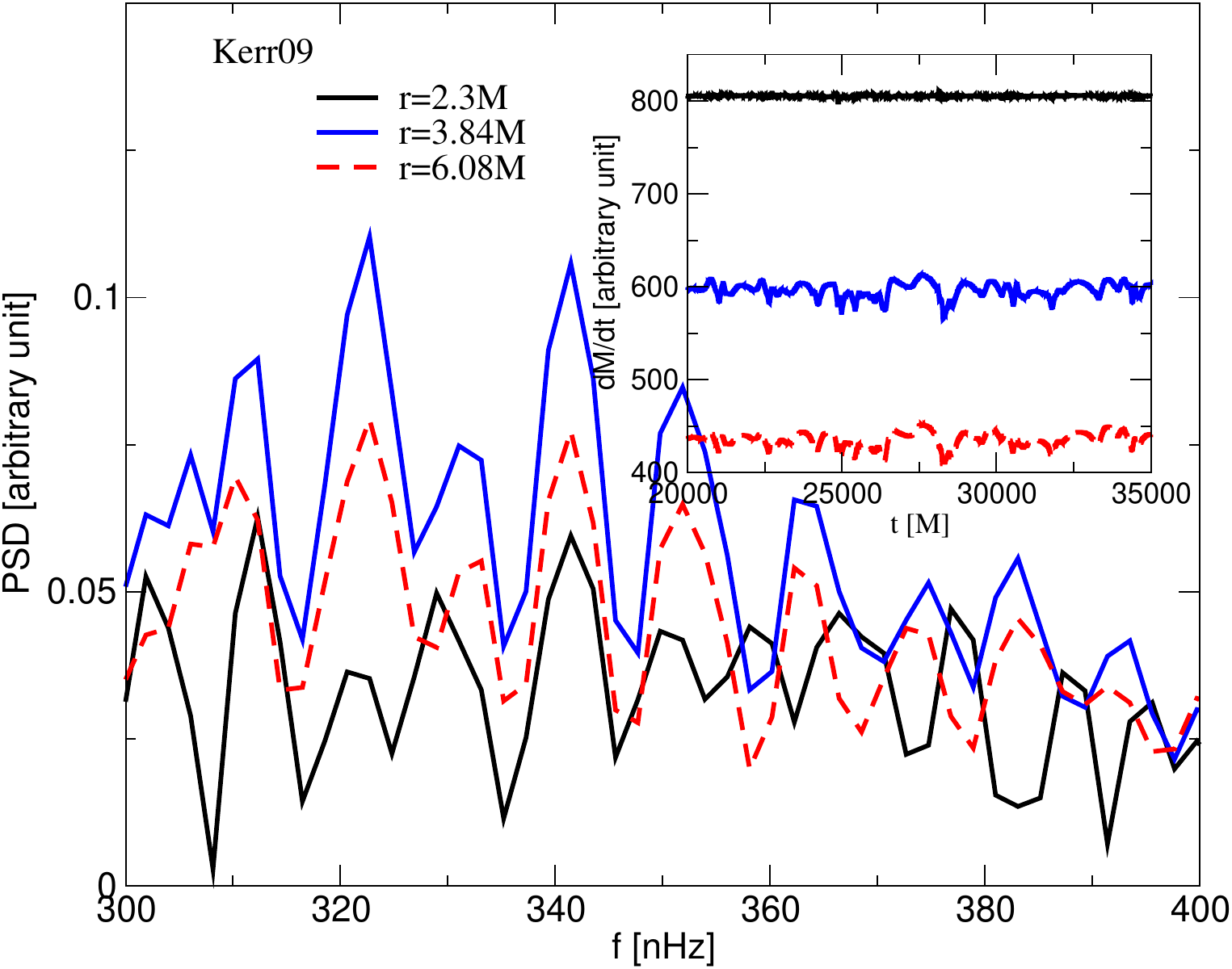,width=3.5cm, height=4.0cm}\\
     \vspace*{0.25cm}
     \psfig{file=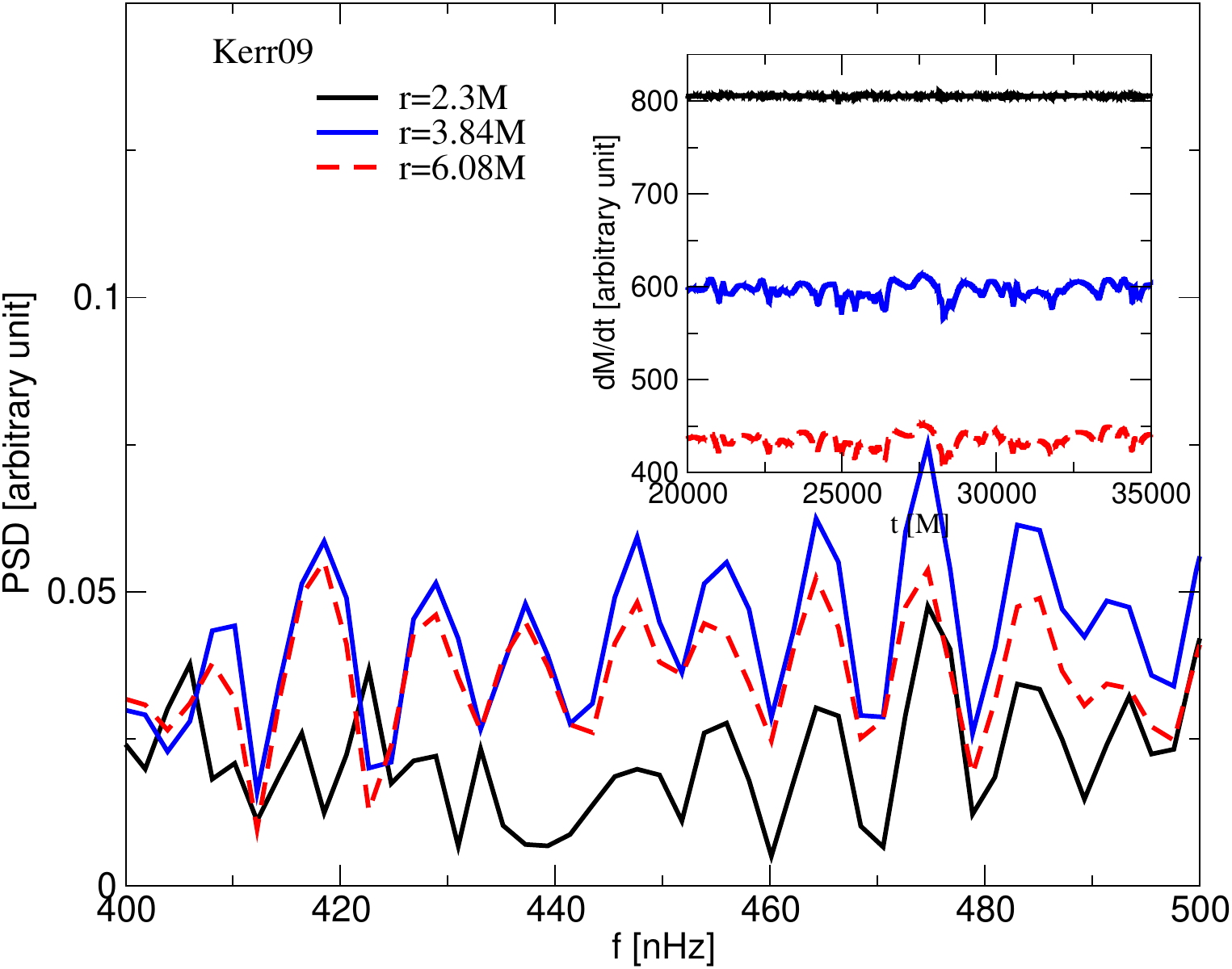,width=3.5cm, height=4.0cm}\hspace*{0.15cm}
     \psfig{file=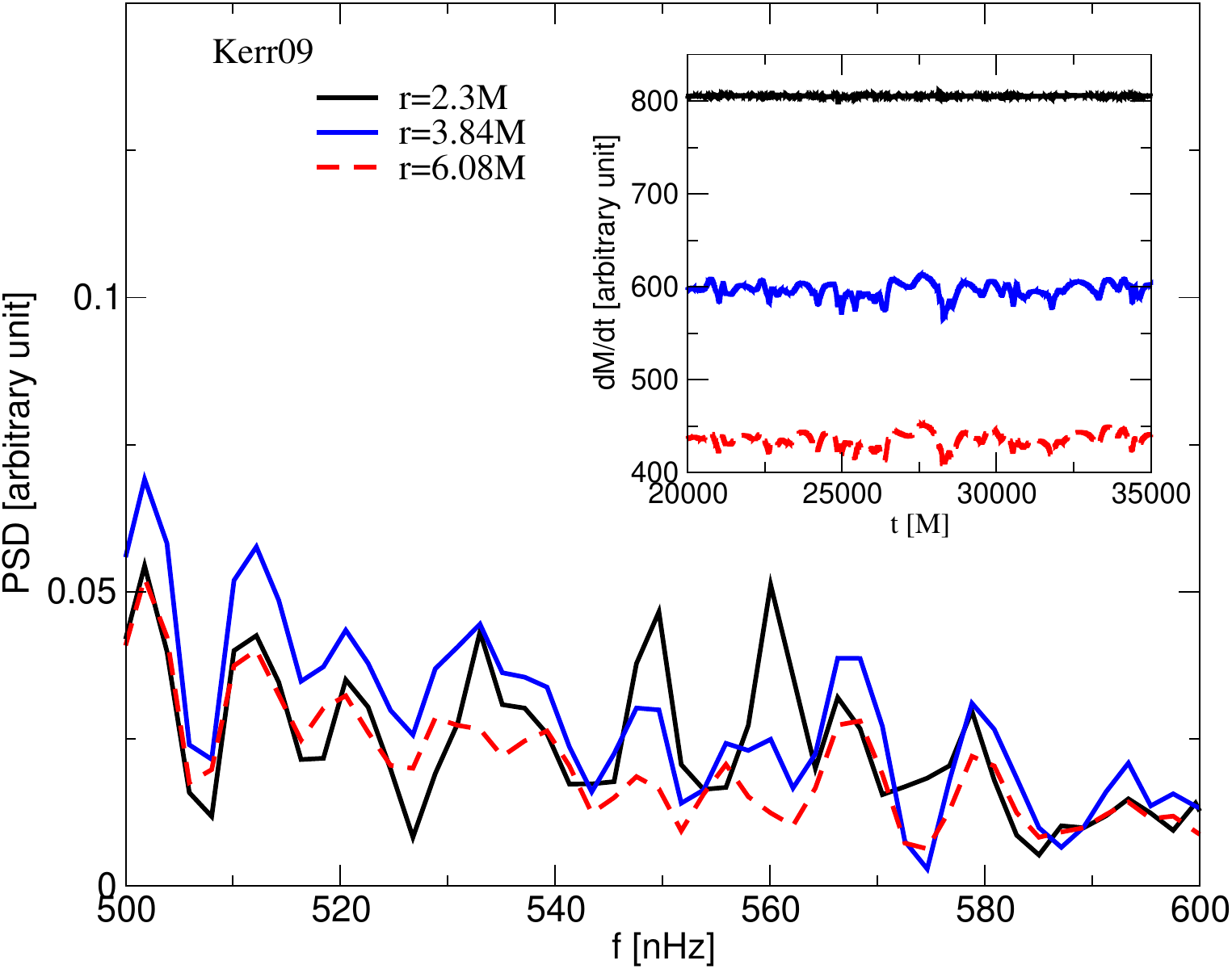,width=3.5cm, height=4.0cm}\hspace*{0.15cm}
     \psfig{file=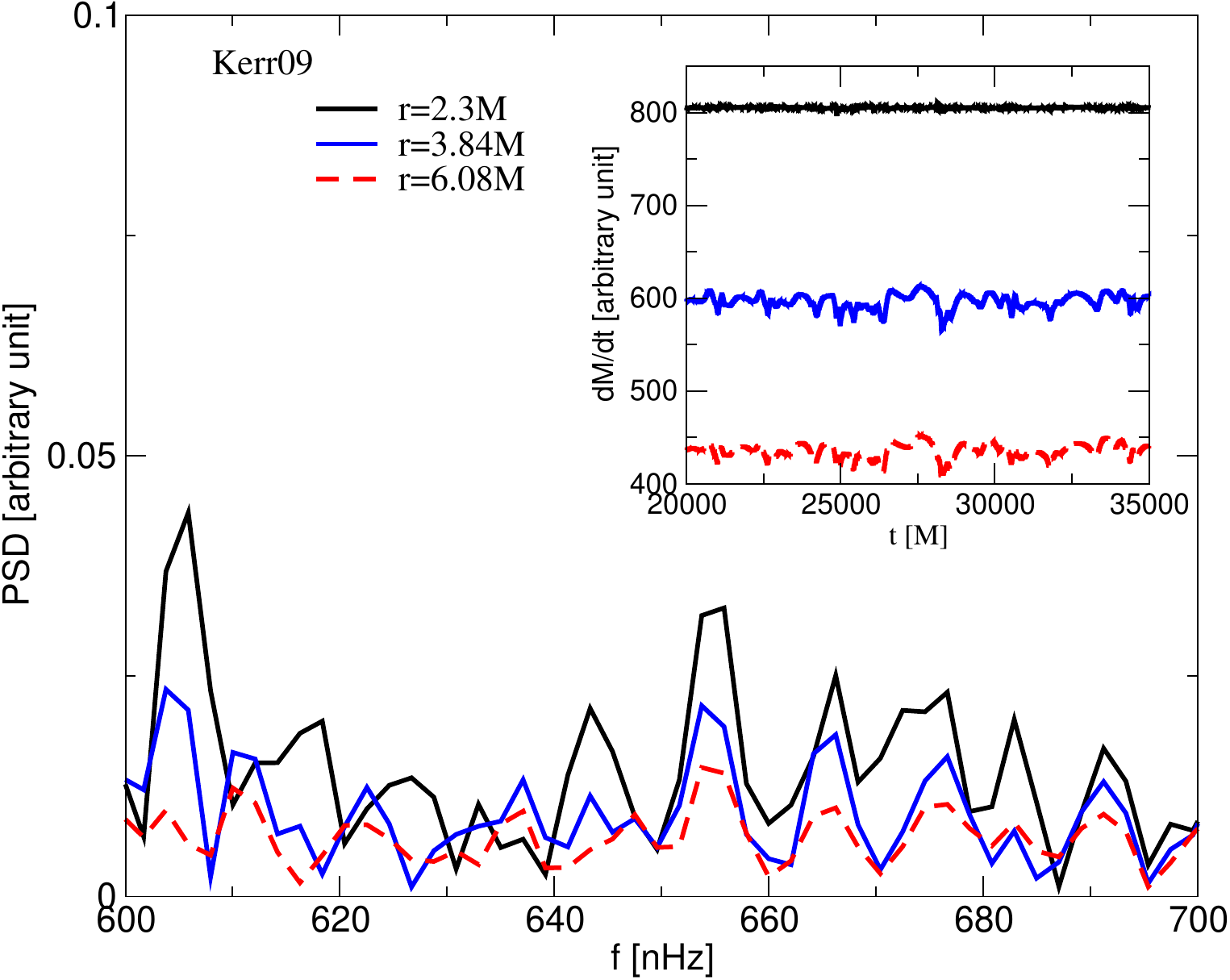,width=3.5cm, height=4.0cm}\hspace*{0.15cm}
     \psfig{file=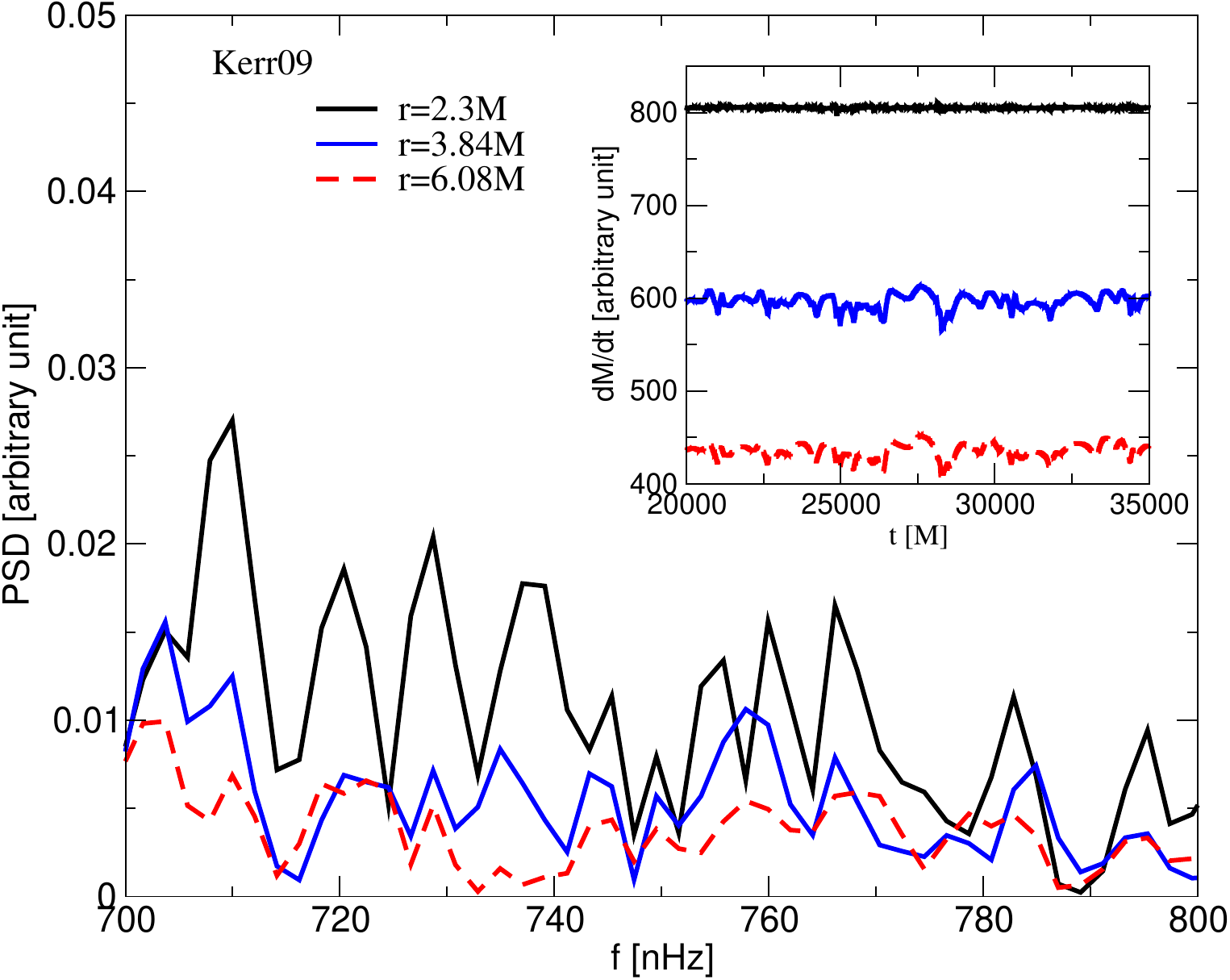,width=3.5cm, height=4.0cm}\\
     \vspace*{0.25cm}
     \psfig{file=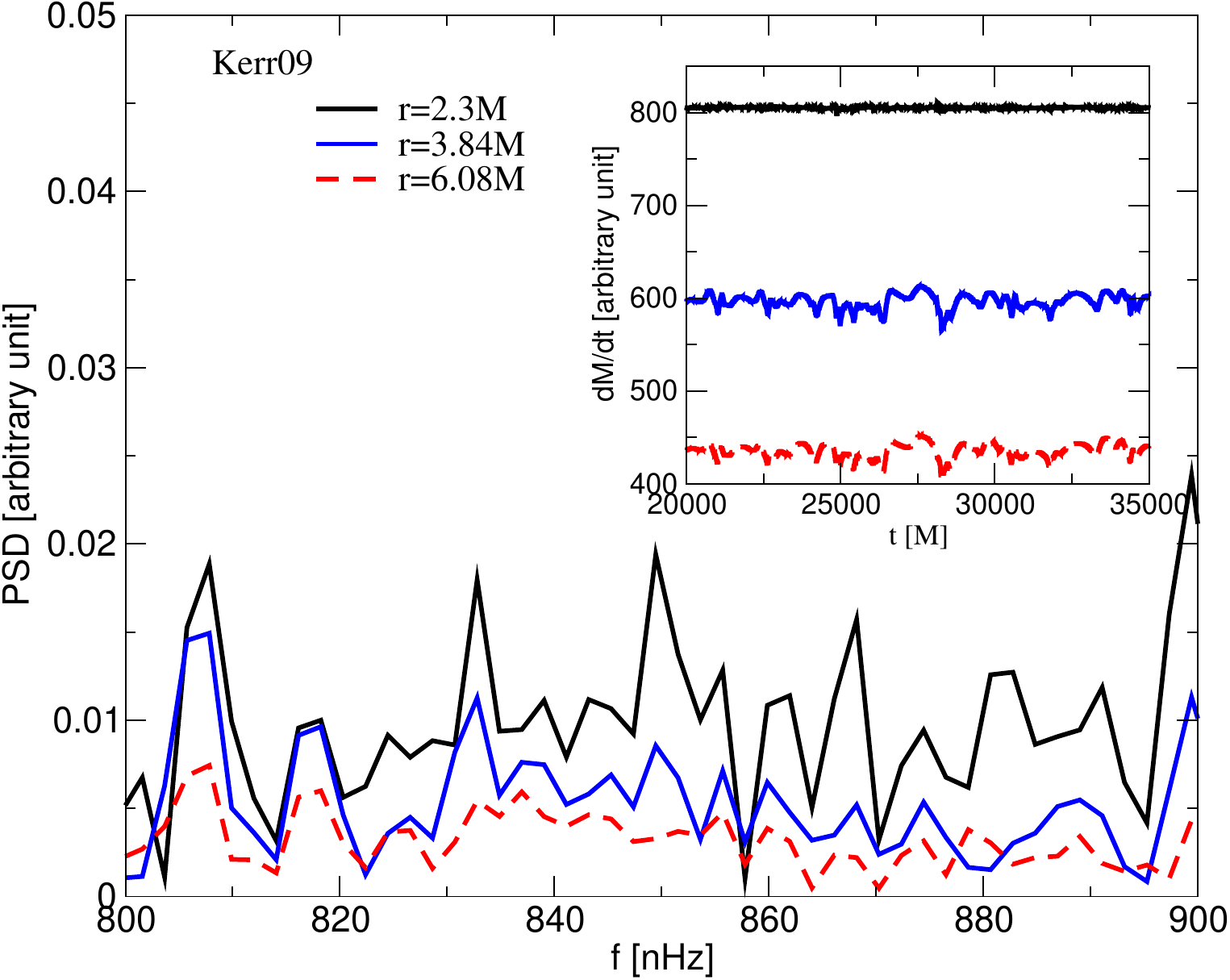, width=3.5cm, height=4.0cm}\hspace*{0.15cm}
     \psfig{file=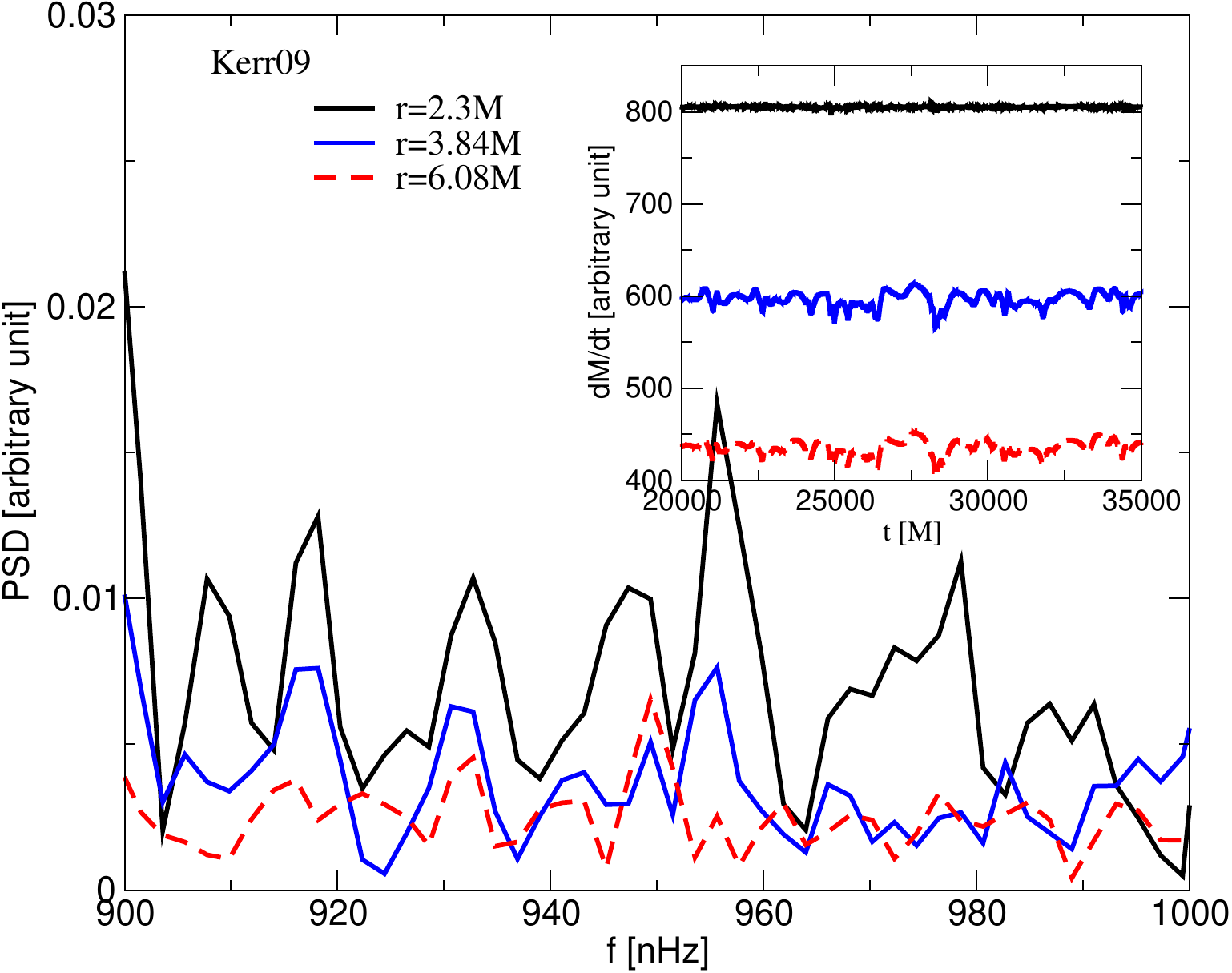,width=3.5cm, height=4.0cm}\hspace*{0.15cm}
     \psfig{file=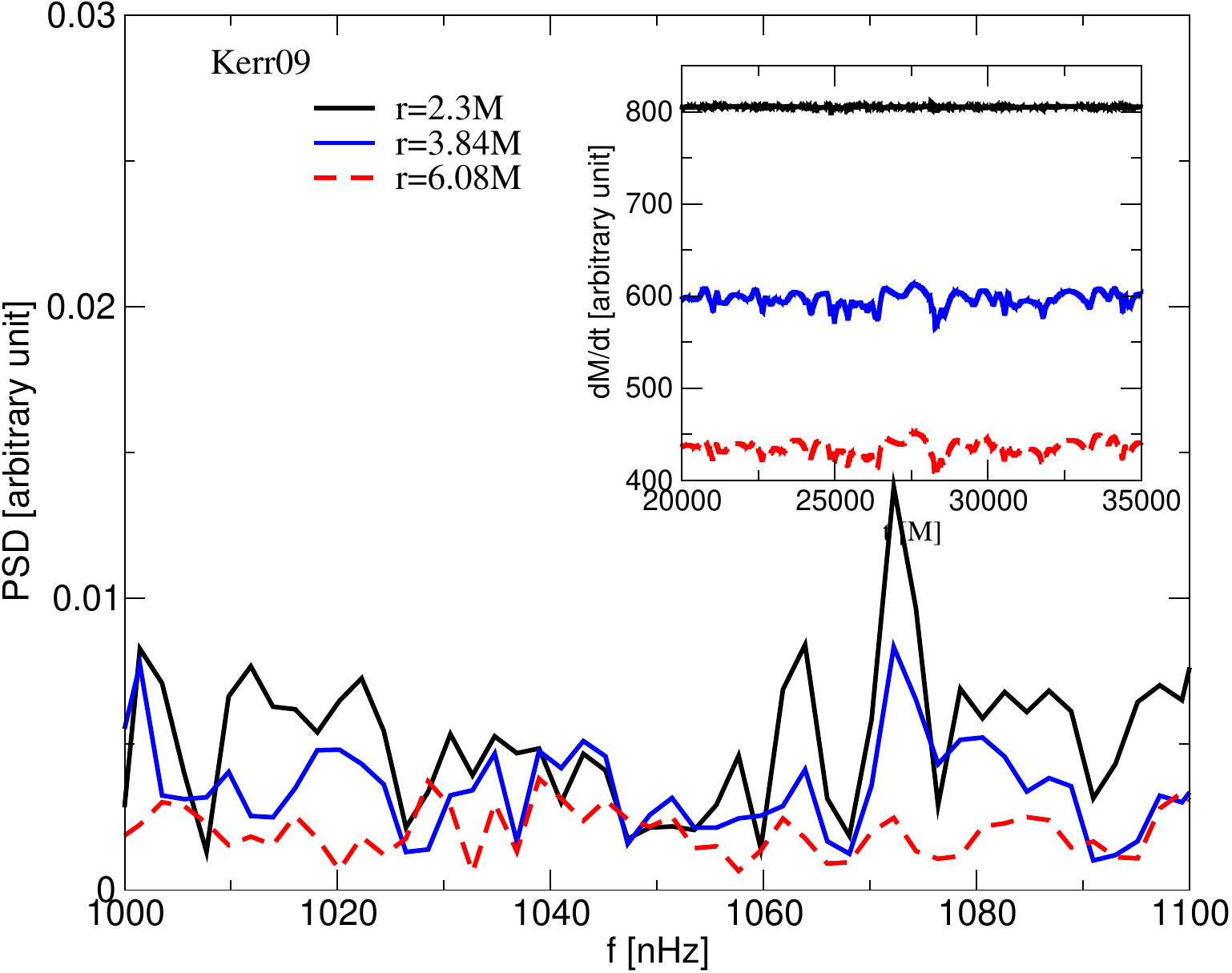,width=3.5cm, height=4.0cm}\hspace*{0.15cm}
     \psfig{file=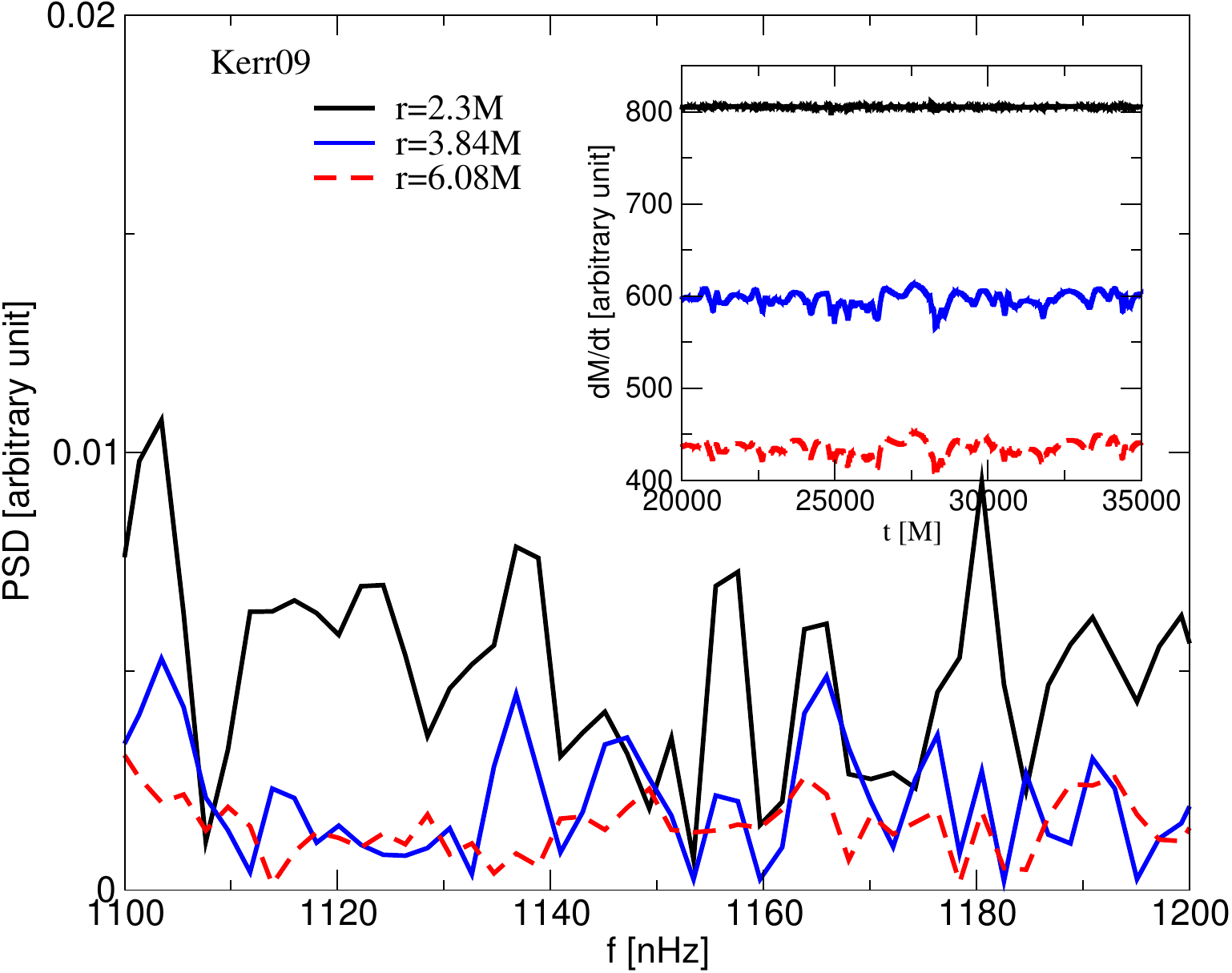,width=3.5cm, height=4.0cm}\\
     \vspace*{0.25cm}
     \psfig{file=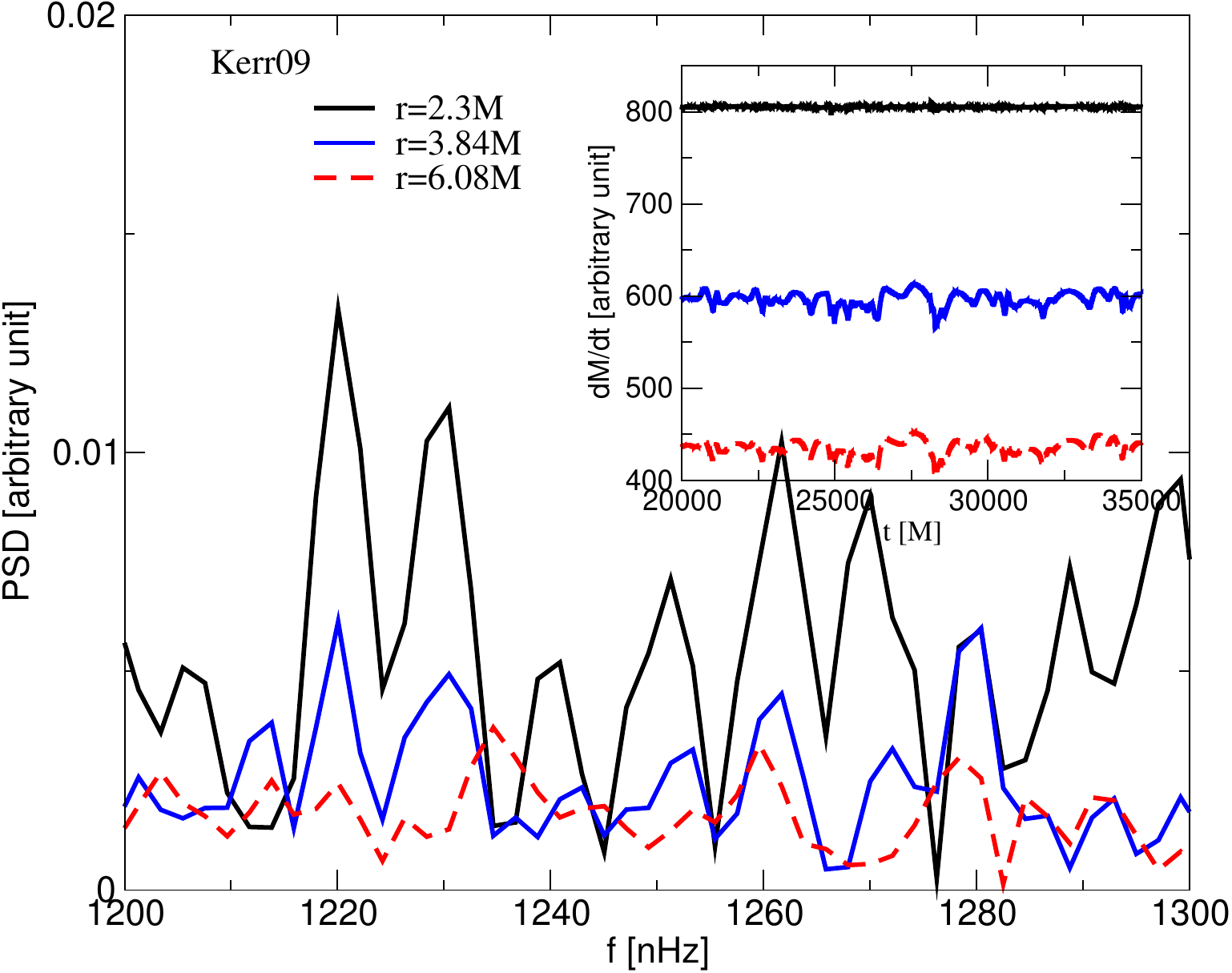,width=3.5cm, height=4.0cm}\hspace*{0.15cm}
     \psfig{file=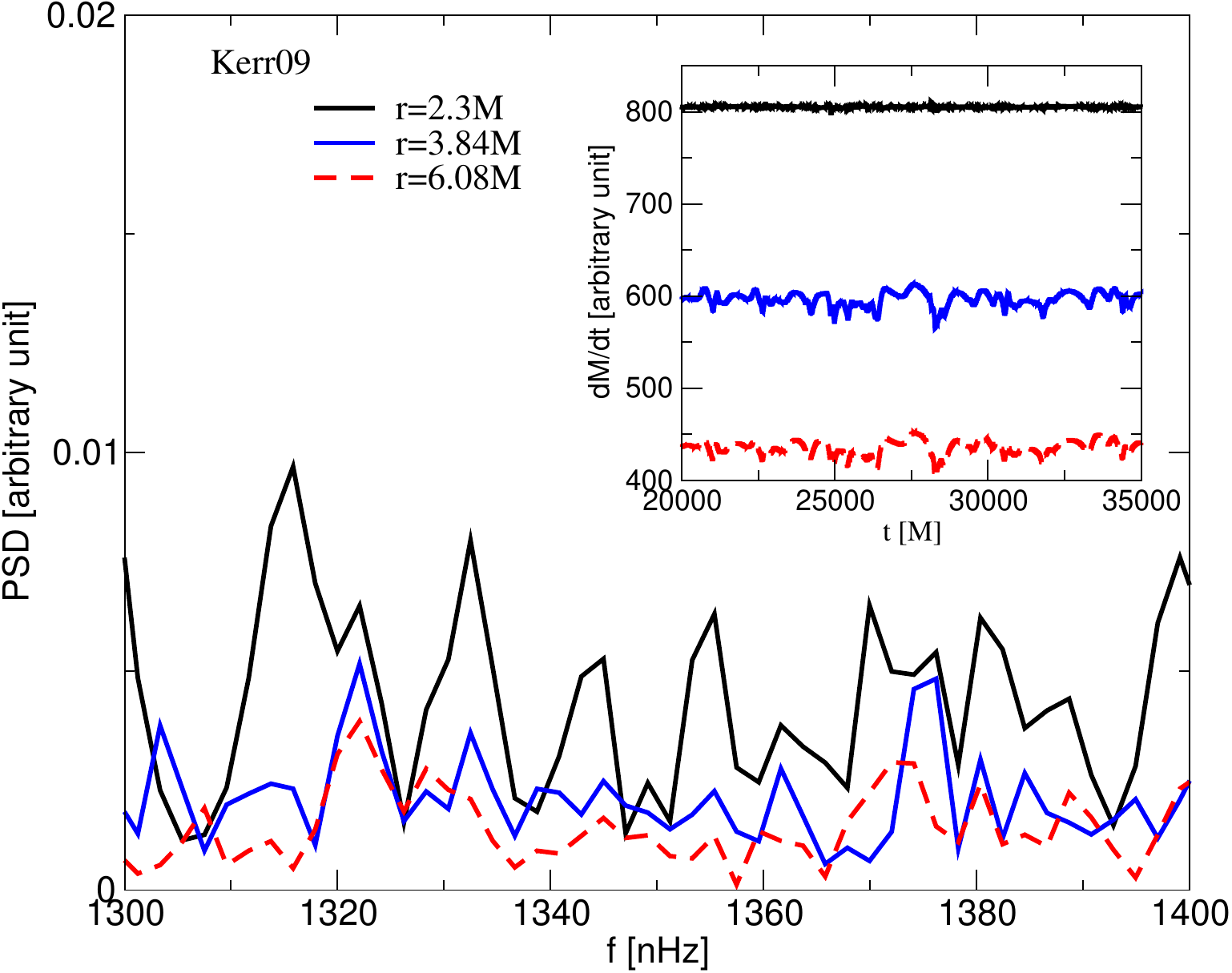,width=3.5cm, height=4.0cm}\hspace*{0.15cm}
     \psfig{file=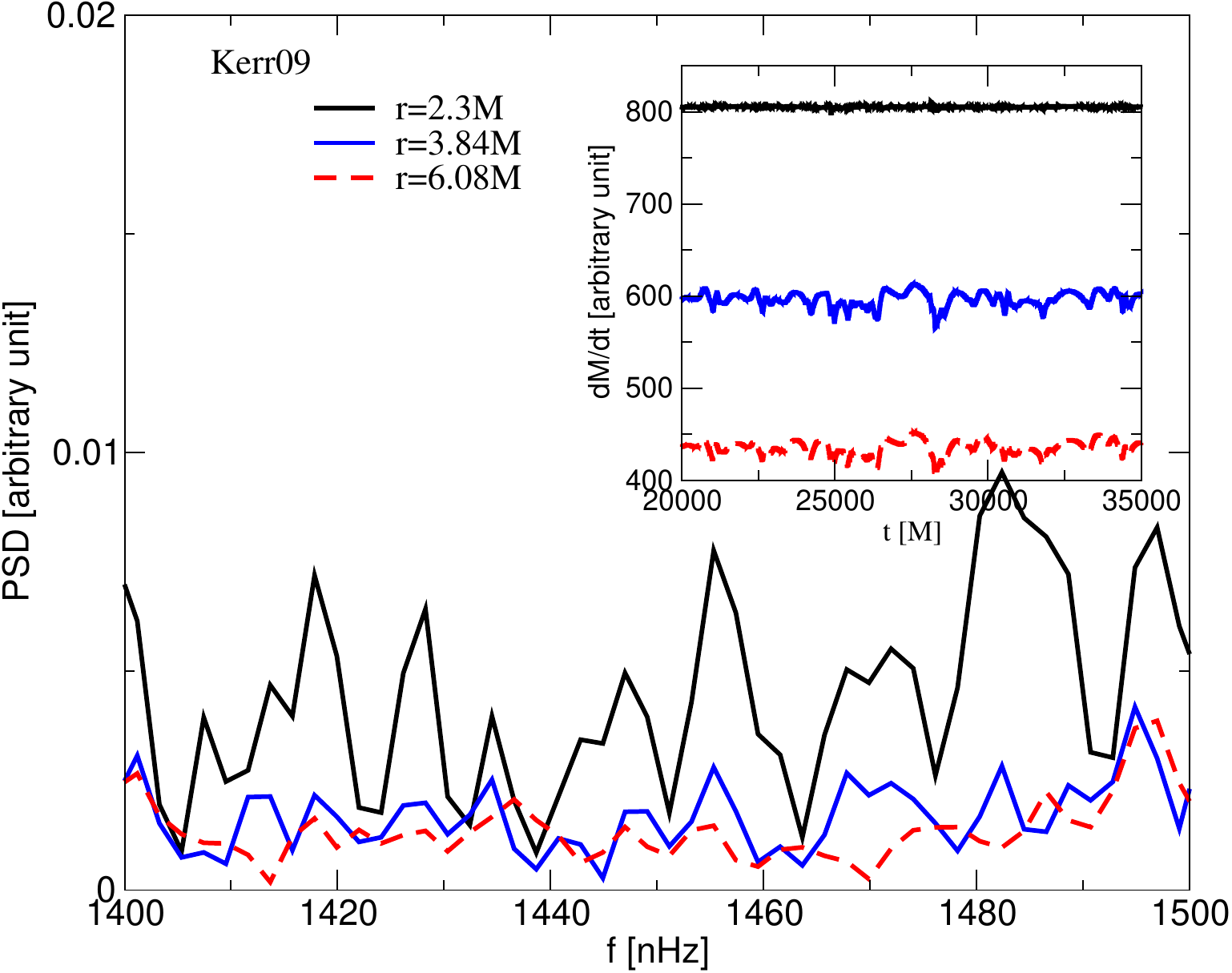,width=3.5cm, height=4.0cm}\hspace*{0.15cm}
     \psfig{file=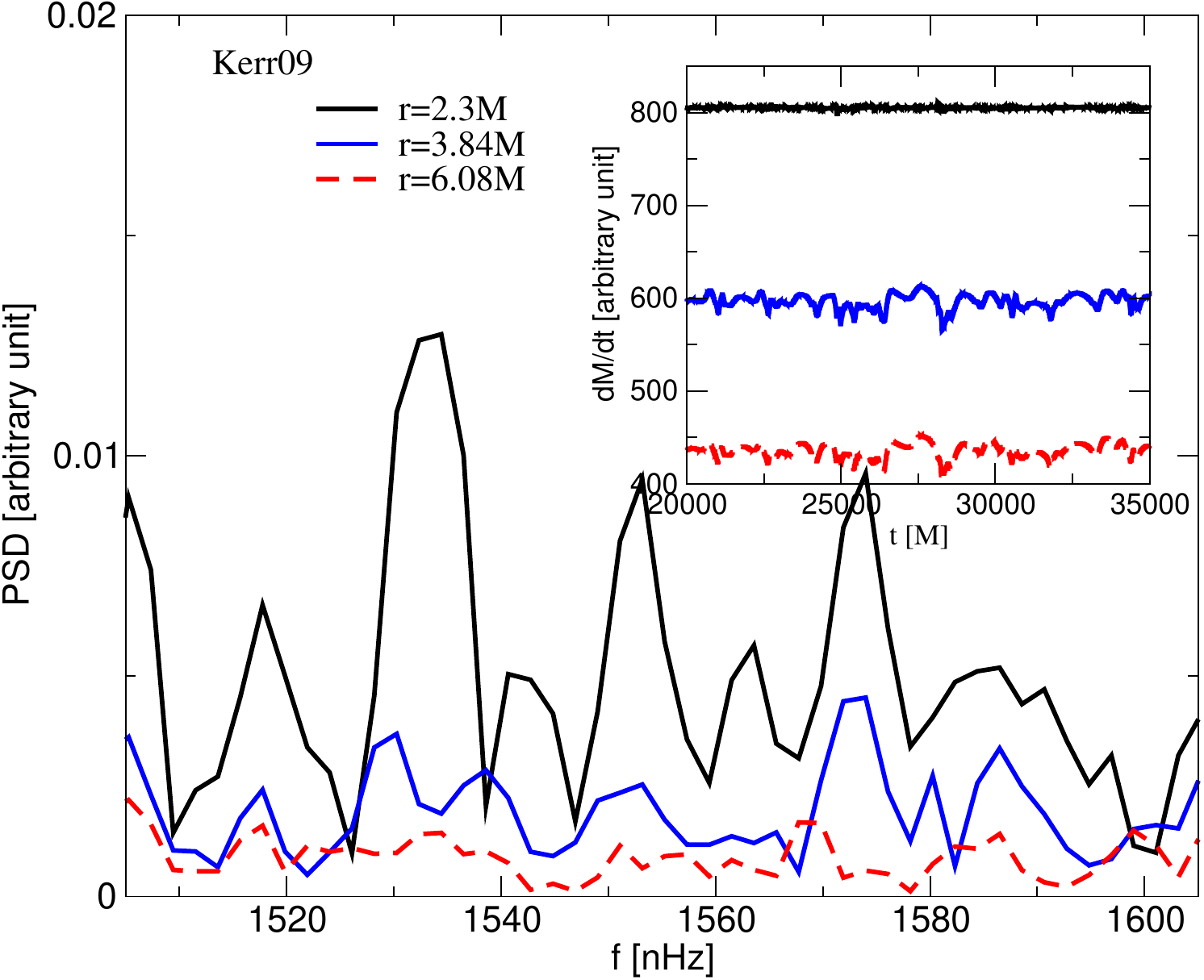,width=3.5cm, height=4.0cm}\\
     \vspace*{0.25cm}     
    \caption{Same as Fig.\ref{Kerr09}, but PSD analysis for model Kerr09 at different frequency ranges.
    }
\vspace{1cm}
\label{Kerr09_Diff_window}
\end{figure*}

In Fig.\ref{Kerr09}, the numerical simulations have identified QPO frequencies around the $M87^*$ Kerr black hole for two different spin parameters, $a/M=0.4$ and $a/M=0.9$. As shown in the left column of Fig.\ref{Kerr09}, low-frequency QPOs in the nHz range are observed for both spin parameters, with the lowest QPO frequency around 11 nHz. Figure \ref{Kerr09_Diff_window} shows that these frequencies also occur in the $\mu$ Hz range, indicating that both low and high-frequency QPOs can be observed from this source. The right columns show that the QPO frequencies are persistent for both spin parameters. PSD analysis for long and short-term data confirms that the same frequencies occur in both cases, indicating that these frequencies are persistent. Additionally, as shown in Fig.\ref{Kerr09} for $a/M=0.4$, the same frequencies with the same amplitude are observed at $r=2.3M$ and $r=6.08M$. However, for $a/M=0.9$, although peaks occur at the same frequency, the amplitude of the frequency at $r=2.3M$ is very low, making it less observable due to the suppression of low QPO frequencies by the Lense-Thirring effect. However, as shown in Fig.\ref{Kerr09_Diff_window},
the modes at  $r=2.3M$ become more observable and prominent in the high-frequency bands in the $\mu$ Hz range, which is consistent with the frequencies resulting from the Lense-Thirring effect discussed in Section \ref{Theory_prediction}.
Also, as seen in the left column of Fig.\ref{Kerr09}, although the effect of the rapidly rotating black hole’s spacetime warping on QPO frequencies is very pronounced at $r=2.3M$, this effect is not seen at $r=3.8M$ and $r=6.08M$. Therefore, to conclude that the observed frequency is a result of the Lense-Thirring effect, these emissions must occur within a strong gravitational field and in close proximity to the black hole's horizon. In addition to the Lense-Thirring effect, QPO frequencies are generated by the radial and angular oscillations of pressure modes trapped within the shock cone. Numerical simulations show that at least three fundamental mode frequencies occur. Other observed peaks in Figures \ref{Kerr09} and \ref{Kerr09_Diff_window} are known to result from the nonlinear couplings of the fundamental modes \cite{Rezzolla2003MNRAS}. For example, according to our estimates, the frequencies of 11 nHz and 29.5 nHz appeared in the rapidly rotating black hole $a/M=0.9$ case are fundamental frequencies created by pressure modes trapped within the shock cone. Their couplings with each other and with newly formed QPO frequencies have resulted in the creation of new frequencies. These include 11+29.5 $\approx$  42, 11 + 2$\times$ 29.2 $\approx$ 69, 29.5+42+11 $\approx$ 83.5, etc.

In Fig.\ref{Kerr09_Diff_window}, PSD analyses for different frequency bands are presented for the rapidly rotating black hole at $a/M=0.9$. As shown in this figure, while the frequency window is between $0-100$ nHz , the amplitude of the peak at $r=2.3M$ is very small compared to other locations. The reason for this, as explained in Section \ref{Theory_prediction}, is that in a strong gravitational field, particularly around rapidly rotating black holes, the Lense-Thirring effect is strong, leading to the suppression of pressure modes. As a result, the amplitudes of the QPO peaks at the point  $r=2.3M$  are significantly smaller compared to the peaks formed at locations farther from the strong gravitational field. However, it has been observed that the amplitude of the QPO frequency at $r=2.3M$  increases in the high-frequency band intervals. In the $500-600$ nHz  band, the amplitude of the frequencies at $r=2.3M$  is the same as the frequencies at the other points. Therefore, frequencies resulting from the Lense-Thirring effect are much higher when compared to those generated by pressure modes. Thus, in QPO observations of the $M87^*$ black hole, if different low and high frequencies are observed, low frequencies are generated by the excitation of physical mechanisms around the black hole, while high frequencies are QPOs formed by the spacetime warping around the rapidly rotating black hole."

  Lastly, as shown at  $a/M=0.4$ in Fig.\ref{Kerr09}, when the black hole rotates at a slow to moderate angular momentum, the effect of the Lense-Thirring on QPO oscillations is not very significant. Therefore, the peaks at each $r$ point have the same amplitude. However, the Lense-Thirring effect becomes prominent in the strong gravitational field at $a/M=0.9$. As previously discussed and seen in Fig.\ref{Kerr09_Diff_window}, while the QPO frequency amplitudes at $r=2.3M$ are low at lower frequencies, as we move towards higher frequency bands, similar QPO frequency amplitudes are observed at all $r$ points. At even higher frequency bands, i.e., after 1000 nHz, we can say that the Lense-Thirring effect has a dominant influence on the observable QPOs. This is because the QPO frequency amplitudes at $r=2.3M$ are more pronounced compared to those at other points in the strong gravitational field. Therefore, numerical models predict that the highest observable QPO frequency for the $M87^*$ black hole using Kerr gravity could be at the $\mu$ Hz level.


\subsection{QPO frequencies of $M87^*$ from the observed shadow values given in Table \ref{Inital_Con_1}}
\label{Num_QPO1}

In this section, using the parameters of the hairy Kerr black hole consistent with the observed shadow of the $M87^*$ black hole as calculated in Ref.\citep{Afrin_2021} and given in Table \ref{Inital_Con_1}, we reveal the structure and behavior of the shock cone formed around this black hole through BHL accretion. The parameters of the hairy Kerr black hole are calculated using the observed radius of the photon sphere $R_s$ and the deviation of the left edge of the shadow from the circle boundary $\delta_s$. Additionally, we numerically calculate the resulting QPO frequencies. Thus, we provide a prediction regarding the QPOs that might be observed in the strong gravitational field of the $M87^*$ black hole in the future.

The formation of a shock cone around the black hole through BHL accretion and the revelation of the Lense-Thirring effect caused by the warping of spacetime in the strong gravitational field significantly impacts the formation and excitation of QPOs, as well as contribute to the explanation of many physical phenomena. As shown in Fig.\ref{Color_HKA05} even though we place the inner radius of the computational domain at $r=2.3M$, very close to the black hole's horizon, the inner boundary of the resulting shock cone is also located at the same place with  the computational domain. This confirms that the shock cone is an important physical mechanism in revealing the effects of the strong gravitational field near the black hole's horizon.

\begin{figure*}
  \vspace{1cm}
  \center
  \psfig{file=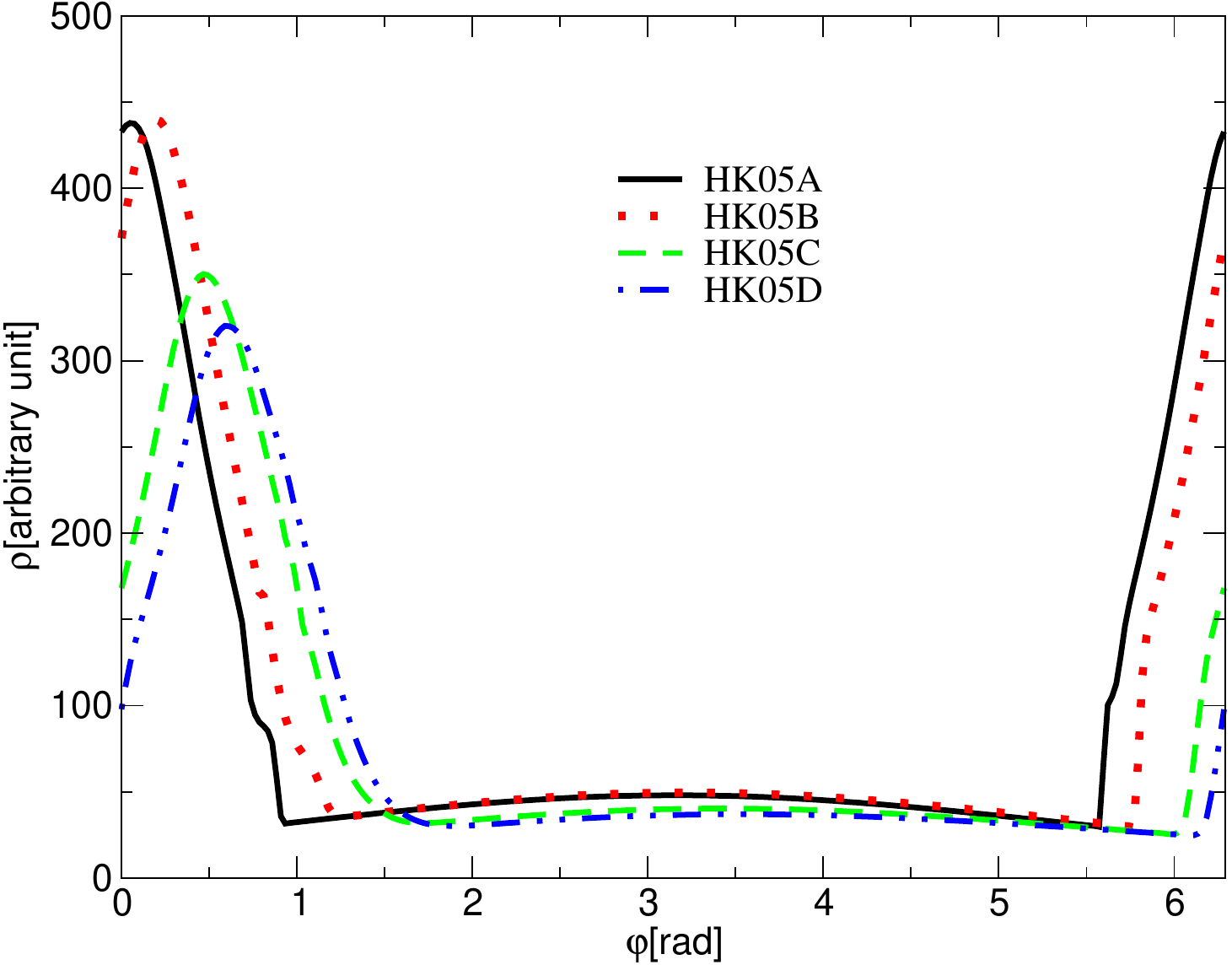,width=7.0cm, height=6.0cm}\hspace*{0.15cm}
  \psfig{file=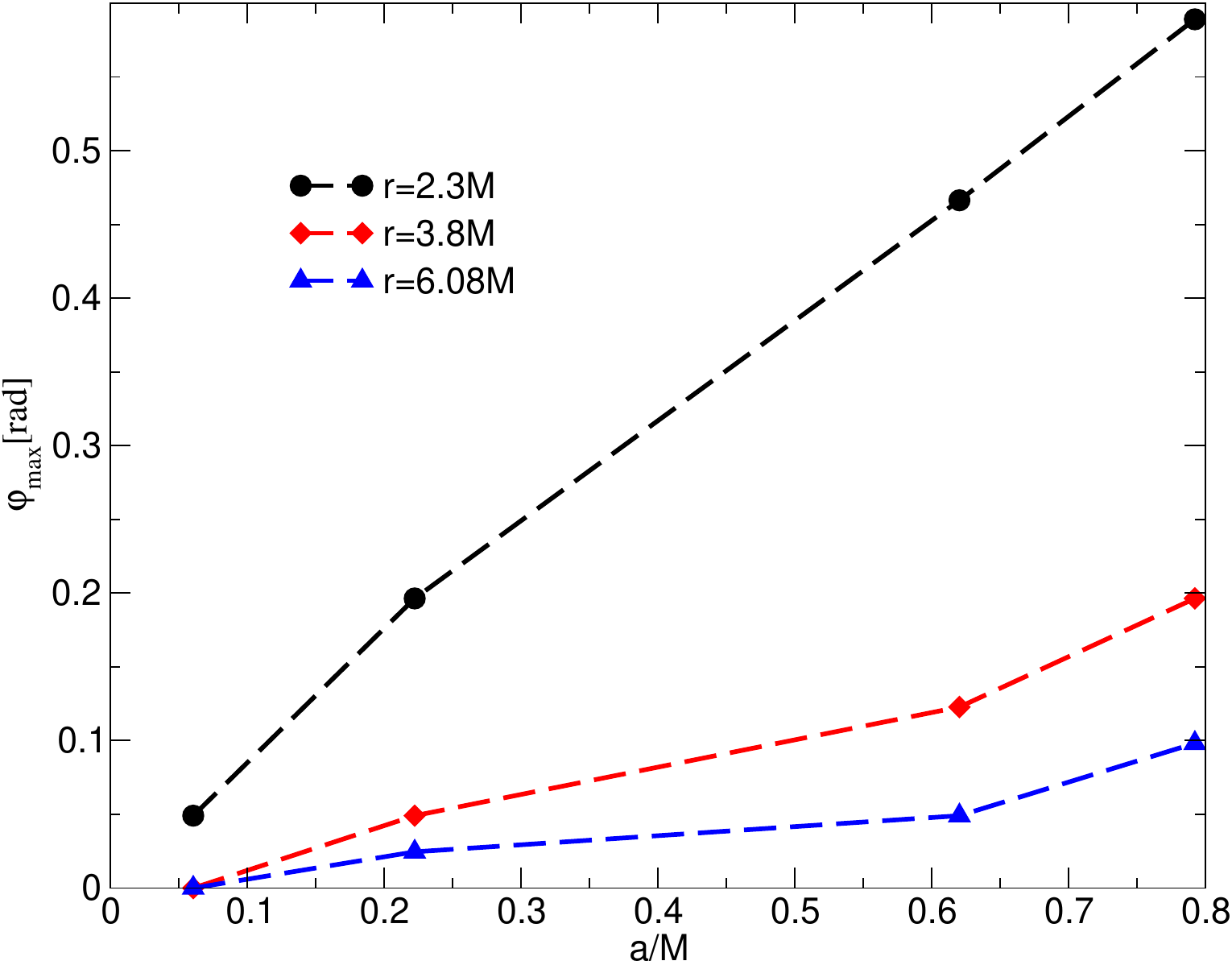,width=7.0cm, height=6.0cm}\\
 \vspace*{0.5cm}     
  \psfig{file=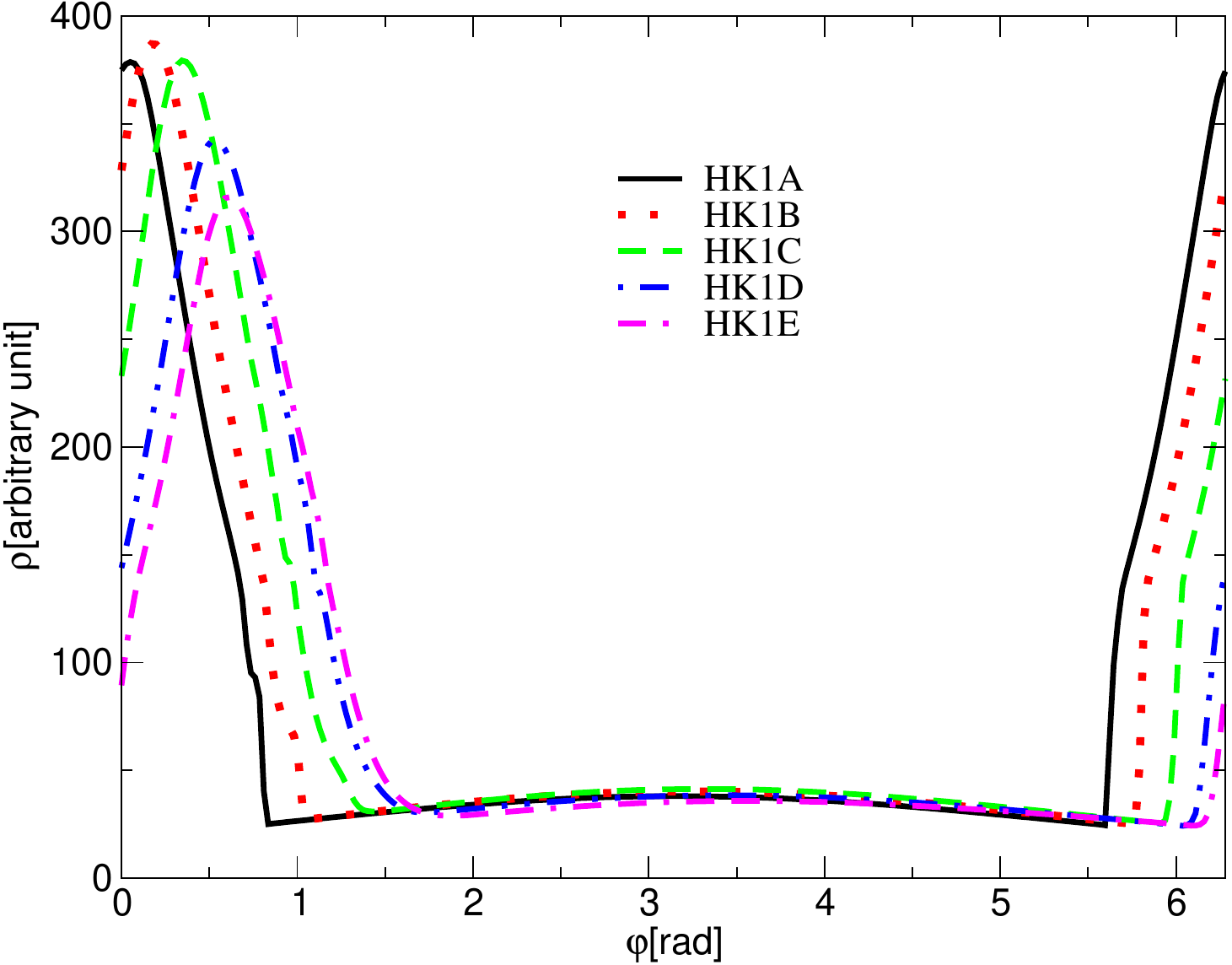, width=7.0cm, height=6.0cm}\hspace*{0.15cm}
  \psfig{file=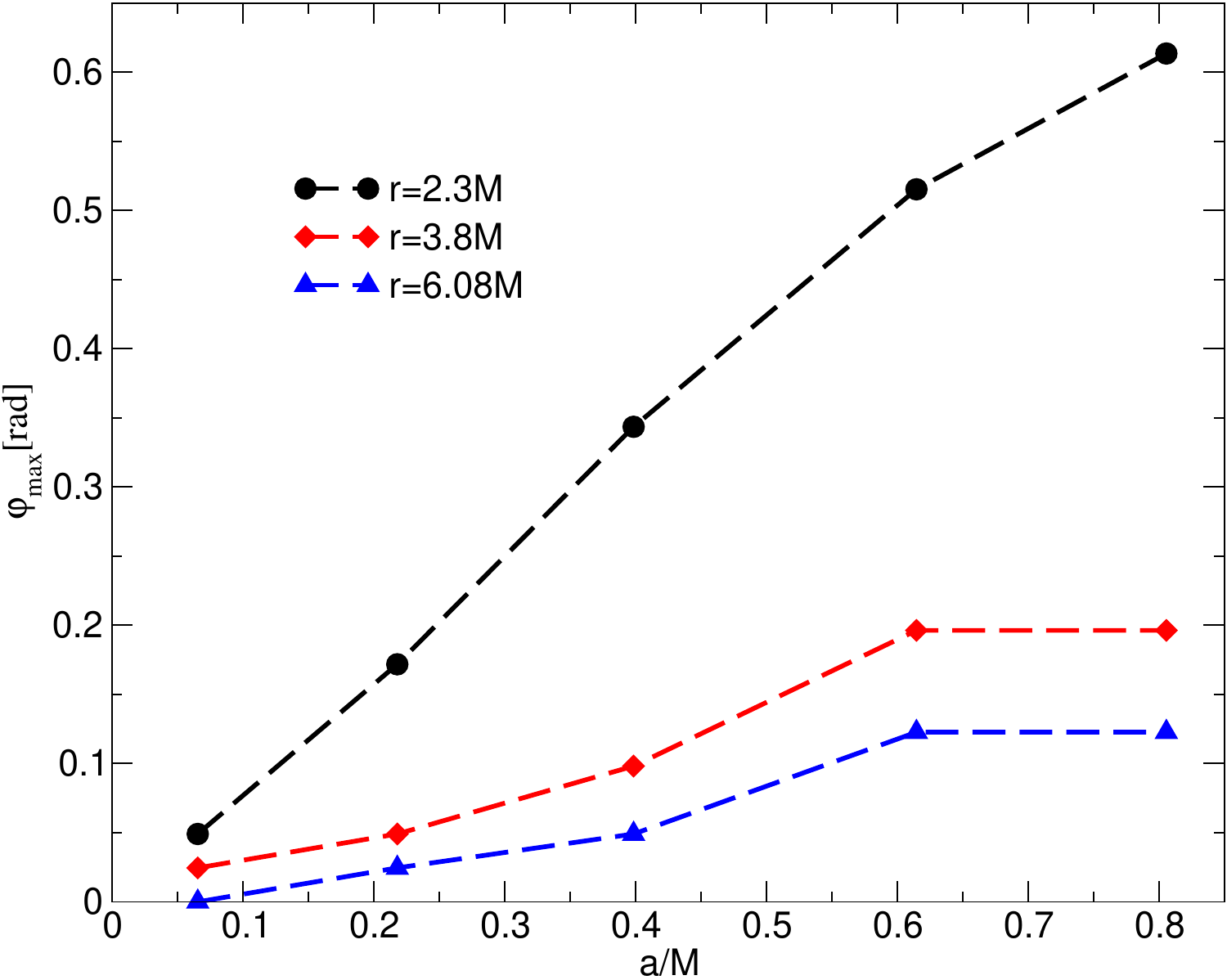,width=7.0cm, height=6.0cm}\\
  \caption{In the left column, the angular variation of the rest-mass density at the closest point to the black hole's horizon, at $r=2.3M$, is shown for the radius of the observable circle (photon sphere) $R_s$ and the deviation of the left edge of the shadow from the circle boundary $\delta_s$. The right column shows the variation of the point of maximum rest-mass density inside the cone at different radial positions according to the black hole's spin parameter $a/M$. For the initial values given in Table \ref{Inital_Con_1}, the top panels show results for $\eta=0.5$, while the bottom panels show results for $\eta=1$.
 }
\vspace{1cm}
\label{DensHK05}
\end{figure*}

Here, using the initial values in Table \ref{Inital_Con_1}, we reveal the changes in the shock cone formed around the $M87^*$ black hole due to the warping of spacetime in the strong gravitational field and determine the QPO frequencies. Fig.\ref{DensHK05}, using the initial values in Table \ref{Inital_Con_1}, shows how the intensity of the warping of the shock cone near the black hole  horizon changes according to the given parameters. In the top row of Fig.\ref{DensHK05}, the deviation of the hairy Kerr metric from the Kerr metric is given for $\eta=0.5$, while in the bottom row, this parameter is $\eta=1$. The left column of Fig.\ref{DensHK05} shows the variation of the the rest-mass density in the angular direction $\varphi$ at $r=2.3M$, while the right column shows the intensity of the warping of the shock cones formed at different radial points with the warping of spacetime. As understood from Table \ref{Inital_Con_1}, each model used in numerical calculations has different black hole spin and hair parameters.

In the left column of Fig.\ref{DensHK05}, at $r=2.3M$, as the black hole's spin parameter increases, the intensity of the warping of spacetime increases, and the maximum density of the resulting shock cone shifts from  $\varphi=0$ rad to $\varphi=1$ rad. This demonstrates the strength of the Lense-Thirring effect. Similarly, the graphs in the right column show how the angular location of the maximum the rest-mass density of the shock cone changes with the black hole's spin parameter at the closest point to the black hole's horizon and at different points (in strong gravitational fields). Although the locations $r=2.3M$ and $r=3.8M$ are not very different from each other, the effect of the warping of spacetime on the warping of the shock cone is very strong at $r=2.3M$. The warping of the shock cone increases with the increase in the black hole's spin parameter, as expected. Therefore, if the effect of the strong gravitational field is to be revealed and the impact of the Lense-Thirring phenomenon on the resulting QPOs is to be understood, the observed data must be very close to the black hole's horizon.


\begin{figure*}
  \vspace{1cm}
  \center
  \psfig{file=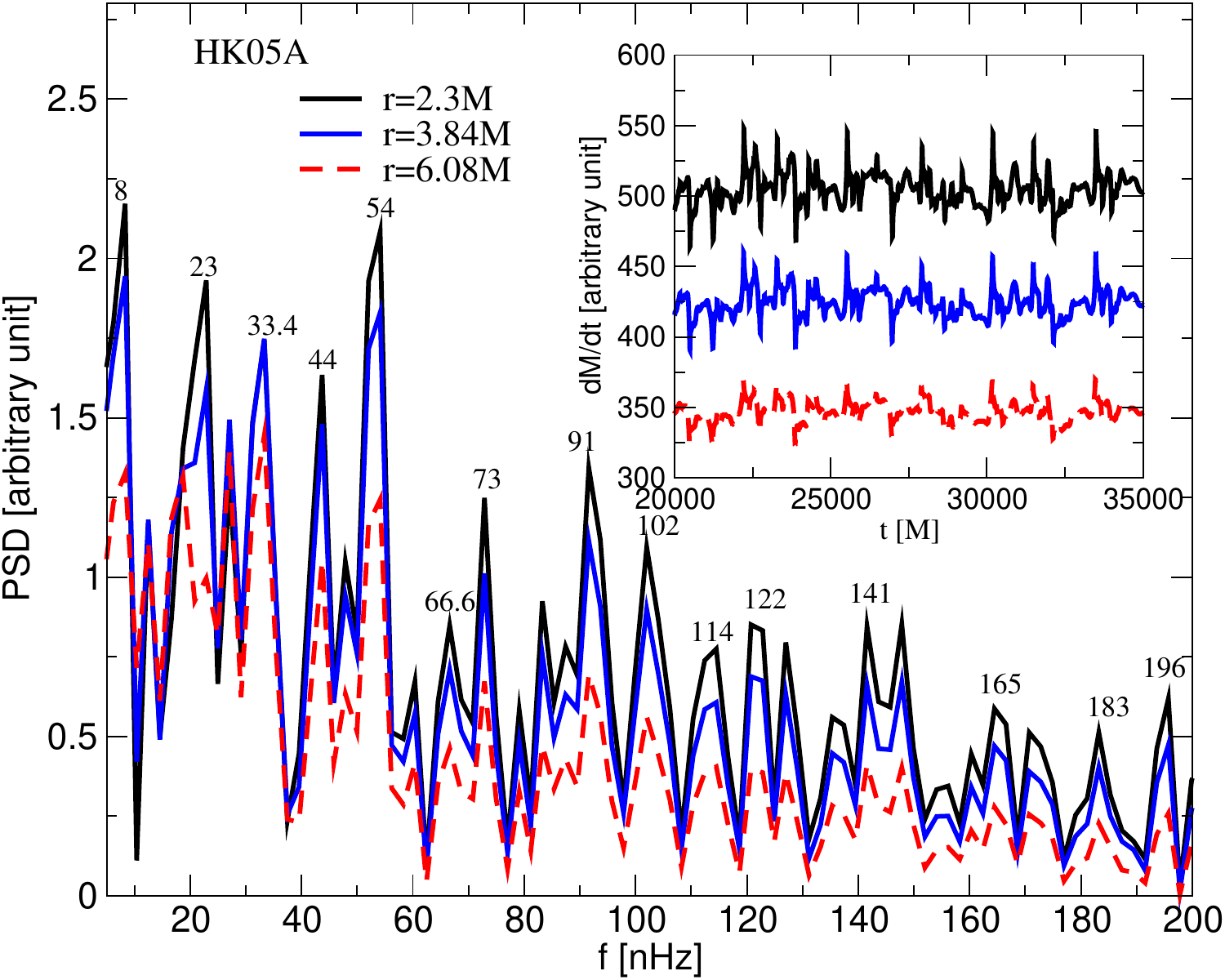,width=7.5cm, height=7.5cm}\hspace*{0.15cm}
  \psfig{file=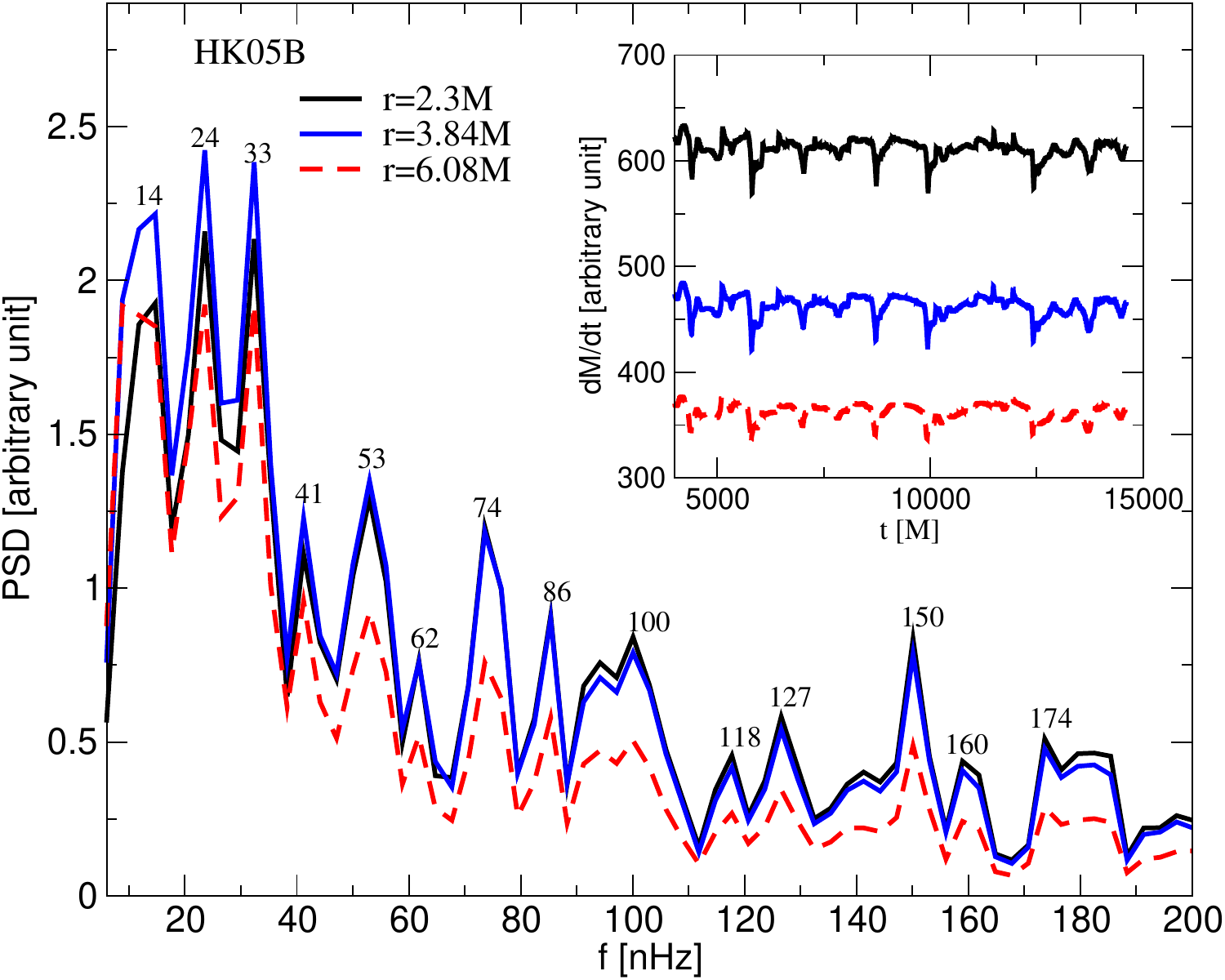,width=7.5cm, height=7.5cm}\\
  \vspace*{0.3cm}
  \psfig{file=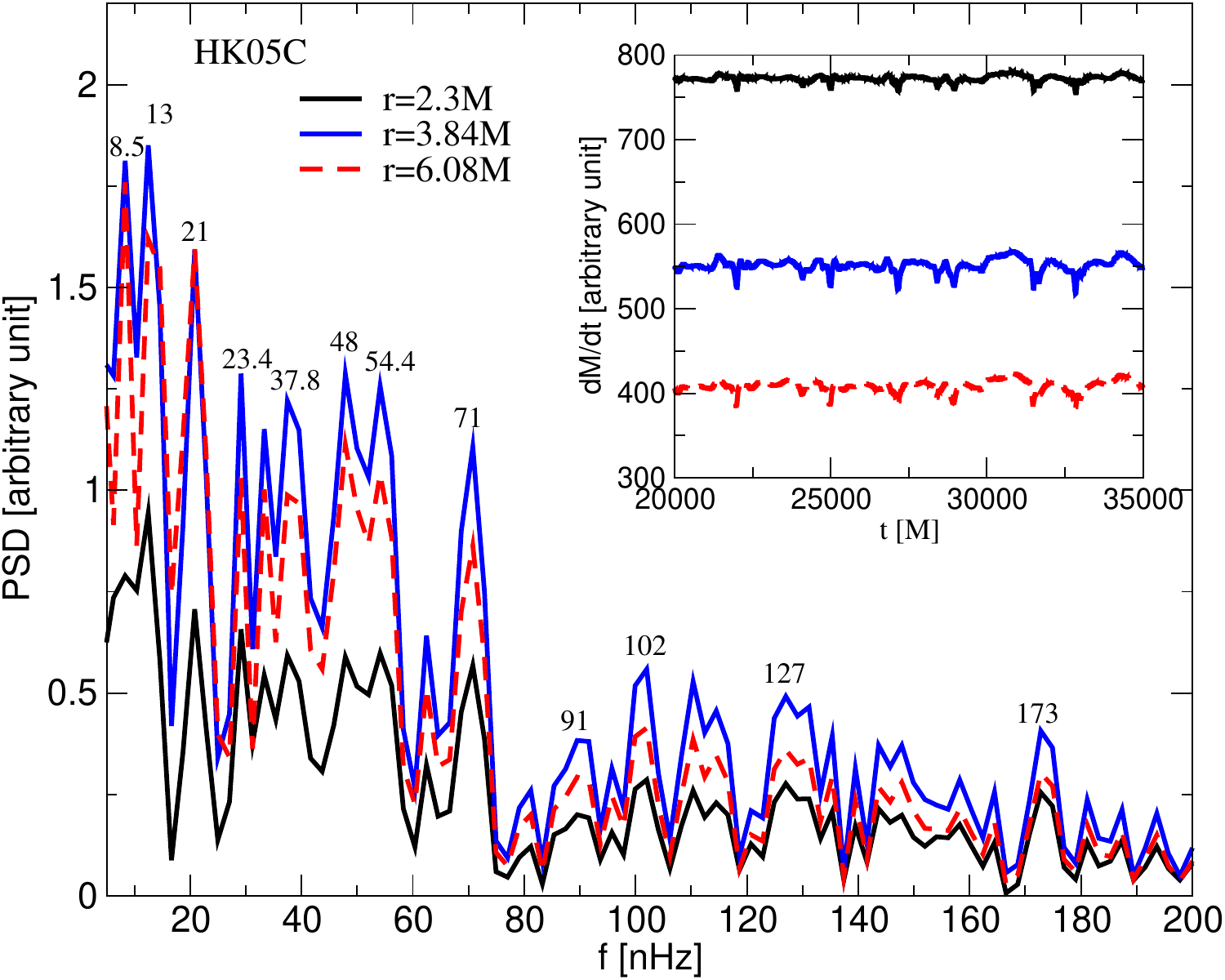,width=7.5cm, height=7.5cm}\hspace*{0.15cm}
  \psfig{file=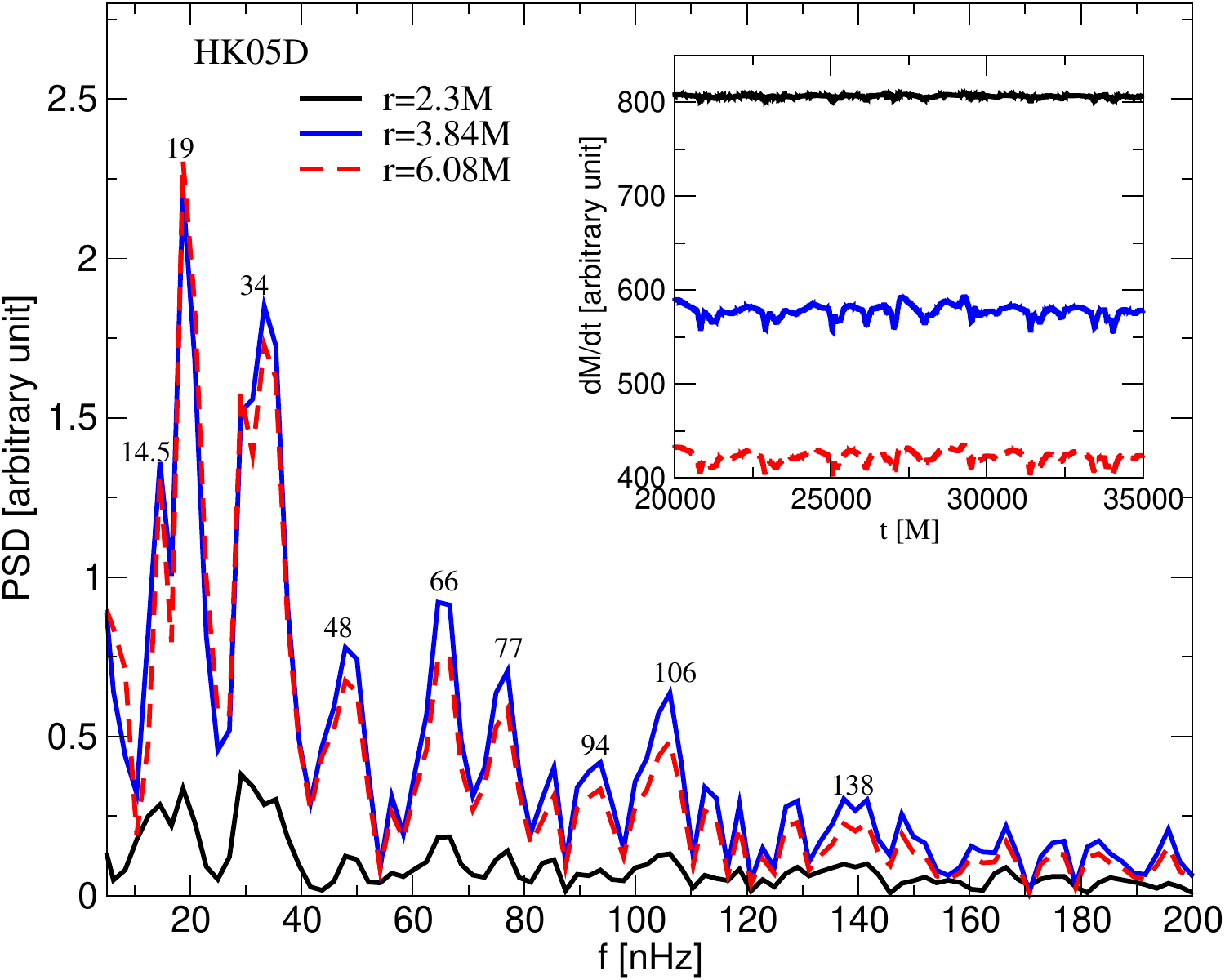,width=7.5cm, height=7.5cm}\\  
     \caption{PSD analysis around the Hairy Kerr black hole in a strong gravitational field  at $r=2.3M$, $r=3.84M$, and $r=6.08M$ for the radius of the  observable circle (photon sphere) $R_s$,  the deviation of the left edge of the shadow from the circle boundary $\delta_s$ , and $\eta=0.5$. It shows PSD analysis results obtained from the mass accretion rates calculated at different radii long after the disk has reached the steady state for the models given in Table \ref{Inital_Con_1}.
    }
\vspace{1cm}
\label{PSDHK05_1}
\end{figure*}

\begin{figure*}
  \vspace{1cm}
  \center
  \psfig{file=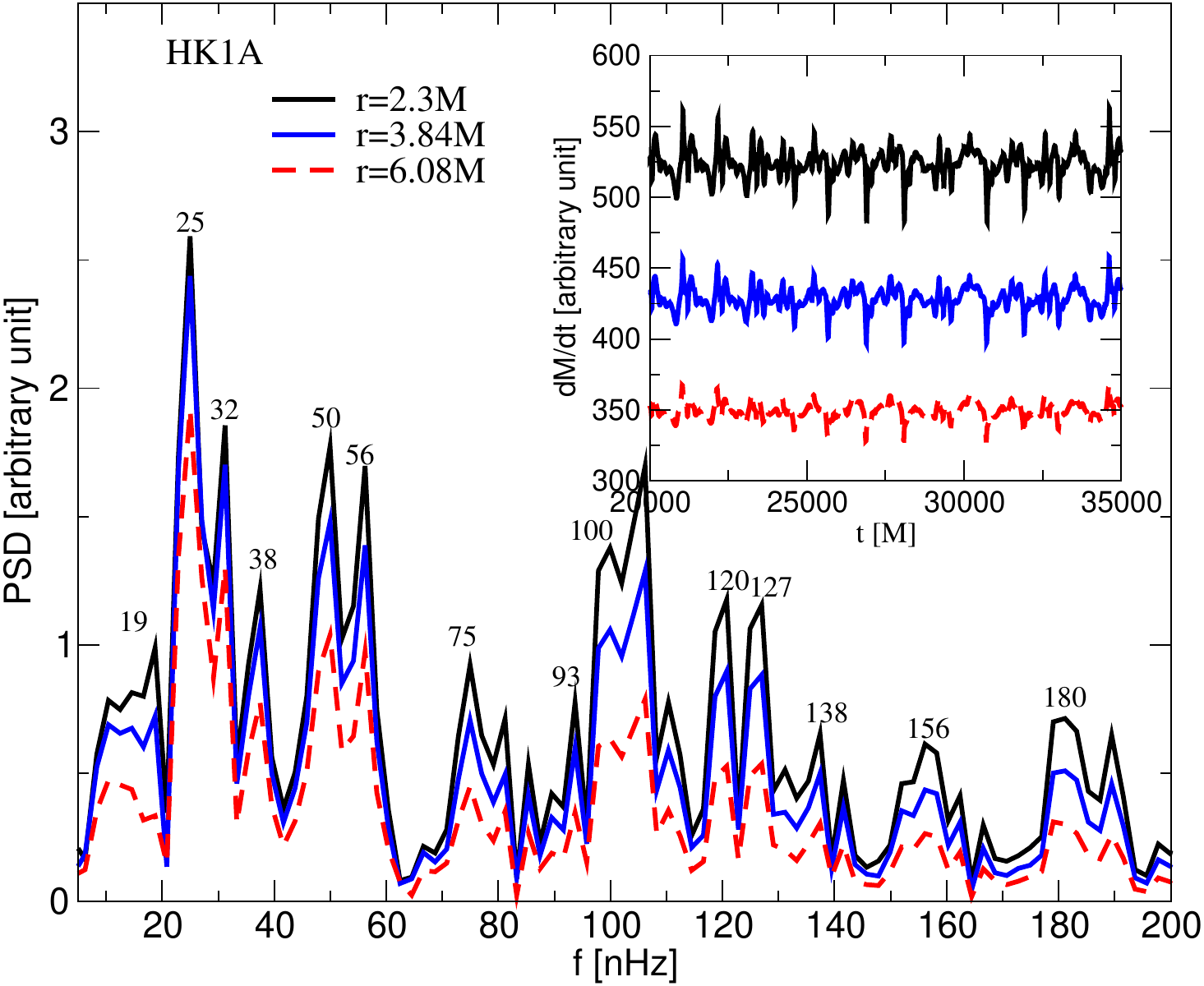,width=7.5cm, height=7.0cm}\hspace*{0.15cm}
  \psfig{file=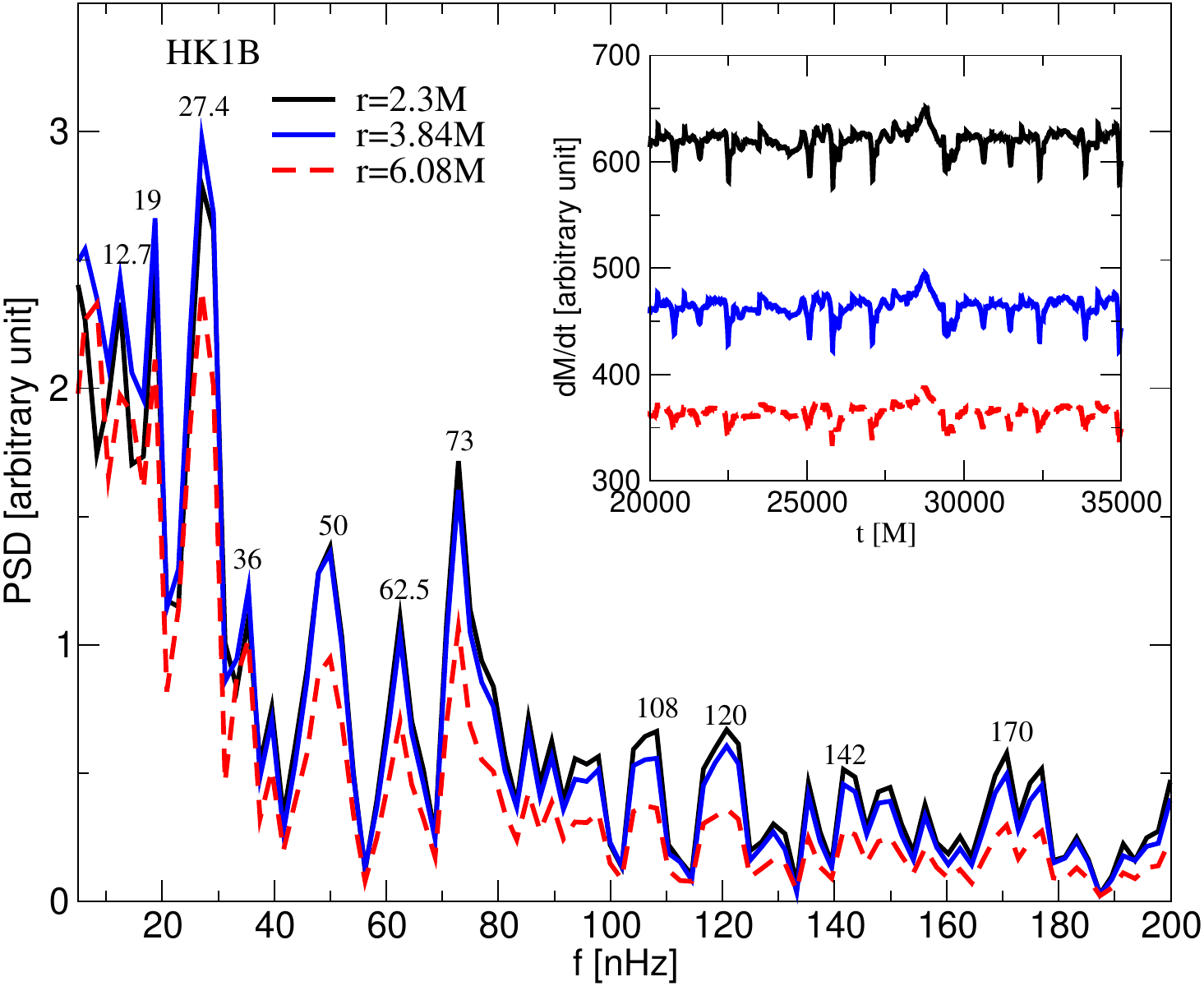,width=7.5cm, height=7.0cm}\\
  \vspace*{0.3cm}
  \psfig{file=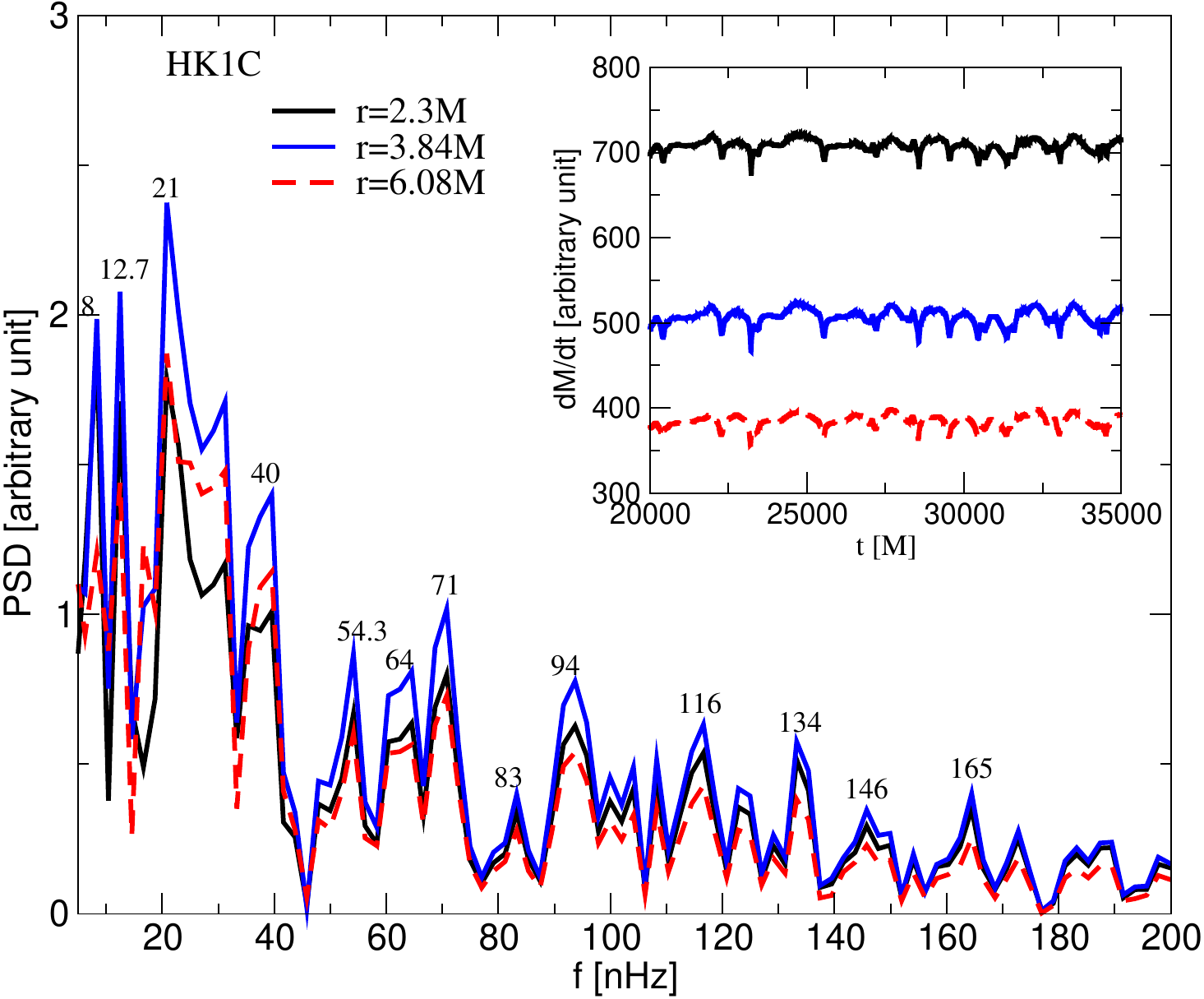,width=7.5cm, height=7.0cm}\hspace*{0.15cm}
  \psfig{file=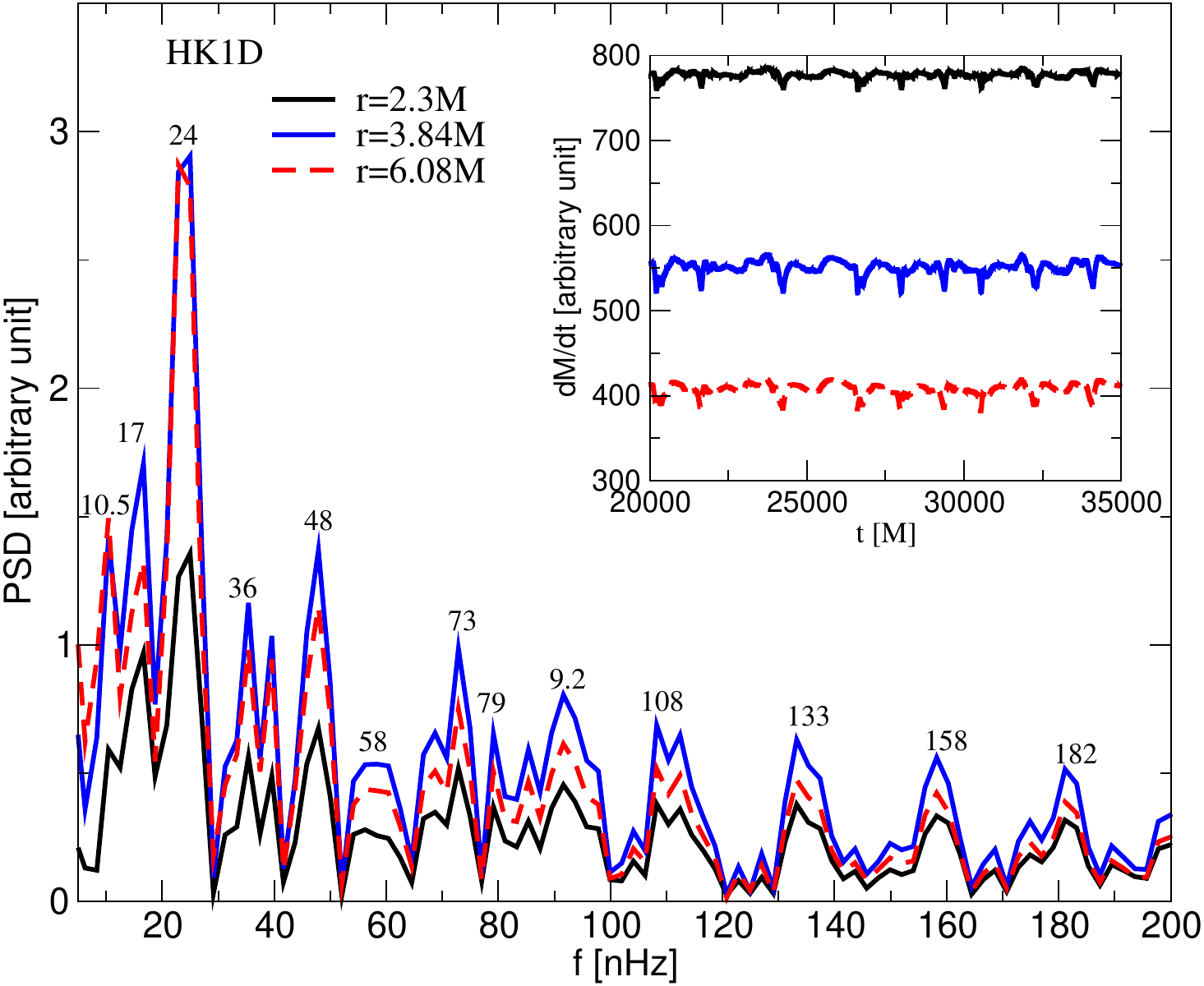,width=7.8cm, height=7.0cm}\\
  \vspace*{0.3cm}
  \psfig{file=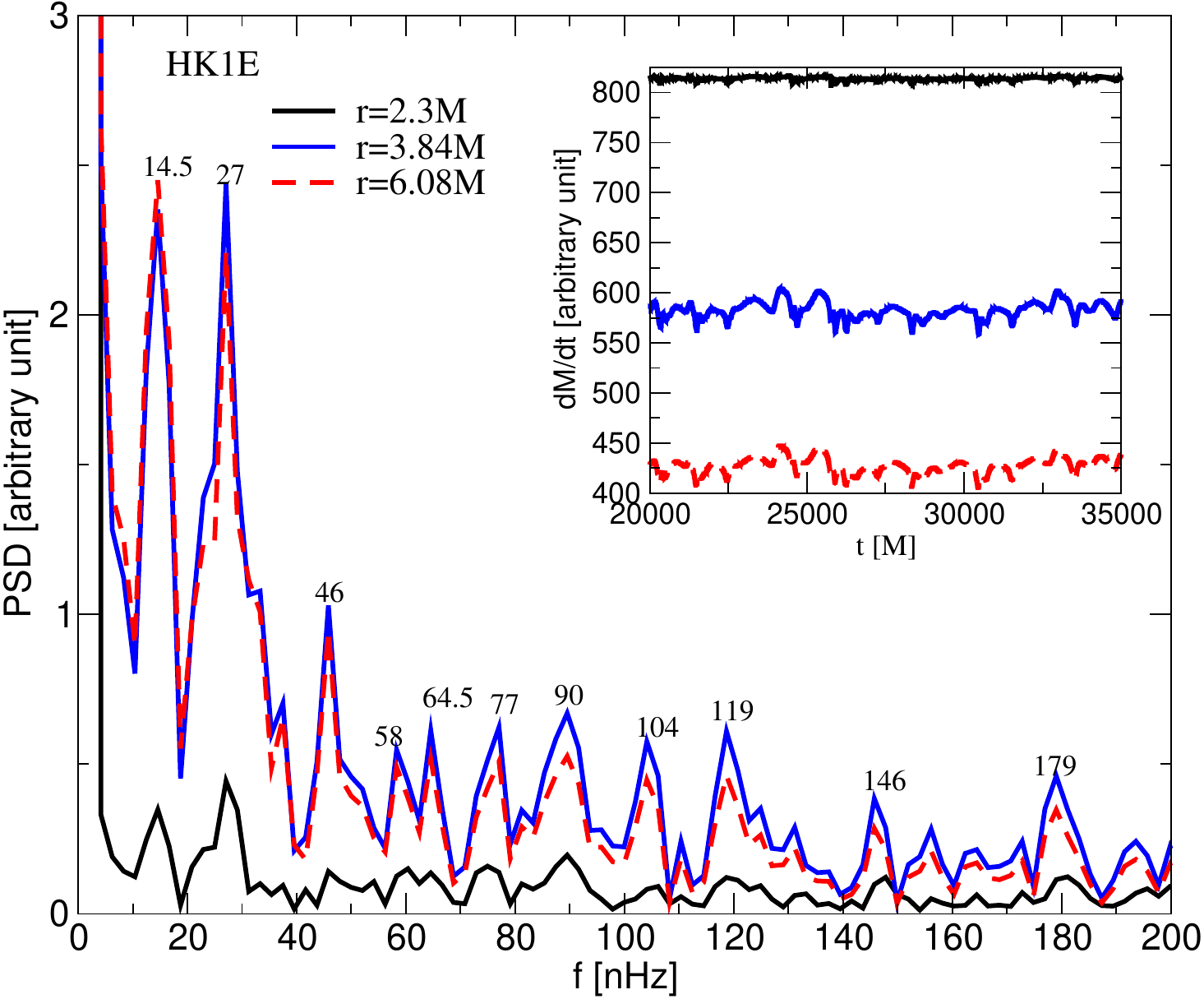,width=7.5cm, height=7.0cm}\\  
     \caption{Same as Fig.\ref{PSDHK05_1} but for $\eta=1$.
    }
\vspace{1cm}
\label{PSDHK1_1}
\end{figure*}

In Figures \ref{PSDHK05_1} and \ref{PSDHK1_1}, we numerically identified QPO frequencies for the cases given in Fig.\ref{DensHK05} using the initial conditions given in Table \ref{Inital_Con_1}, through PSD analysis. Fig.\ref{PSDHK05_1} shows QPO behavior for a metric deviation value of $\eta=0.5$, while Fig.\ref{PSDHK1_1} shows the variation in oscillation peaks for $\eta=1$. Generally speaking, as theoretically expected, the hair parameter of the black hole results in more complex QPO frequencies and more peaks compared to the Kerr black hole model. This complexity arises from the additional instability introduced by the hair parameter.

As seen in Fig.\ref{PSDHK05_1}, each model results in complex and numerous QPO frequencies. PSD analyses are calculated separately at $r=2.3$M, $r=3.84$M, and $r=6.08$M, and the results are shown in the same graph for comparison. As seen in Fig.\ref{PSDHK05_1}, the QPO frequencies formed at different points within the cone produce the same frequencies. Thus, from an observational perspective, if sufficient technical equipment and sensitivity are achieved, it is  possible to observe these frequencies of $M87^*$.
  The graph also shows that the observed frequencies create resonance states; for example, the $3:2$ ratio appears in frequencies such as $33.4:23$ and $53:33$, while the $2:1$ ratio is observed in frequencies like $62:33$, $66:34$, $48:23.4$, and $66.6:33.4$. The $3:1$ ratio occurs in frequencies like $23:8$, $41:14$, and $23.4:8.5$. All these ratios occur with an approximate $5\%$ margin of error. Upon closer examination, similar resonance states, such as $4:1$, $4:3$, etc. are also observed.
Furthermore, in the same graph, when the black hole's spin parameter is $a/M=0.7923$ (HK05D), i.e., when the black hole starts to spin rapidly, it is observed that although the QPO frequencies formed at different locations are the same, their amplitudes differ. This is because the rapidly rotating black hole suppresses the QPO modes formed by pressure variations within the cone due to the Lense-Thirring effect, thus reducing the amplitude of the QPO frequencies.

While the low-frequency QPOs in the nHz range are observed around slowly and moderately rotating black holes, the high-frequency QPOs are observed in rapidly rotating black holes. This situation is explained in detail below. The lowest observed frequency in these models is around 8 nHz. As seen in Fig.\ref{PSDHK1_1}, the deviation ratio of the Hairy Kerr metric from the Kerr metric, $\eta=1$, shows differences in QPO frequencies compared to the $\eta=0.5$ case. This is an expected situation. Increasing the modified parameter of gravity affects the oscillation properties of the disk and thus the physical results of the black hole-disk interaction, causing QPO frequencies to change. This change affects not only the formed QPO frequencies but also their amplitudes. In the $\eta=1$ case, since the frequency amplitude is greater, their observability is higher. Additionally, in the HK1E model in Fig.\ref{PSDHK1_1}, since the black hole's spin parameter is greater than in other models, the modes excited within the disk are suppressed due to the Lense-Thirring effect. Therefore, the likelihood of observing the low-frequency QPOs in the strong gravitational region around the rapidly rotating black holes (very close to the black hole's horizon) is very low. The PSD analyses obtained in Figures \ref{PSDHK05_1} and \ref{PSDHK1_1}, similar to the Kerr black hole case, are obtained from two different data sets: short-term and long-term. According to the obtained results, it is observed that the frequencies are persistent.

\begin{figure*}
  \vspace{1cm}
  \center
     \psfig{file=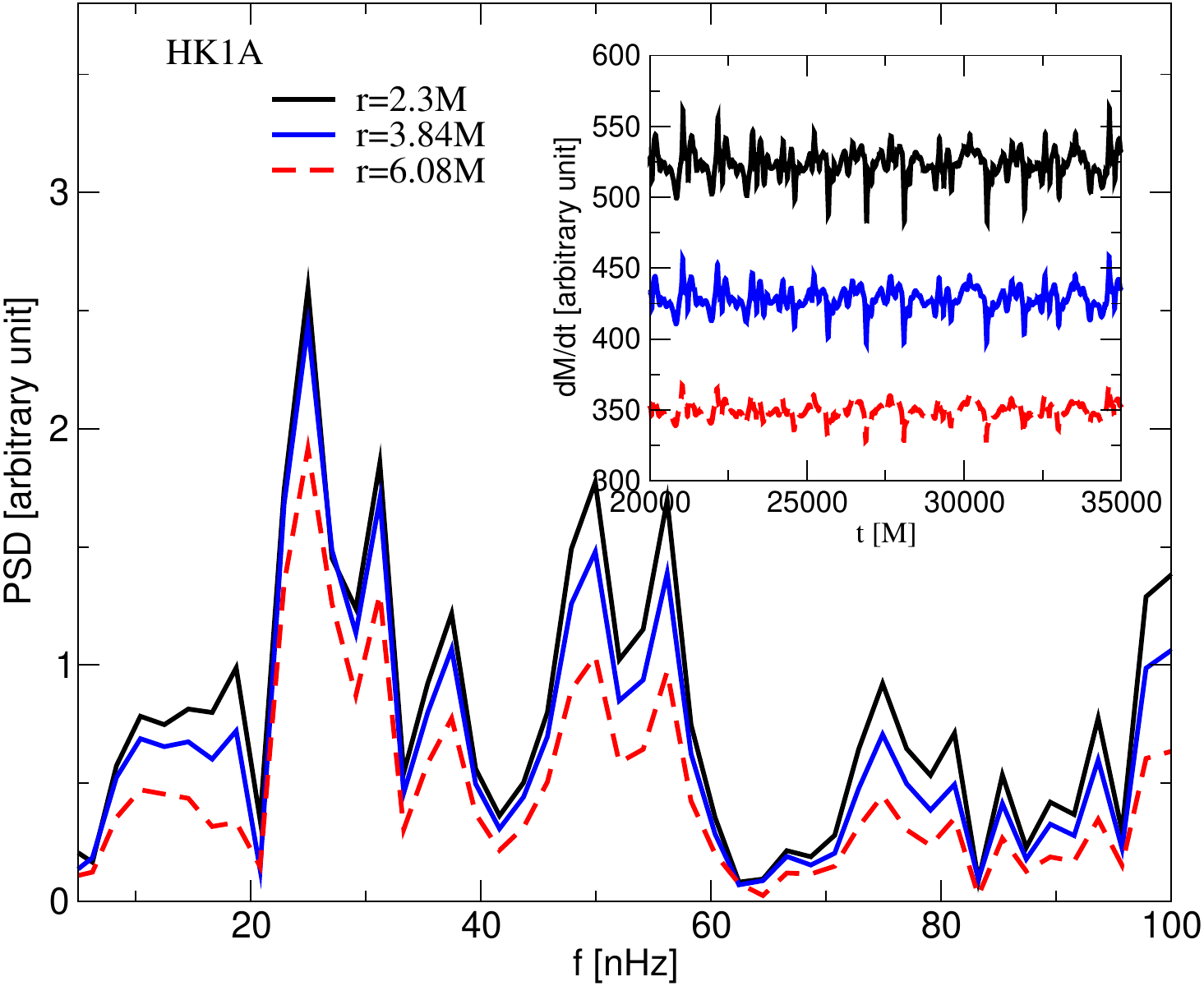,width=3.5cm, height=4.0cm}\hspace*{0.15cm}
     \psfig{file=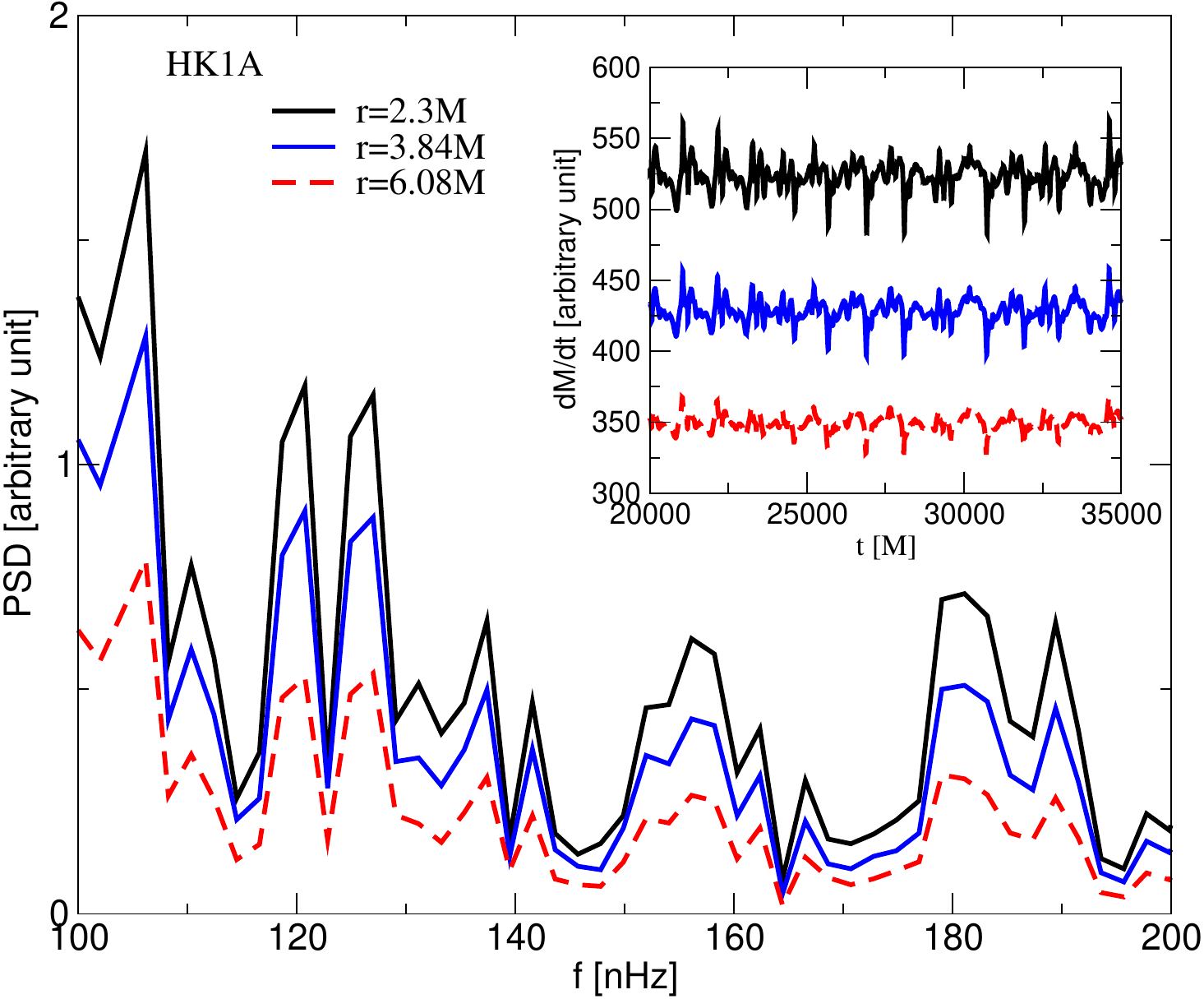,width=3.5cm, height=4.0cm}\hspace*{0.15cm}
     \psfig{file=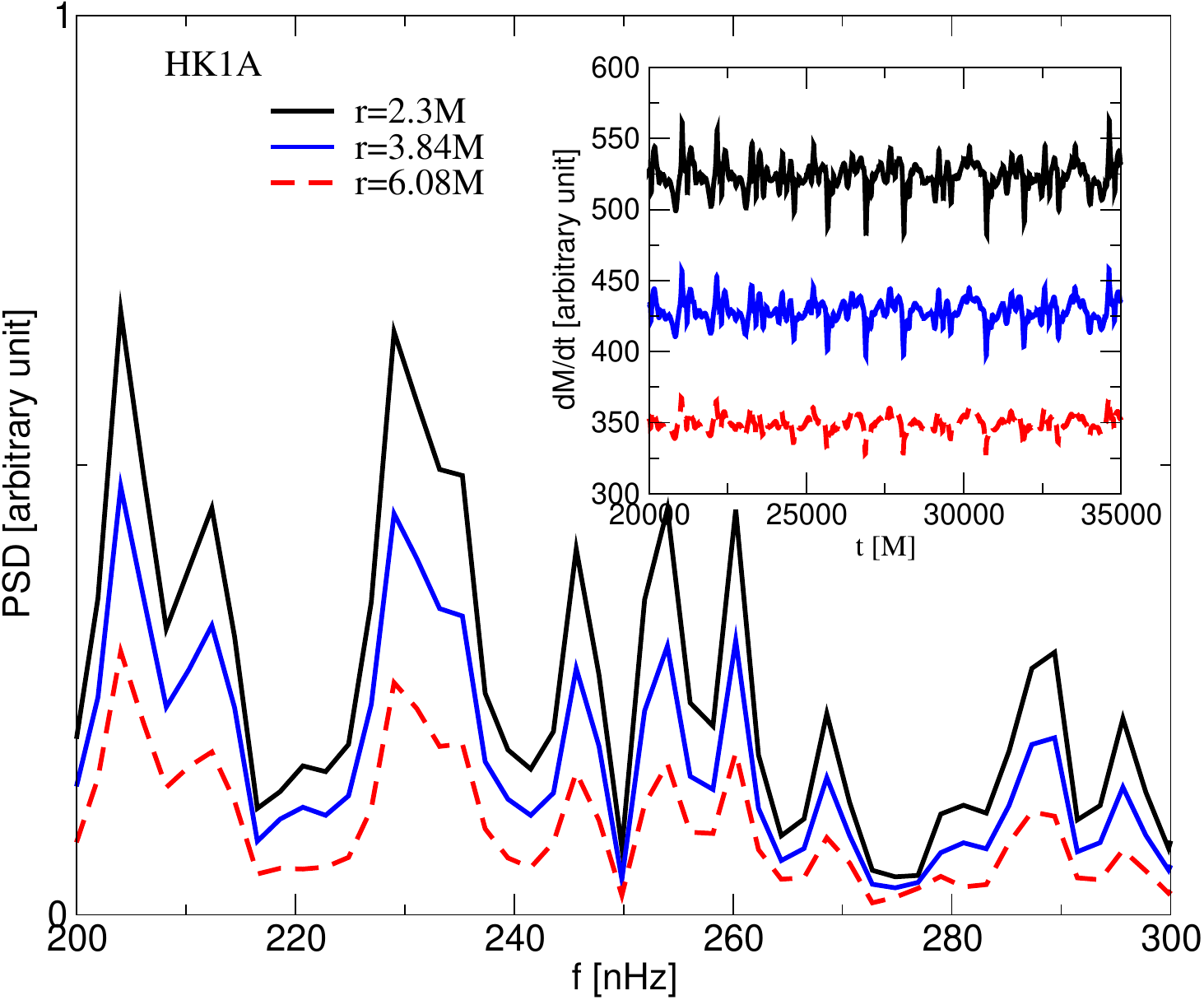,width=3.5cm, height=4.0cm}\hspace*{0.15cm}
     \psfig{file=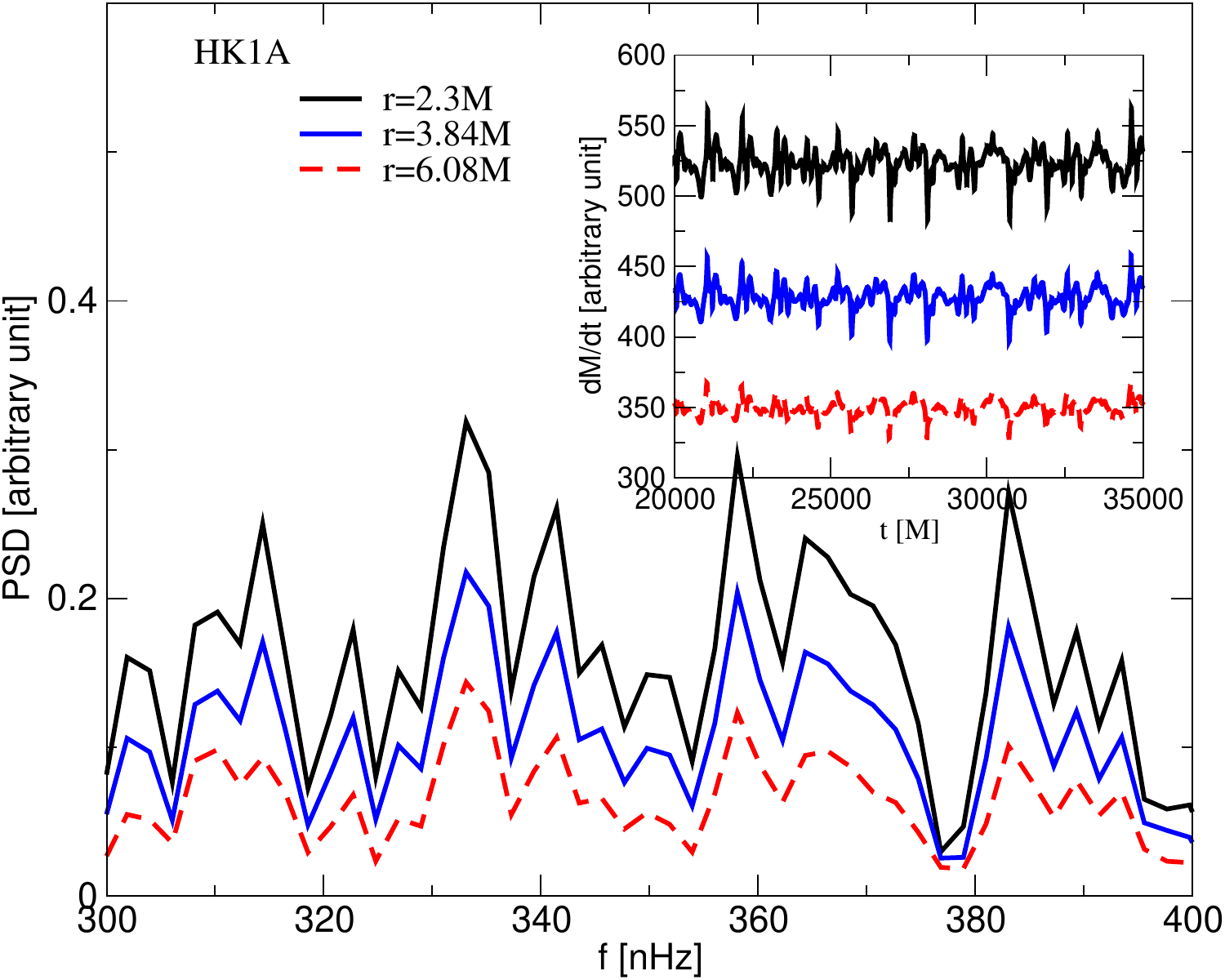,width=3.5cm, height=4.0cm}\\
     \vspace*{0.25cm}
     \psfig{file=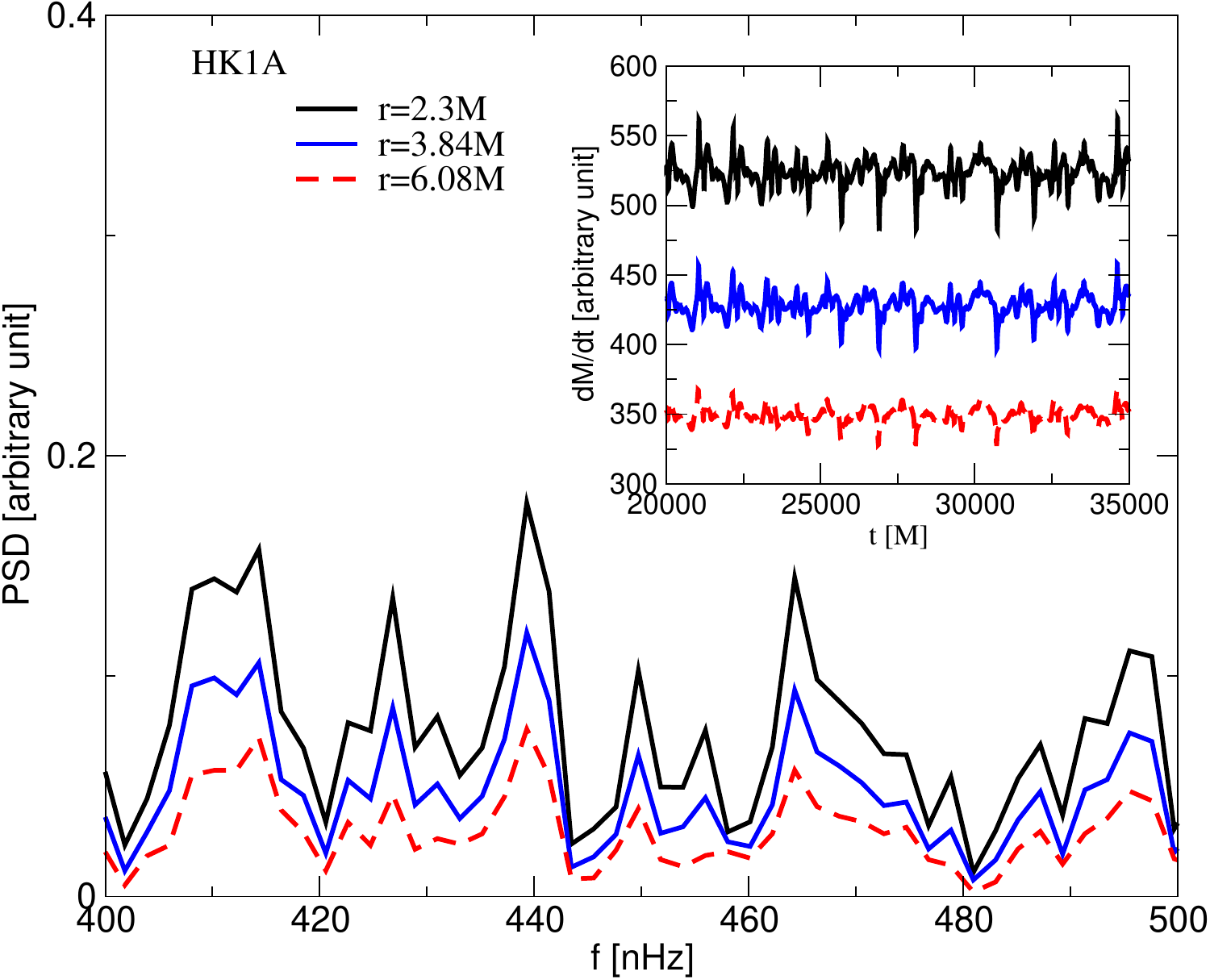,width=3.5cm, height=4.0cm}\hspace*{0.15cm}
     \psfig{file=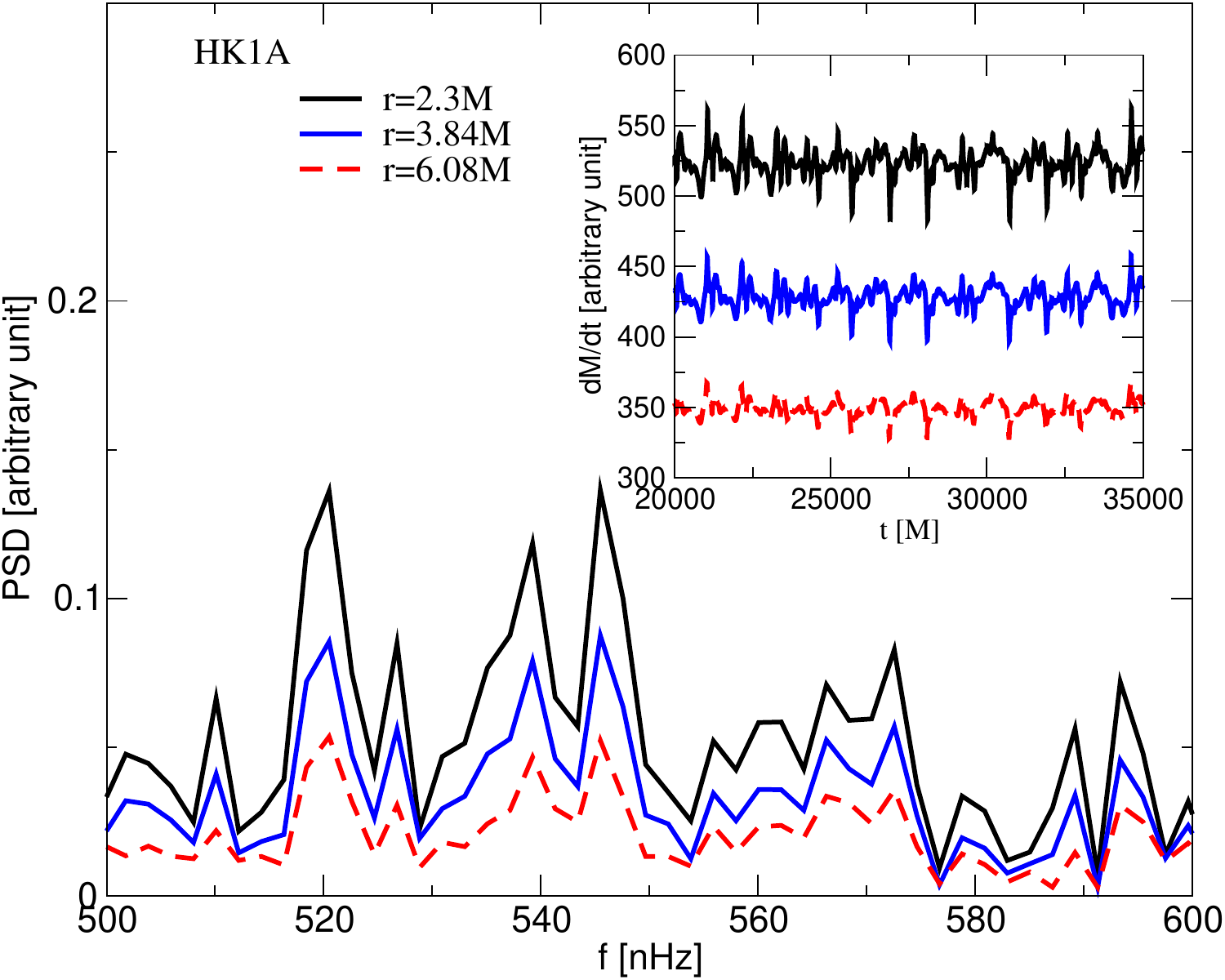,width=3.5cm, height=4.0cm}\hspace*{0.15cm}
     \psfig{file=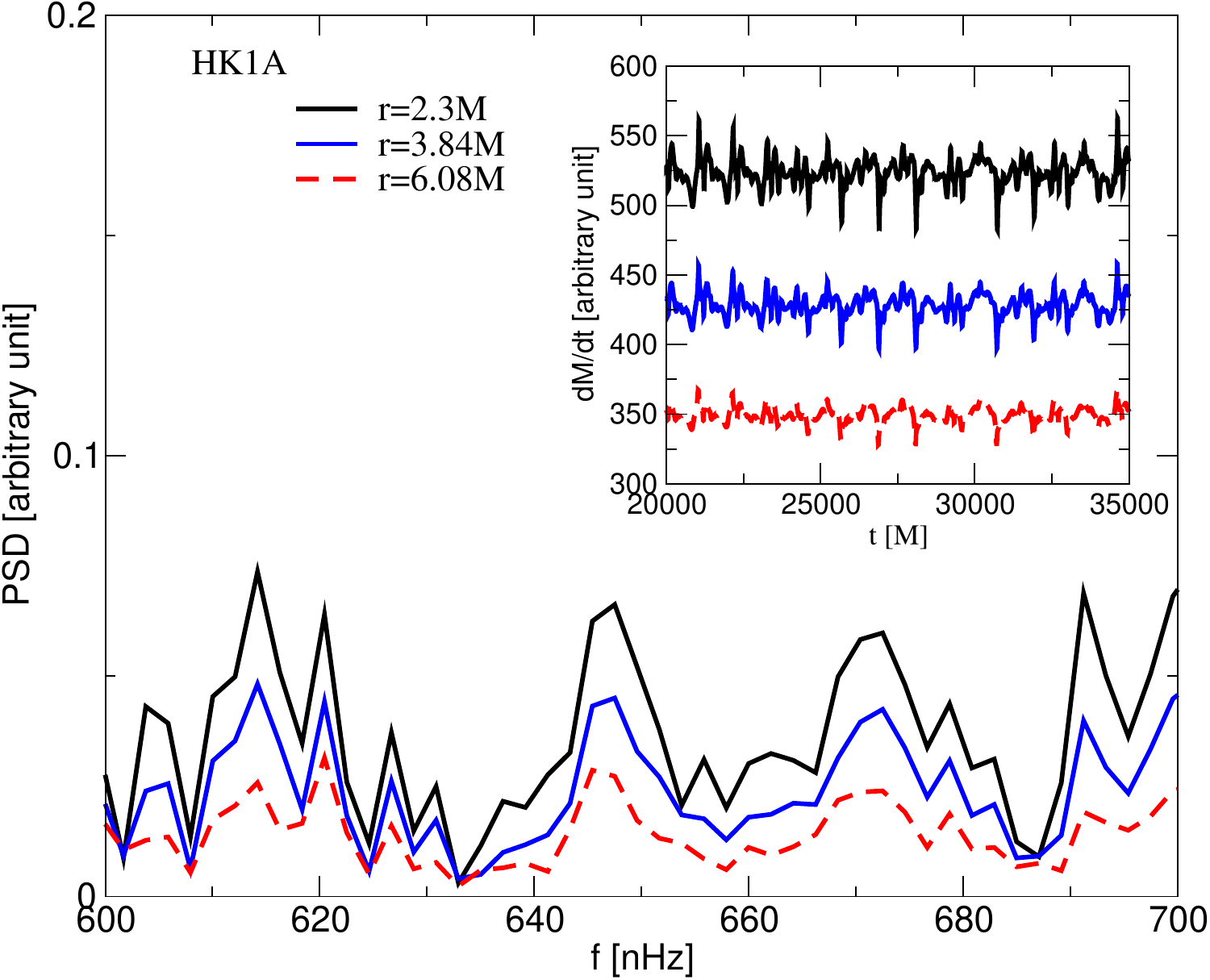,width=3.5cm, height=4.0cm}\hspace*{0.15cm}
     \psfig{file=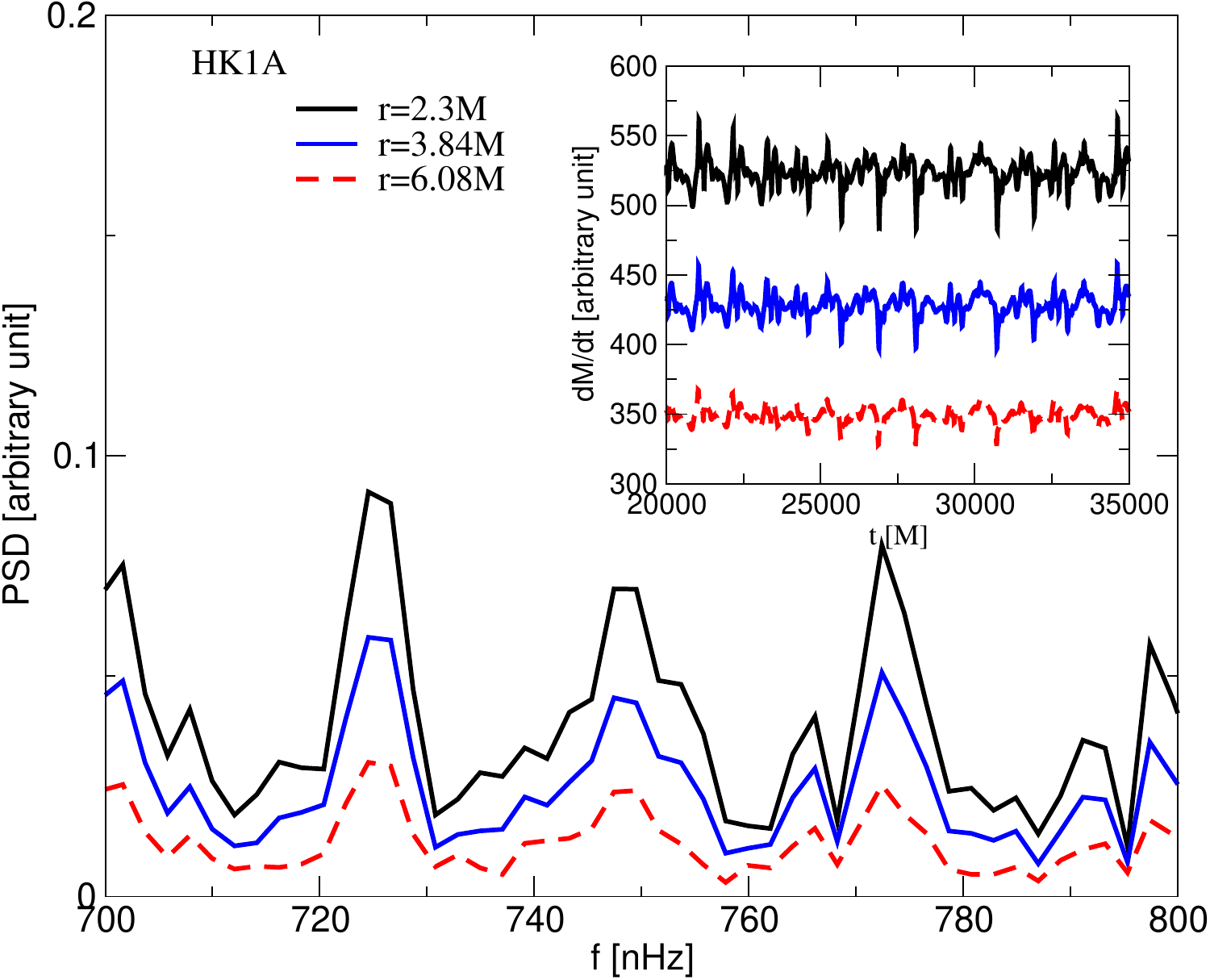,width=3.5cm, height=4.0cm}\\
     \vspace*{0.25cm}
     \psfig{file=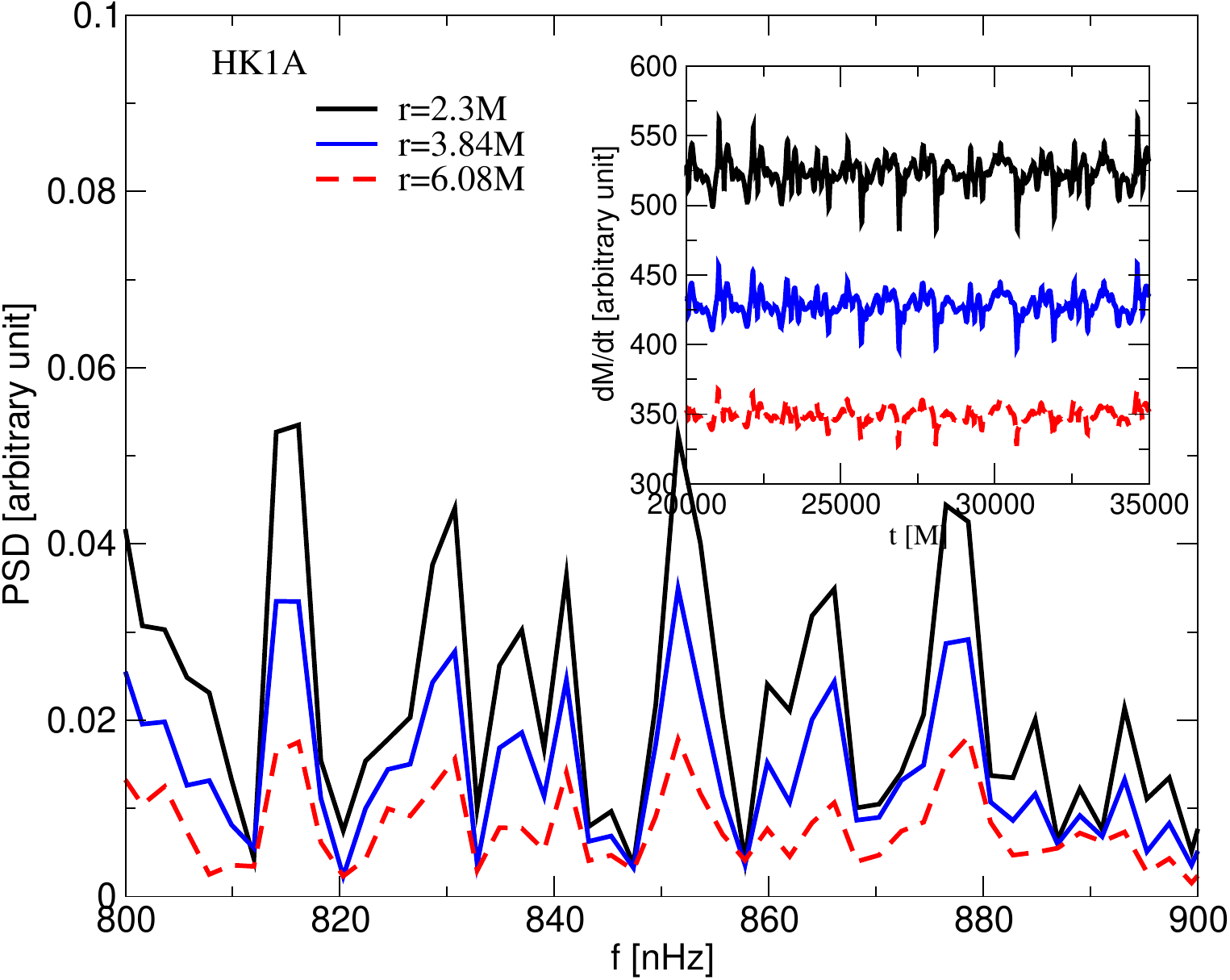, width=3.5cm, height=4.0cm}\hspace*{0.15cm}
     \psfig{file=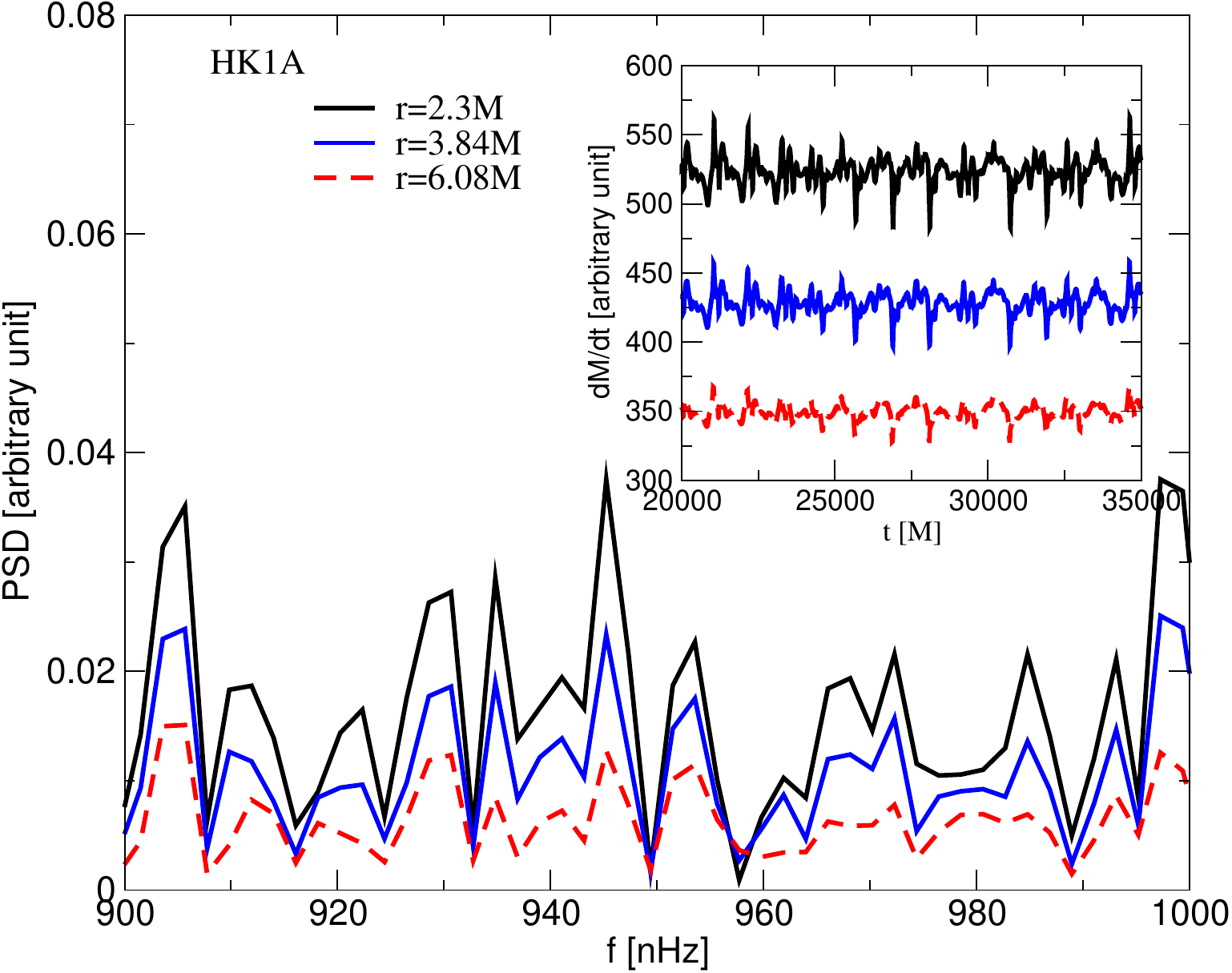,width=3.5cm, height=4.0cm}\hspace*{0.15cm}
     \psfig{file=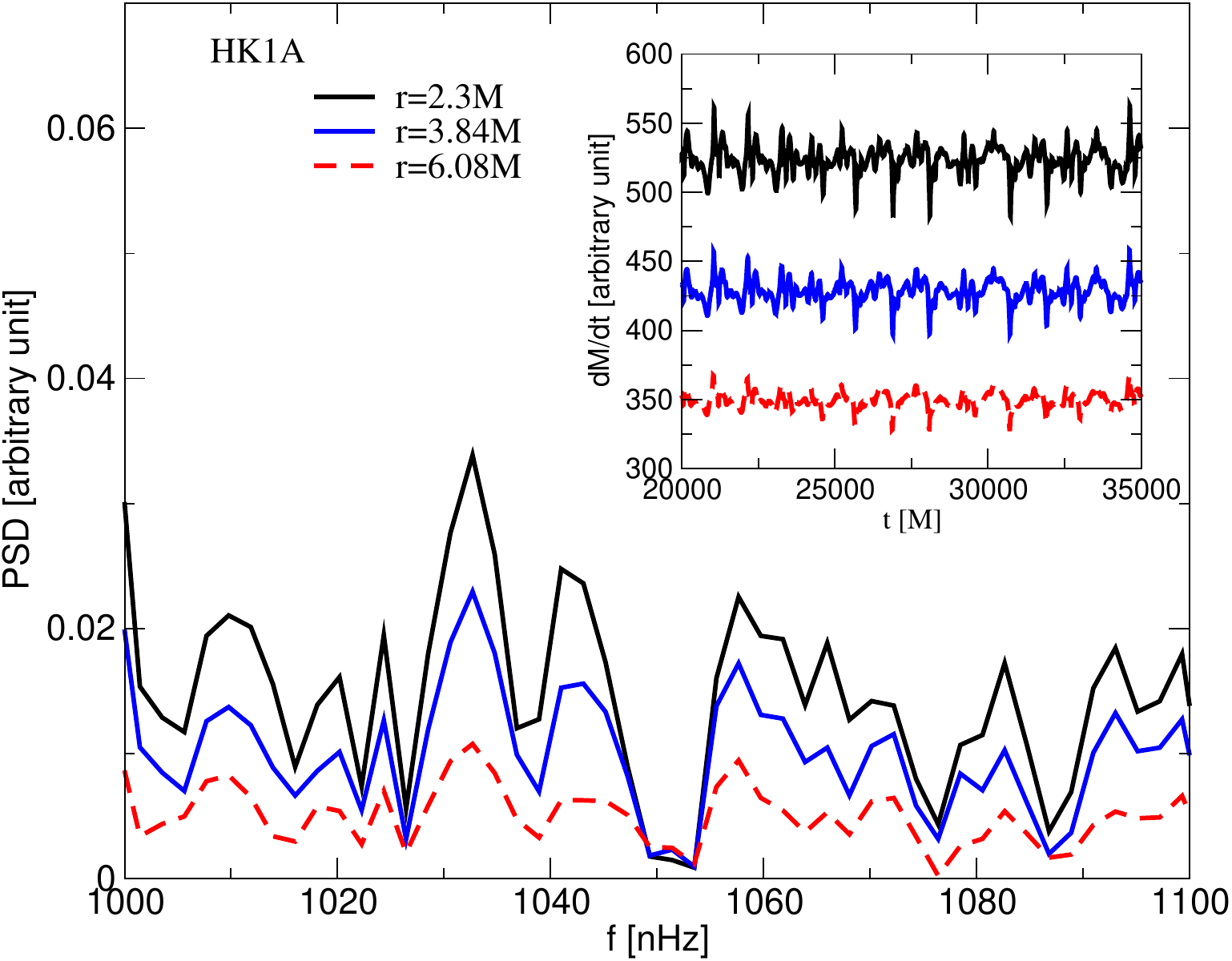,width=3.5cm, height=4.0cm}\hspace*{0.15cm}
     \psfig{file=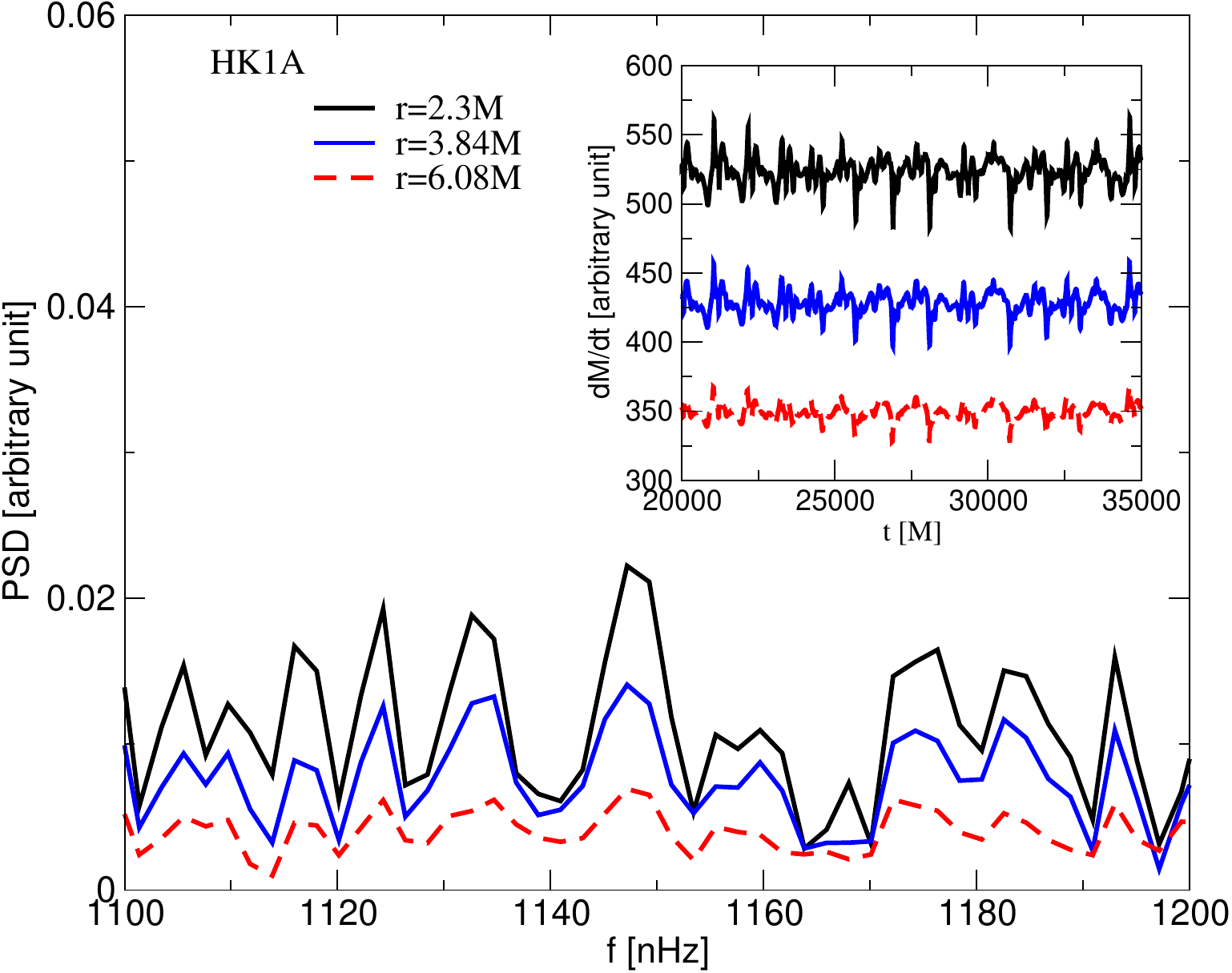,width=3.5cm, height=4.0cm}\\
     \vspace*{0.25cm}
     \psfig{file=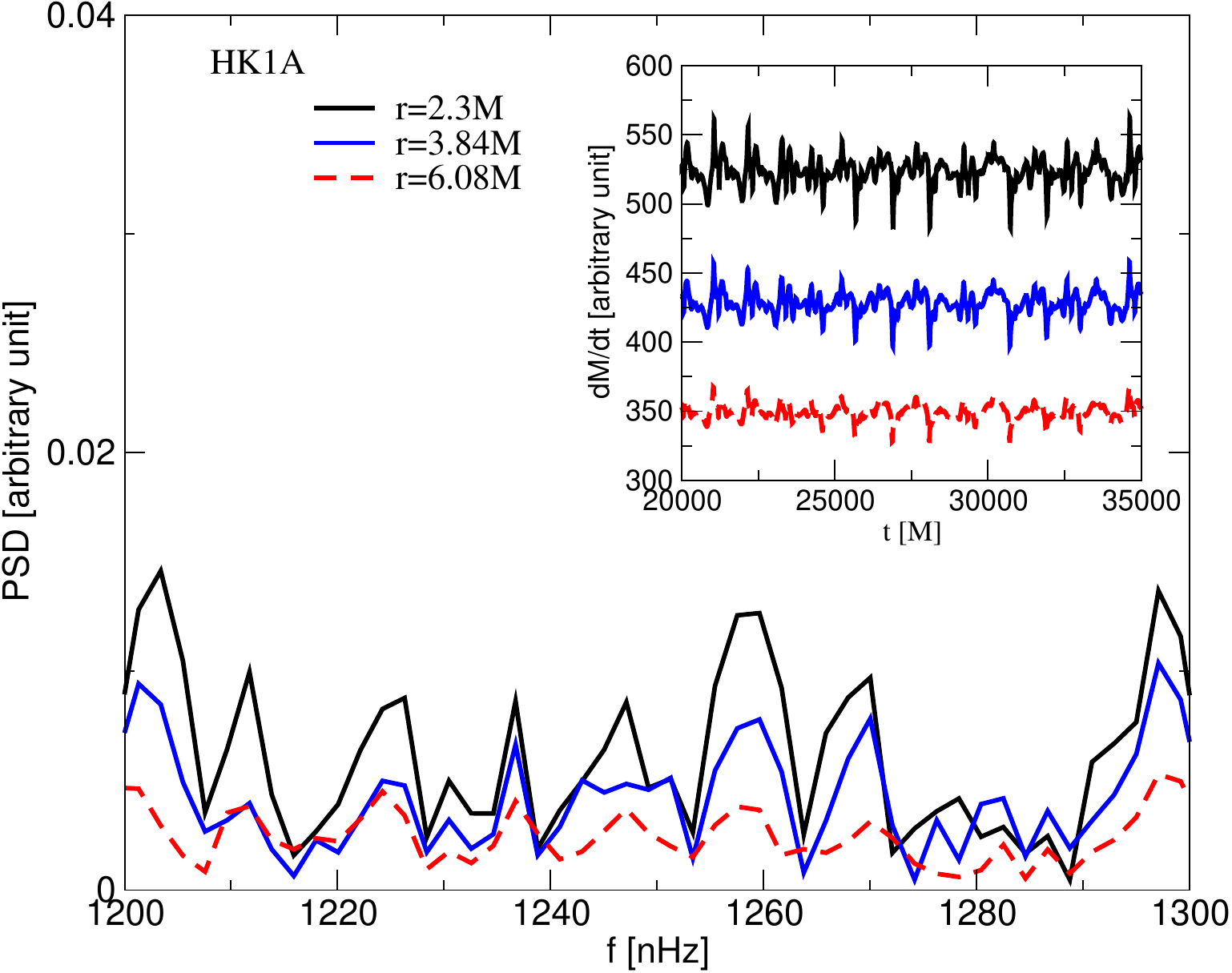,width=3.5cm, height=4.0cm}\hspace*{0.15cm}
     \psfig{file=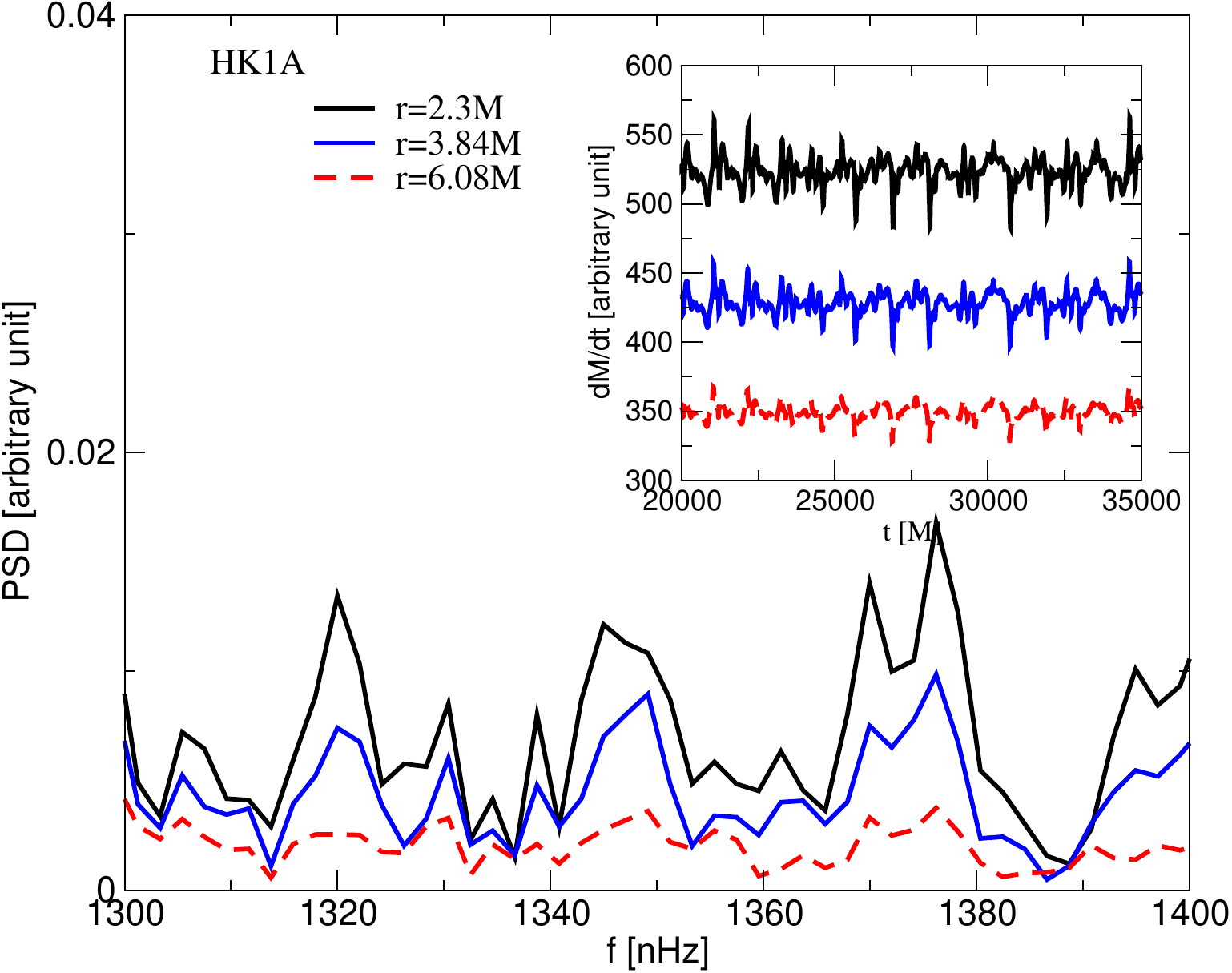,width=3.5cm, height=4.0cm}\hspace*{0.15cm}
     \psfig{file=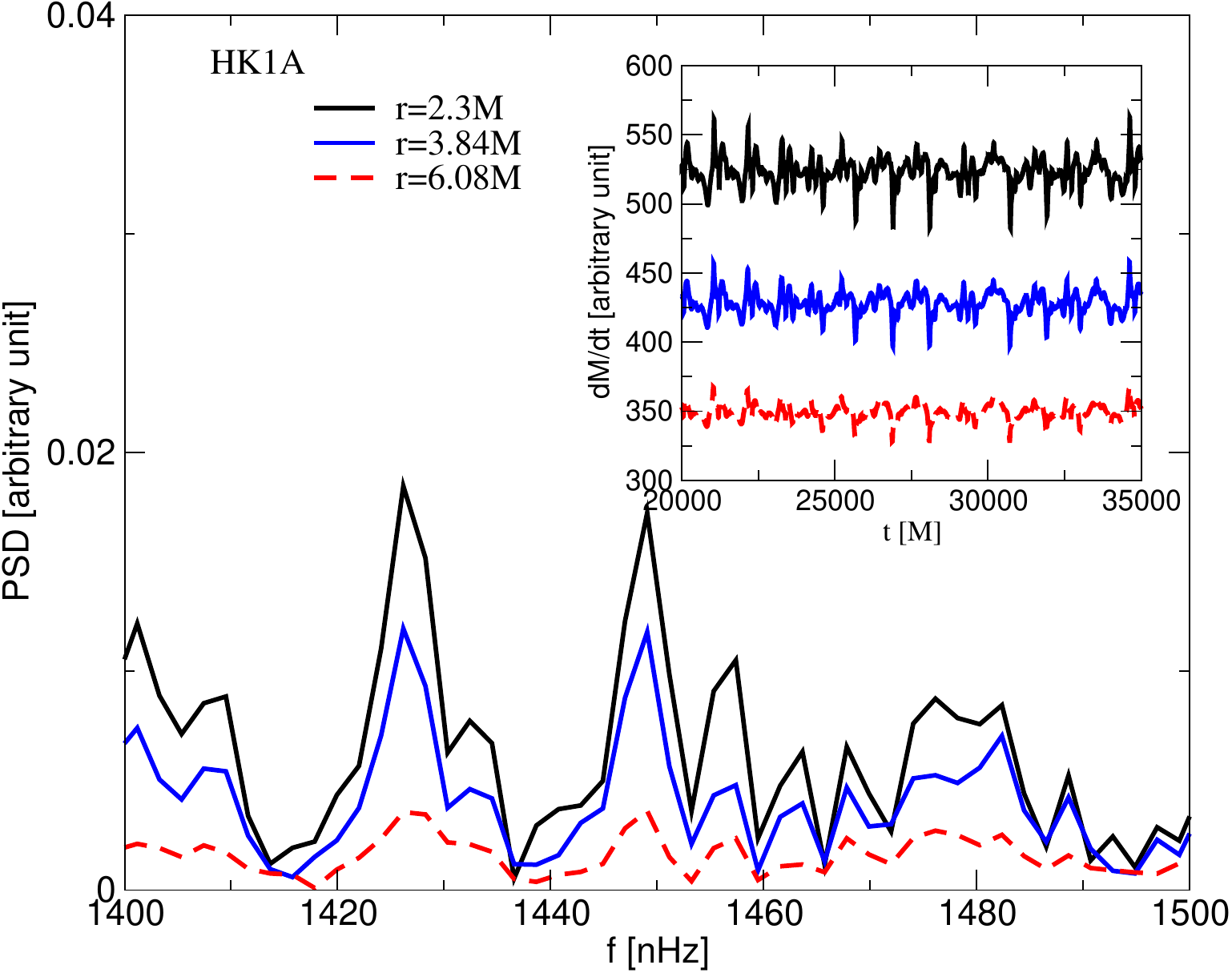,width=3.5cm, height=4.0cm}\hspace*{0.15cm}
     \psfig{file=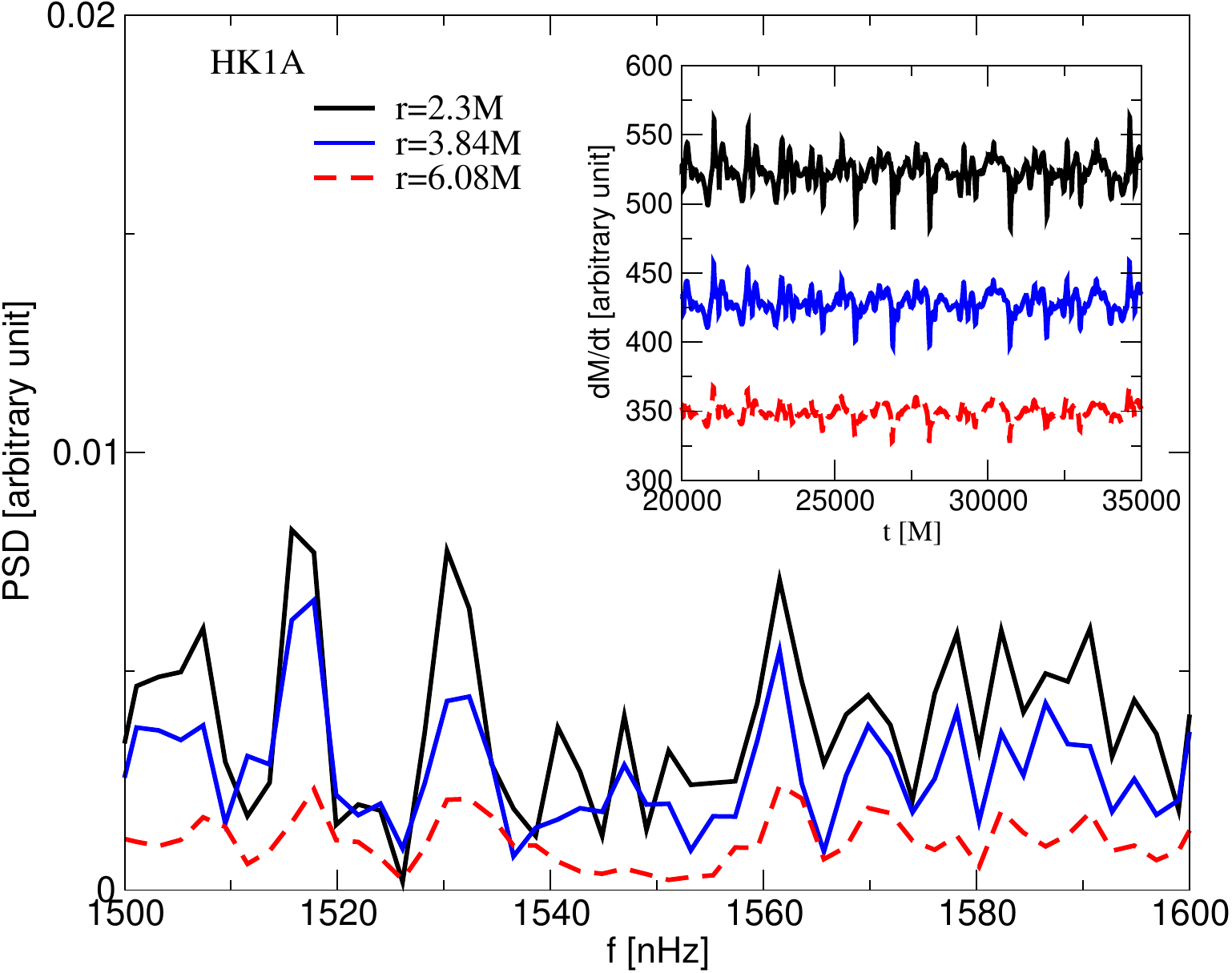,width=3.5cm, height=4.0cm}\\     
    \caption{Same as Fig.\ref{PSDHK1_1}, but PSD analysis for model HK1A at different frequency ranges.
    }
\vspace{1cm}
\label{PSDHKA1A_Diff_window}
\end{figure*}

In the case where the black hole's spin parameter and hair parameter are small, but the deviation parameter $\eta=1$ distinguishes hairy Kerr gravity from Kerr gravity, we plot the PSD analysis seen in Fig.\ref{PSDHKA1A_Diff_window} for the HK1A model across different frequency bands to understand the effect of modified gravity. Since the black hole's spin parameter $a/M=0.0657$ is small, the shock cone is almost not warped around the black hole. Therefore, the Lense-Thirring effect on the resulting QPO frequencies is almost zero. Consequently, in the 0-100 nHz band range, the same frequencies with the same amplitude are numerically observed at all $r$ values.

A careful examination reveals that in this model, the values of frequencies calculated at different locations across the entire frequency band are the same. They also occur with similar amplitudes. Only the amplitudes of the frequencies at $r=6.08M$ are lower compared to the peaks calculated at other locations. This is because the decrease in gravitational effect would automatically reduce the observed frequency's amplitude, as the effect driving these modes on the disk  decreases.
Additionally, the decrease in gravitational effect causes energy dissipation. The reason for this energy dissipation is the strong shock wave present in the formed shock cone. In the presence of shock waves, even in a perfect fluid, there can be irreversible processes that lead to an increase in entropy and effective energy dissipation, although this is not due to viscosity or heat conduction.

When Fig.\ref{PSDHKA1A_Diff_window} is compared with the rapidly rotating Kerr black hole model given in Fig.\ref{Kerr09_Diff_window}, significant differences between the two cases are evident. One of these differences is, as previously mentioned, the Lense-Thirring effect is almost negligible for the case given in Fig.\ref{PSDHKA1A_Diff_window}. Also, for the model in Fig.\ref{PSDHKA1A_Diff_window}, there is a possibility of observing both low and high-frequency QPOs. Therefore, if $M87^*$ is explained with this model, the frequencies observed around $M87^*$ can vary from nHz to mHz. Considering that the Lense-Thirring effect does not create any QPOs, all the QPO frequencies formed here are generated by the excitation of the pressure modes of the matter trapped inside the shock cone.

As previously explained, the radial and angular oscillation of the sound speed generates some of the QPO frequencies (fundamental modes) found in Fig.\ref{PSDHKA1A_Diff_window}, while others have formed new frequencies through the  nonlinear coupling. The probability of forming nonlinear coupling frequencies is higher in Hairy Kerr gravity. The reason is that this gravity has a more complex effect on the formation and oscillation of the disk compared to the Kerr gravity. Therefore, the oscillation modes are affected by hair parameter. Thus, it leads to the formation of a large number of new QPOs through nonlinear couplings. This increases the likelihood of observing more resonance conditions in observations.

\subsection{QPO frequencies of $M87^*$ from the observed shadow values given in Table \ref{Inital_Con_2}}
\label{Num_QPO2}


The characteristics of the shadow of the $M87^*$ black hole have also been revealed by calculating the area ($A/M^2$) and the oblateness ($D$)  of the observed region. Based on these characteristics, Ref.\citep{Afrin_2021} has defined the parameters of the Hairy Kerr black hole that are consistent with the shadow of the $M87^*$ black hole, as given in Table \ref{Inital_Con_2}. In this paper, using these parameters as initial conditions, we have revealed the dynamic structure of the shock cone formed around the $M87^*$ black hole. Generally, it has been observed that the dynamic structure of the resulting cone is almost the same as the physical structure of the cone obtained using the initial conditions in Table \ref{Inital_Con_1}. As seen in Tables \ref{Inital_Con_1} and \ref{Inital_Con_2}, especially the stagnation point and the opening angle of the cone at this point are almost identical in both shadow calculations. Only the time to reach steady-state for each model shows slight differences, indicating that the models have different internal dynamics.

\begin{figure*}
  \vspace{1cm}
  \center
  \psfig{file=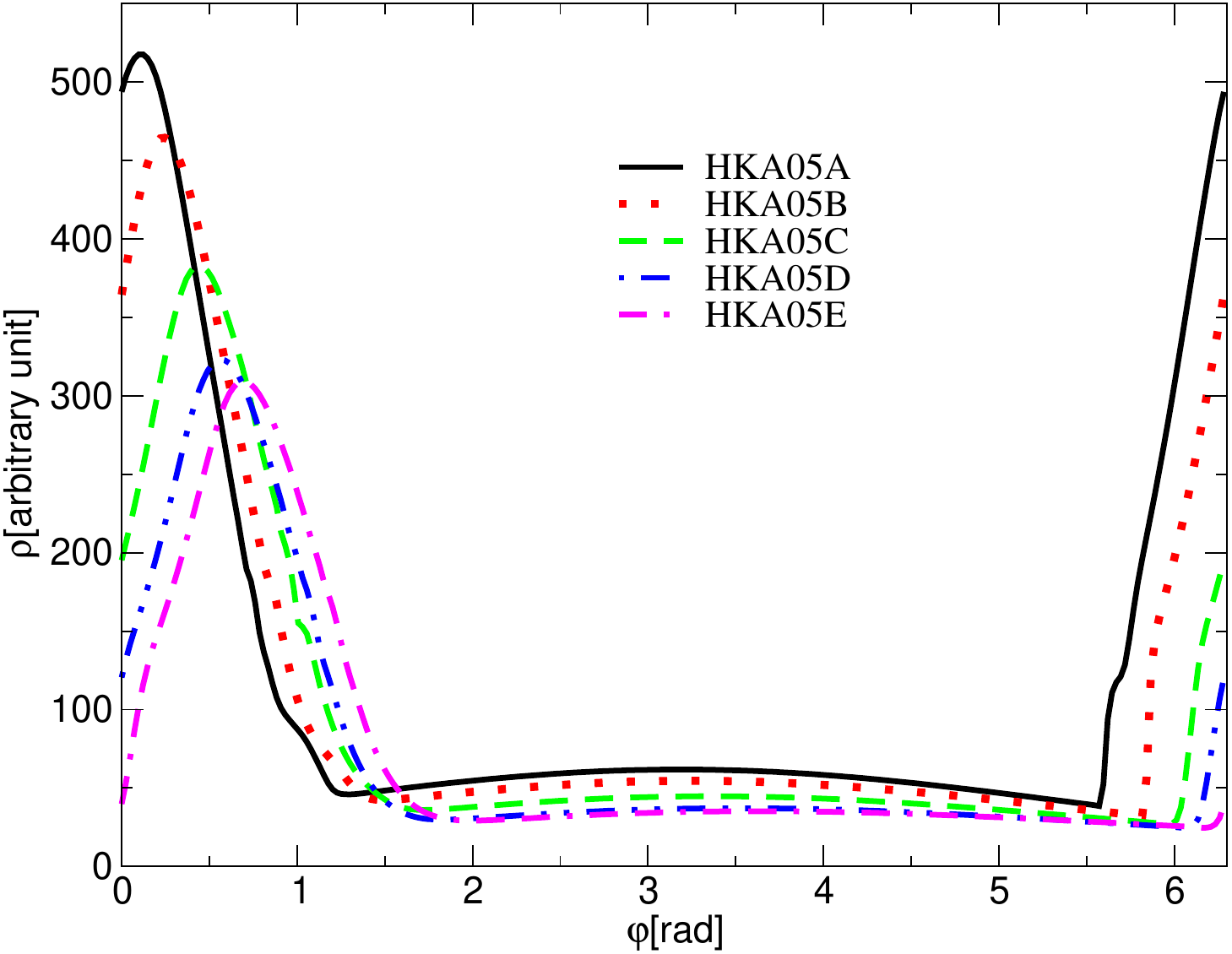,width=7.0cm, height=6.0cm}\hspace*{0.15cm}
  \psfig{file=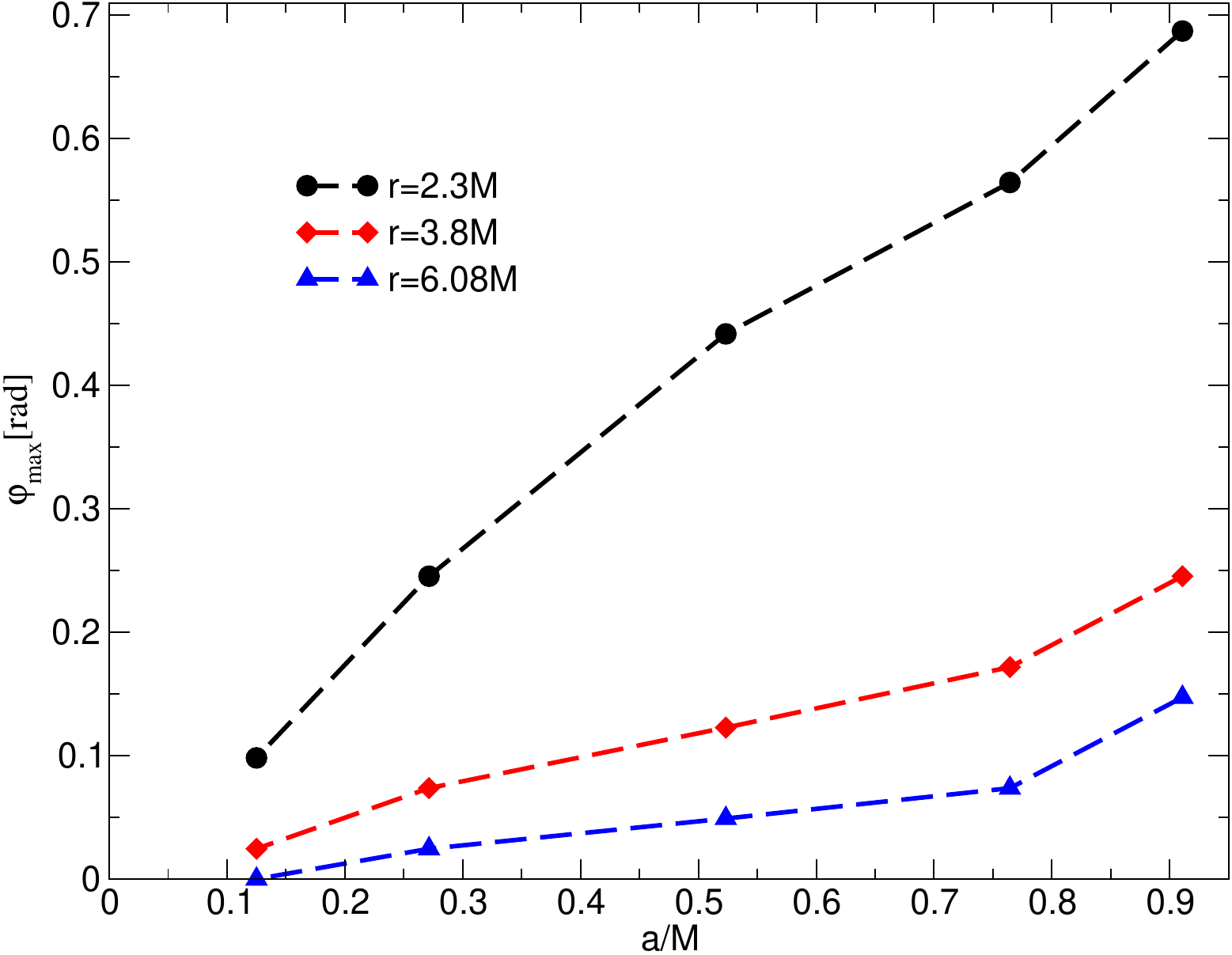, width=7.0cm, height=6.0cm}\\
   \vspace*{0.5cm}   
  \psfig{file=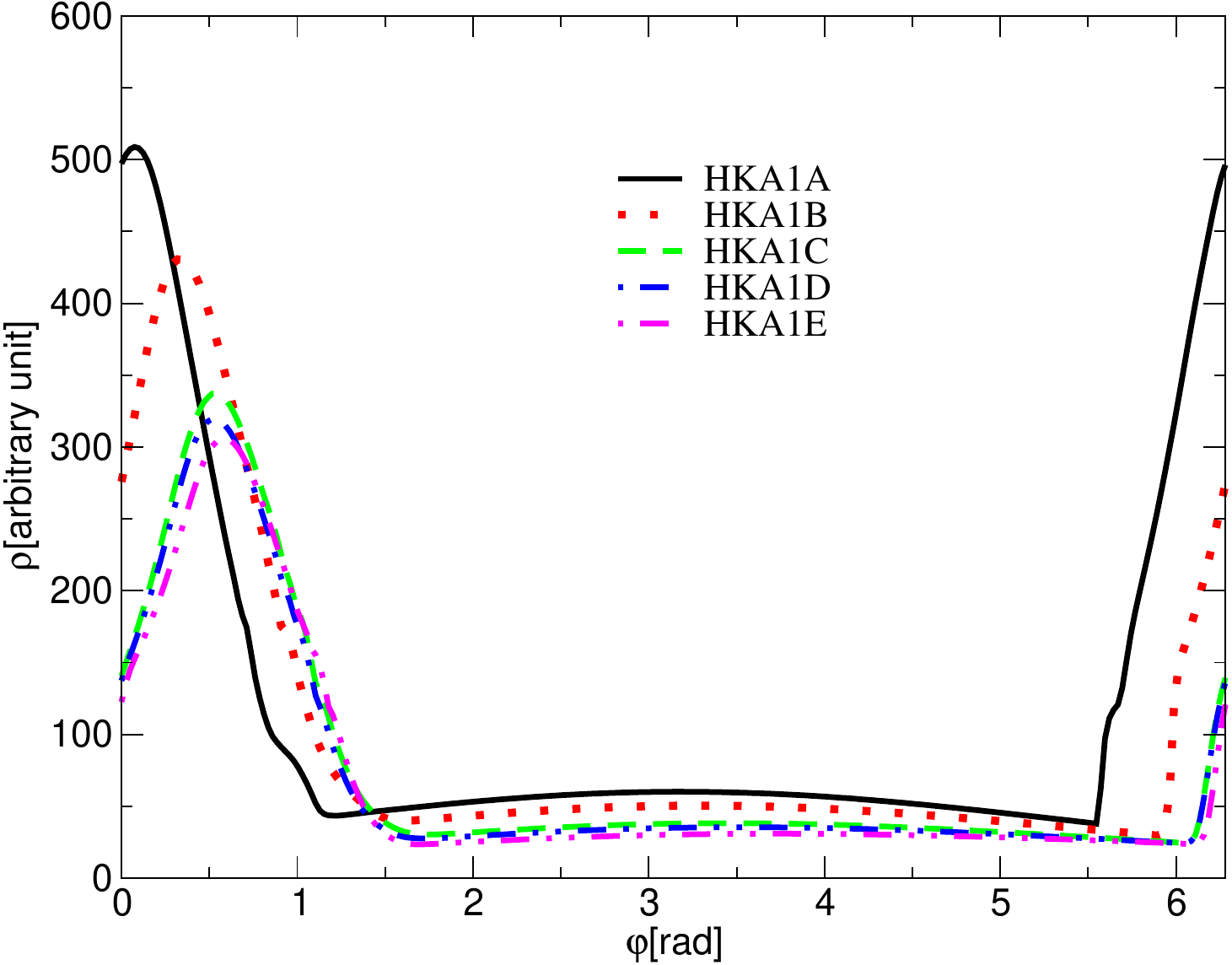, width=7.0cm, height=6.0cm}\hspace*{0.15cm}
  \psfig{file=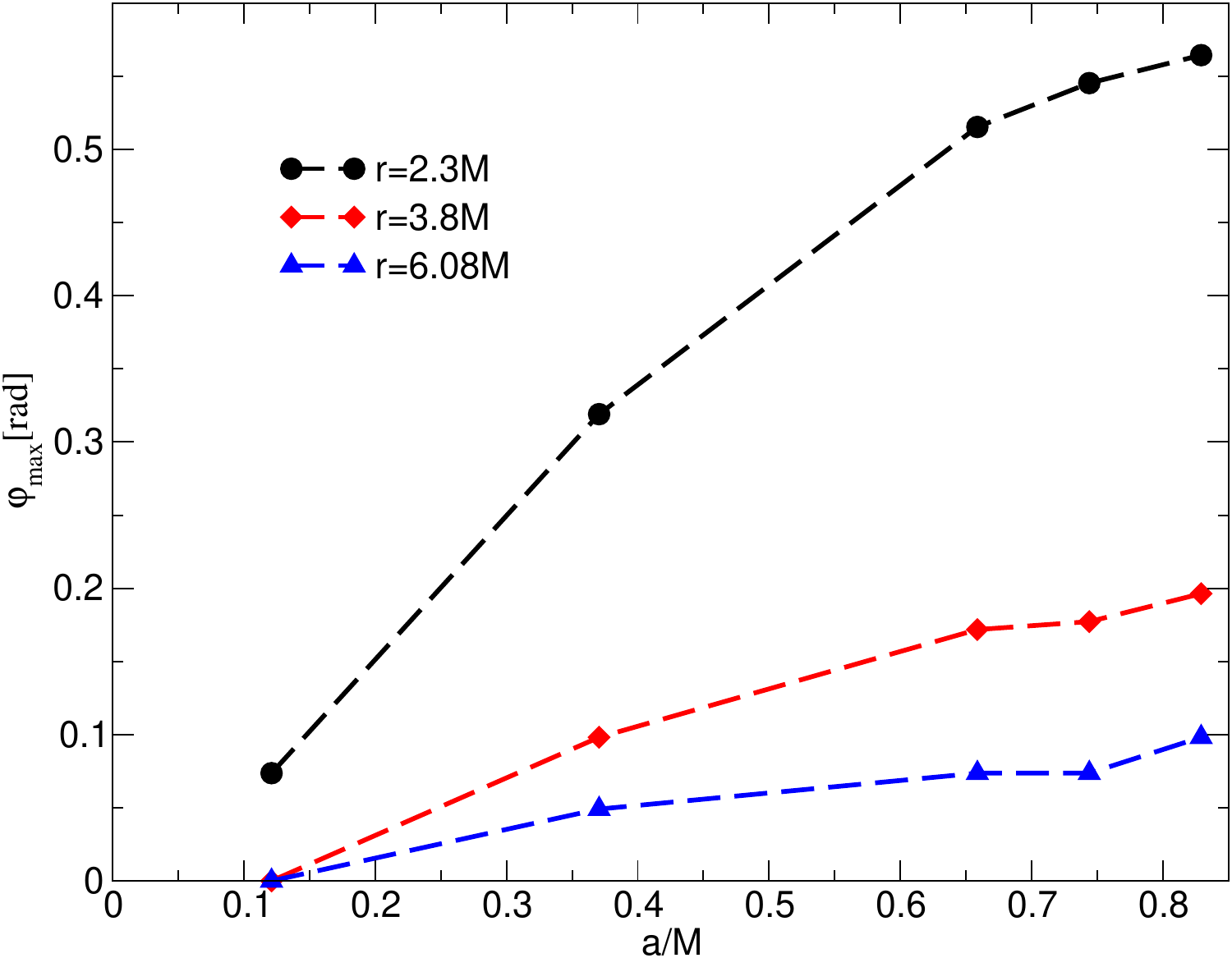, width=7.0cm, height=6.0cm}\\
  \caption{Same as Fig.\ref{DensHK05}, but it is for the known shadow observable area ($A/M^2$) and oblateness ($D$) for the initial values given in Table \ref{Inital_Con_2}.
    }
\vspace{1cm}
\label{DensHKA05}
\end{figure*}

Similarly, as seen in the left column of Fig.\ref{DensHKA05}, the variation of density at $r=2.3M$ in different models and the behavior of the frame-dragging effect according to the black hole's spin parameter given in the right column of the same figure are very similar to those in Fig.\ref{DensHK05}. The effect of the black hole's spin parameter is the same in Figures \ref{DensHK05} and \ref{DensHKA05}. Likewise, the complexity added to the physical system by the deviations from Kerr and hair parameters is also the same. However, there are differences in details. For example, the density of matter accreted inside the cone is higher than the density obtained using the initial values in Table \ref{Inital_Con_2}. This difference in density is expected, as it is related to the geometry of the shadow's surface.

The results found are also theoretically expected. Because the parameters defining the shadow of $M87^*$ given in Tables \ref{Inital_Con_1} and \ref{Inital_Con_2} create a similar gravitational effect to the accretion mechanism around the black hole, the interactions with the surrounding matter create different effects due to the differences in the oblateness and radius of the shadow. Therefore, the effects they add to the resulting cone show differences, leading to different complexities in the physical events within the cone.

\begin{figure*}
  \vspace{1cm}
  \center
  \psfig{file=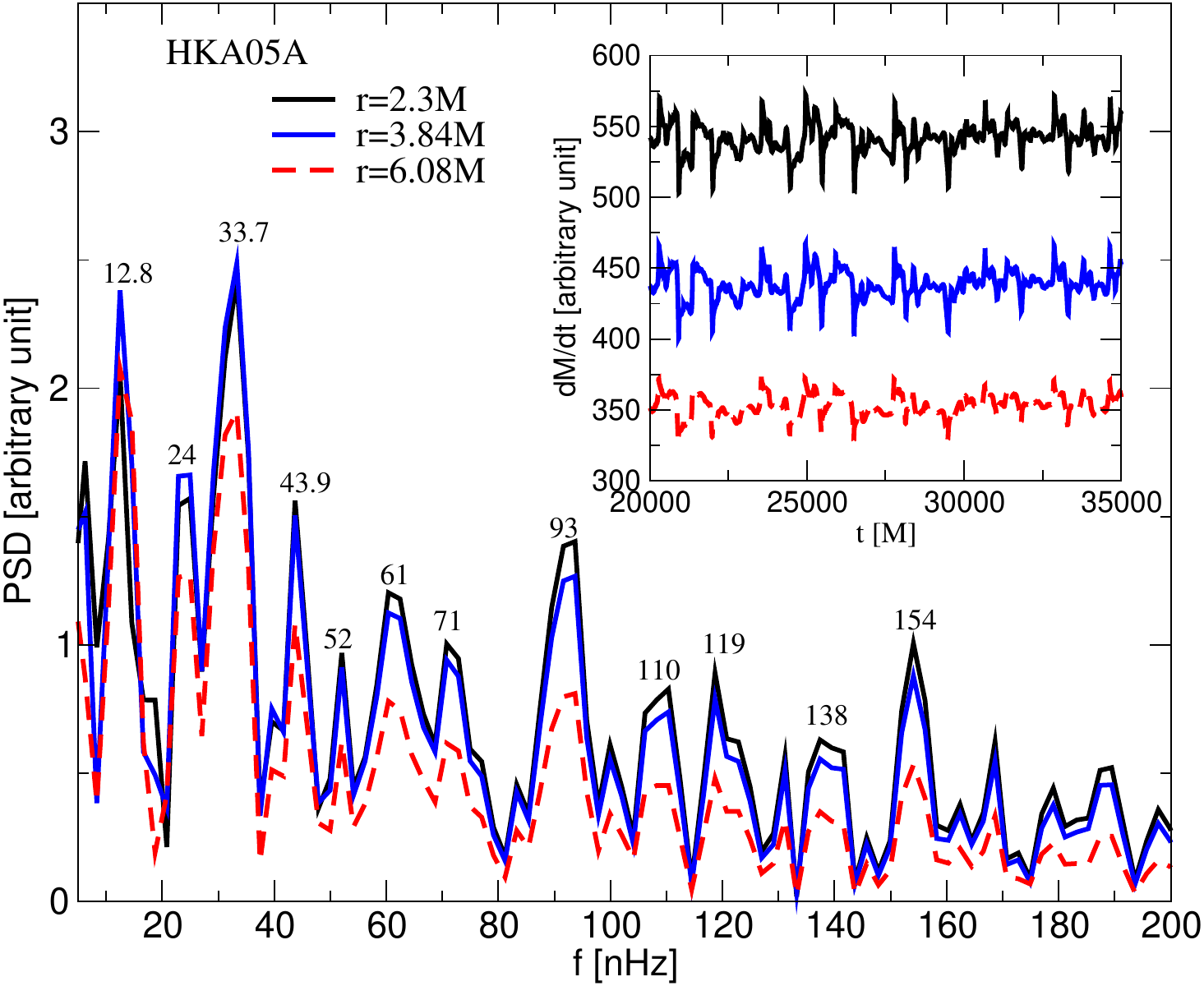,width=7.5cm, height=6.8cm}\hspace*{0.15cm}
  \psfig{file=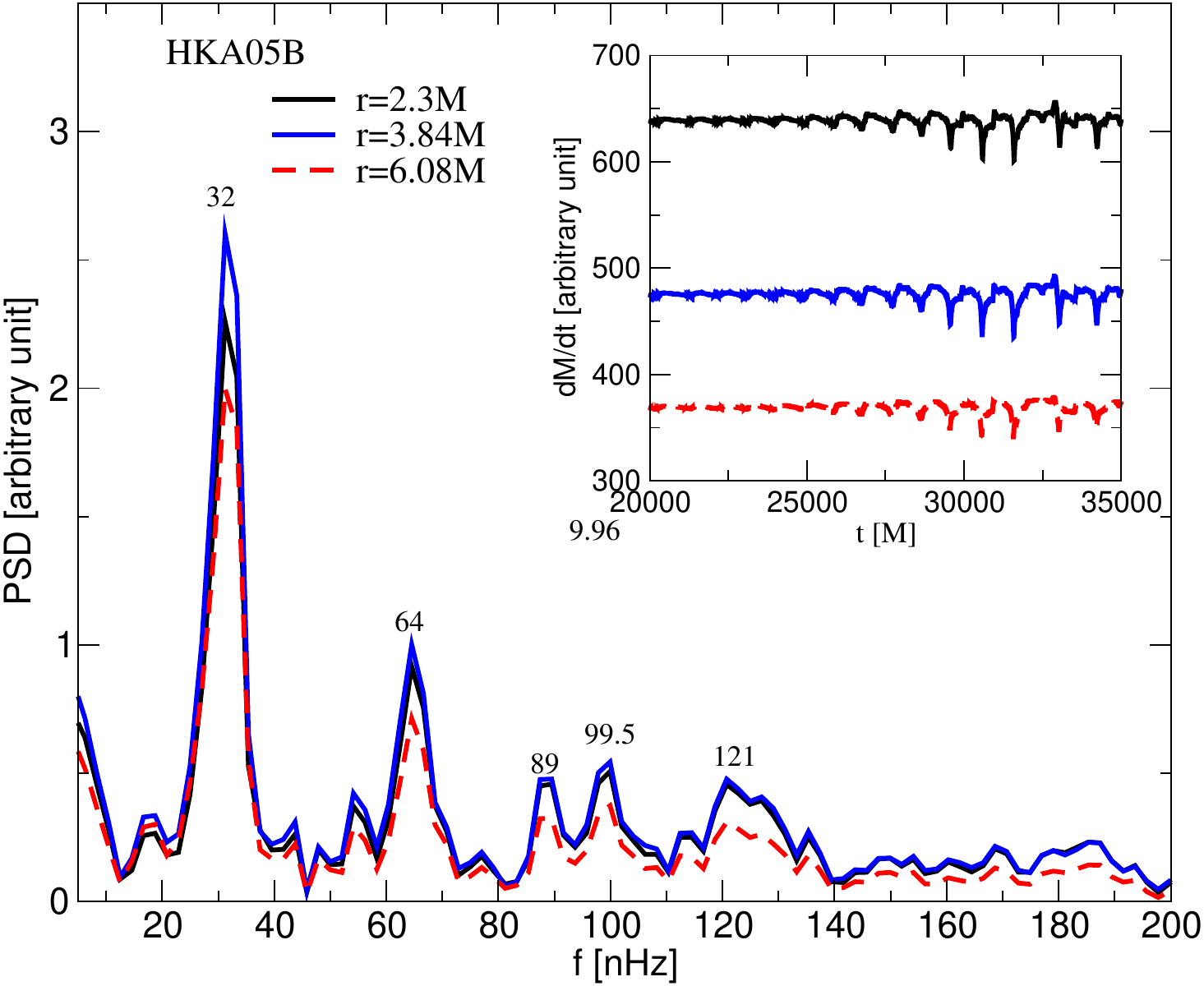,width=7.5cm, height=6.8cm}\\
  \vspace*{0.3cm}
  \psfig{file=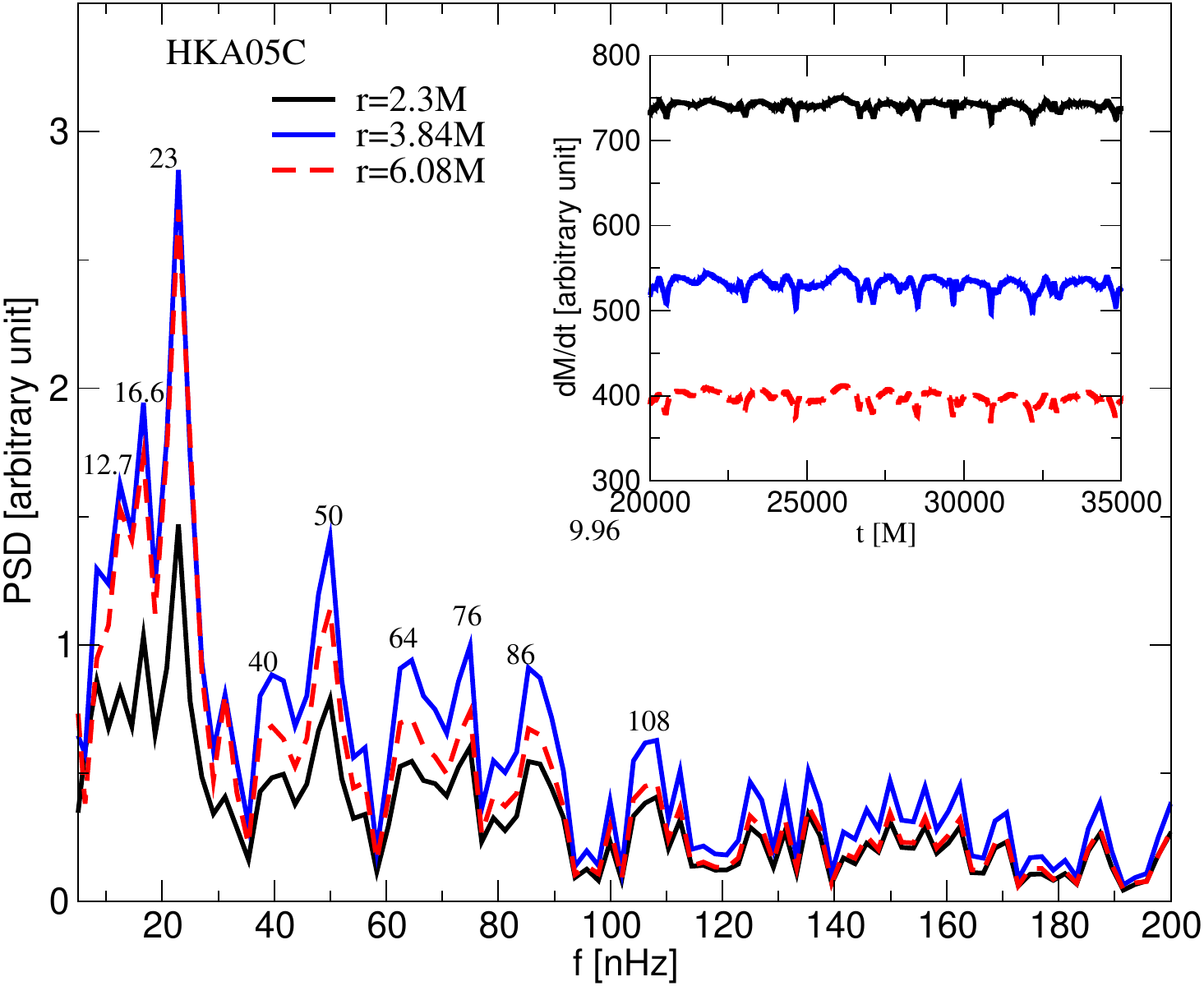,width=7.5cm, height=6.8cm}\hspace*{0.15cm}
  \psfig{file=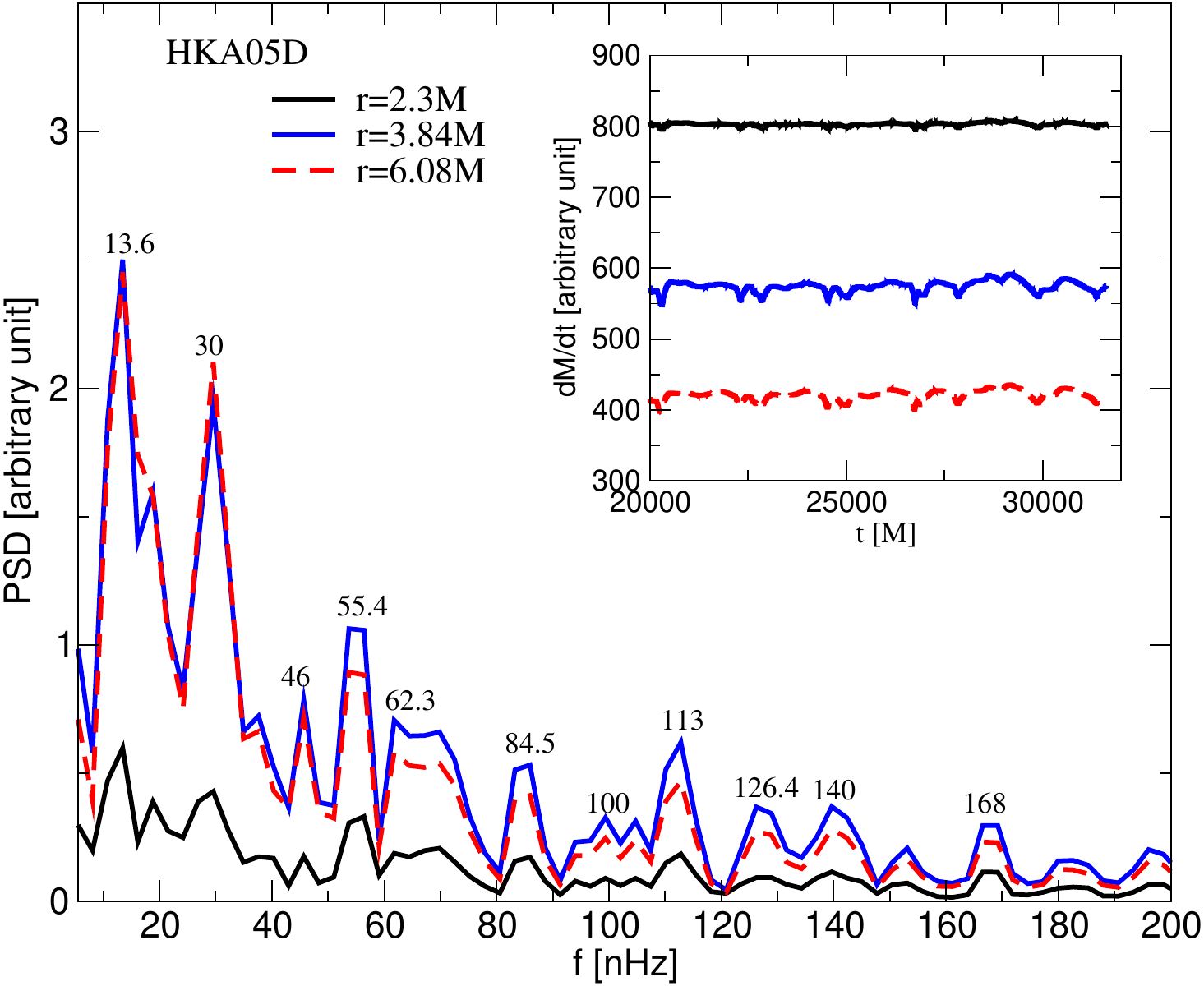,width=7.5cm, height=6.8cm}\\
  \vspace*{0.3cm}
  \psfig{file=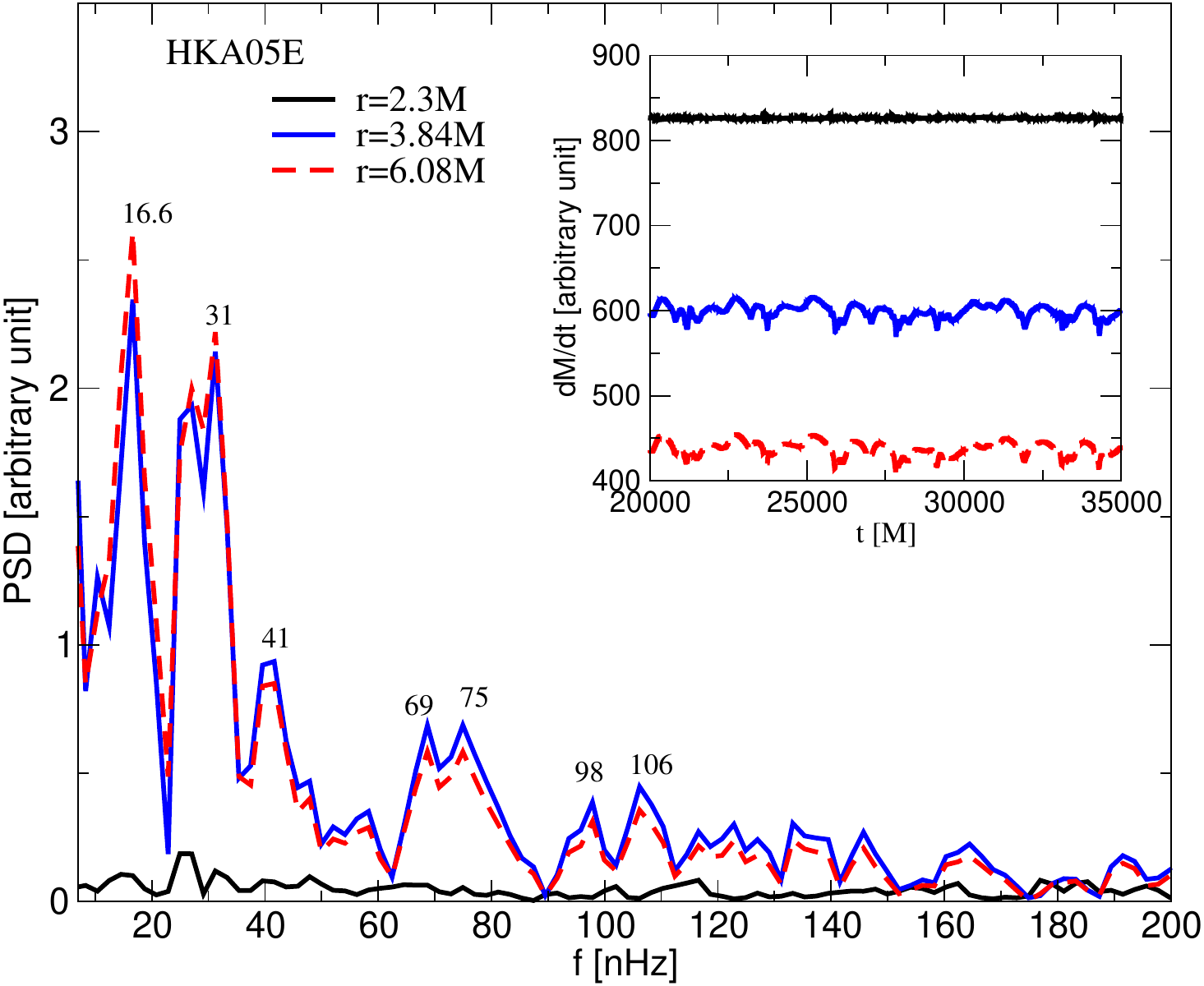,width=7.5cm, height=6.8cm}\\  
      \caption{PSD analysis around the Hairy Kerr black hole in a strong gravitational field  at $r=2.3M$, $r=3.84M$, and $r=6.08M$ for the known shadow observable area ($A/M^2$), oblateness ($D$) , and $\eta=0.5$. It shows PSD analysis results obtained from the mass accretion rates calculated at different radii long after the disk has reached the steady state for the models given in Table \ref{Inital_Con_2}
    }
\vspace{1cm}
\label{PSDHKA05_1}
\end{figure*}

\begin{figure*}
  \vspace{1cm}
  \center
     \psfig{file=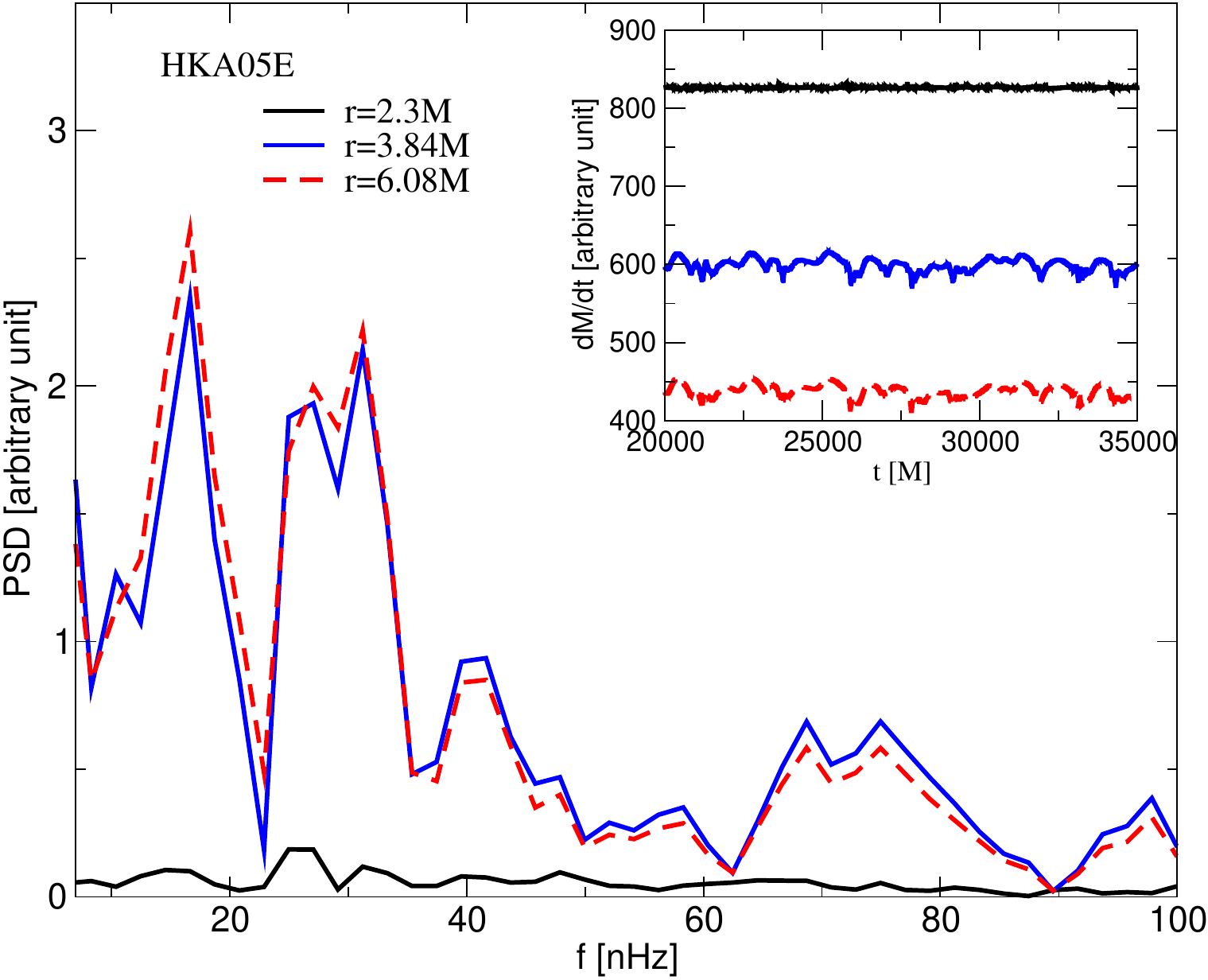,width=3.5cm, height=4.0cm}\hspace*{0.15cm}
     \psfig{file=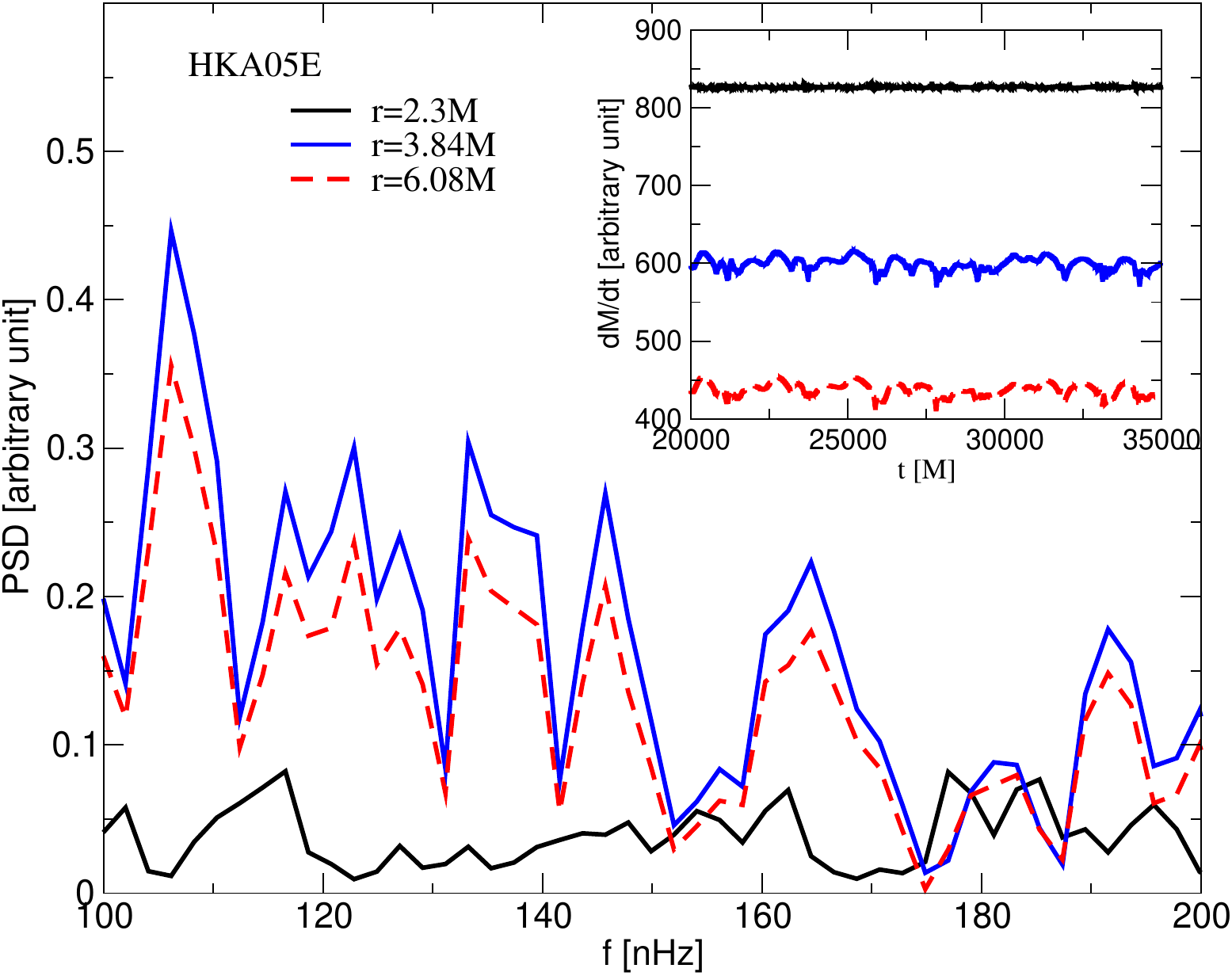,width=3.5cm, height=4.0cm}\hspace*{0.15cm}
     \psfig{file=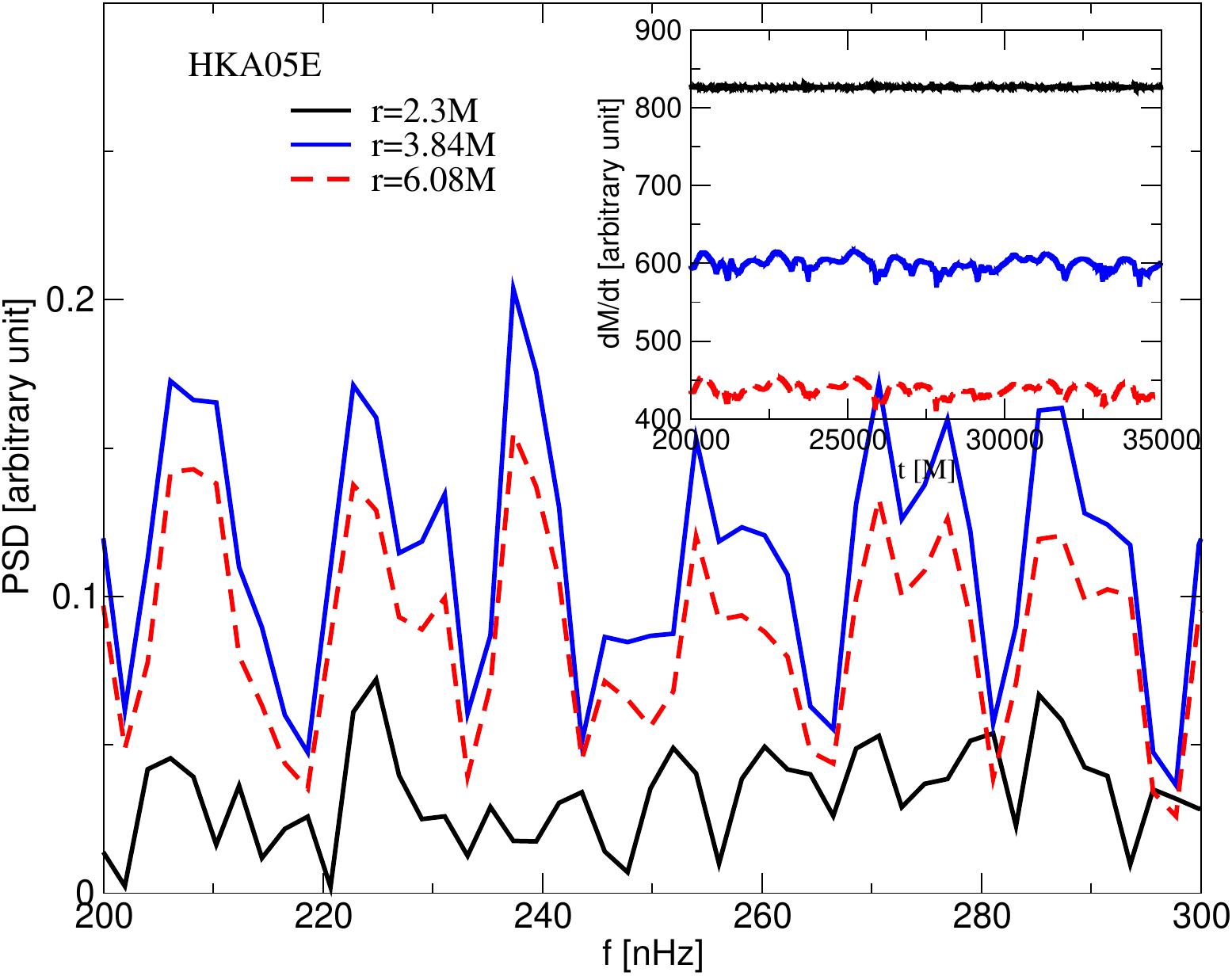,width=3.5cm, height=4.0cm}\hspace*{0.15cm}
     \psfig{file=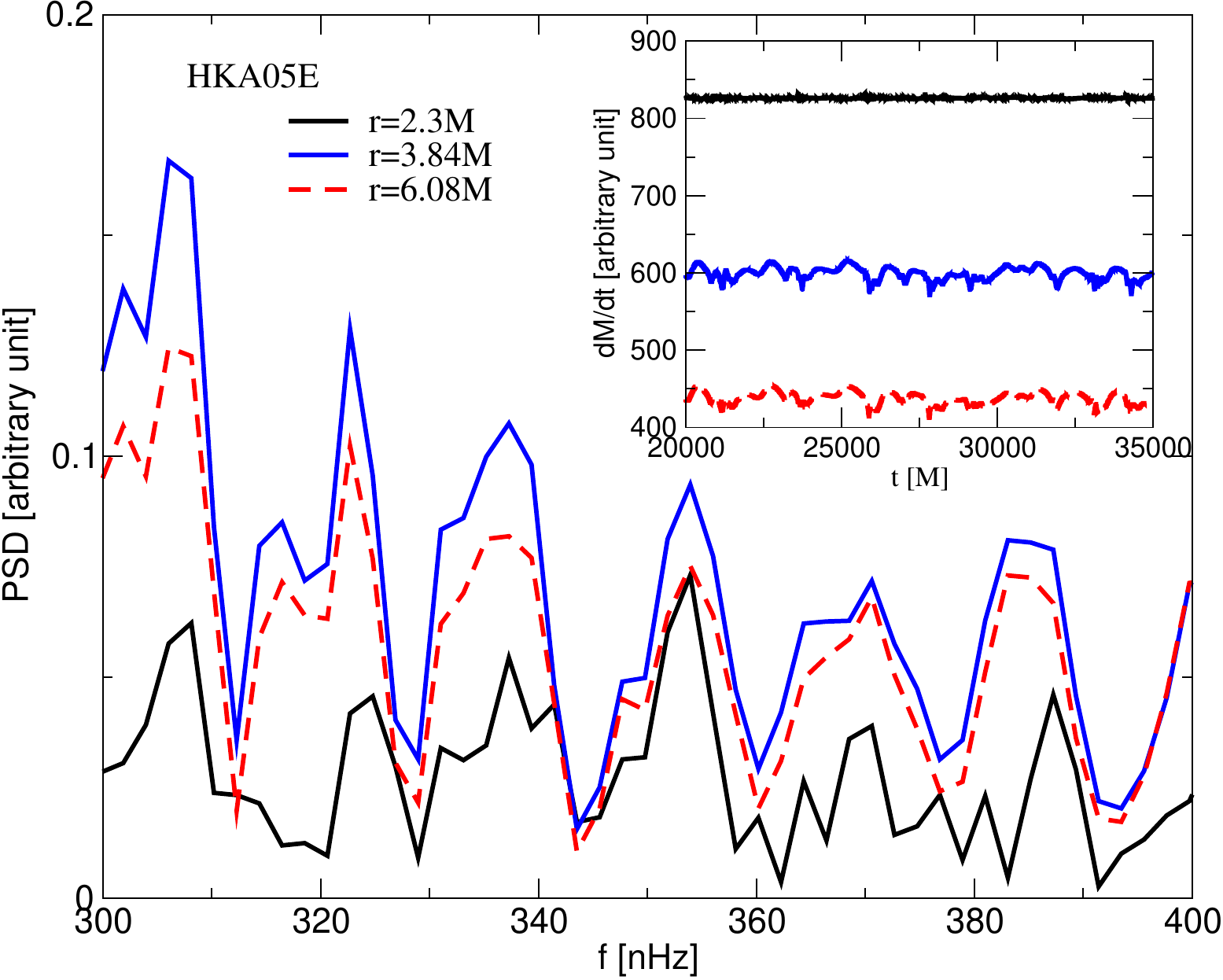,width=3.5cm, height=4.0cm}\\
     \vspace*{0.25cm}
     \psfig{file=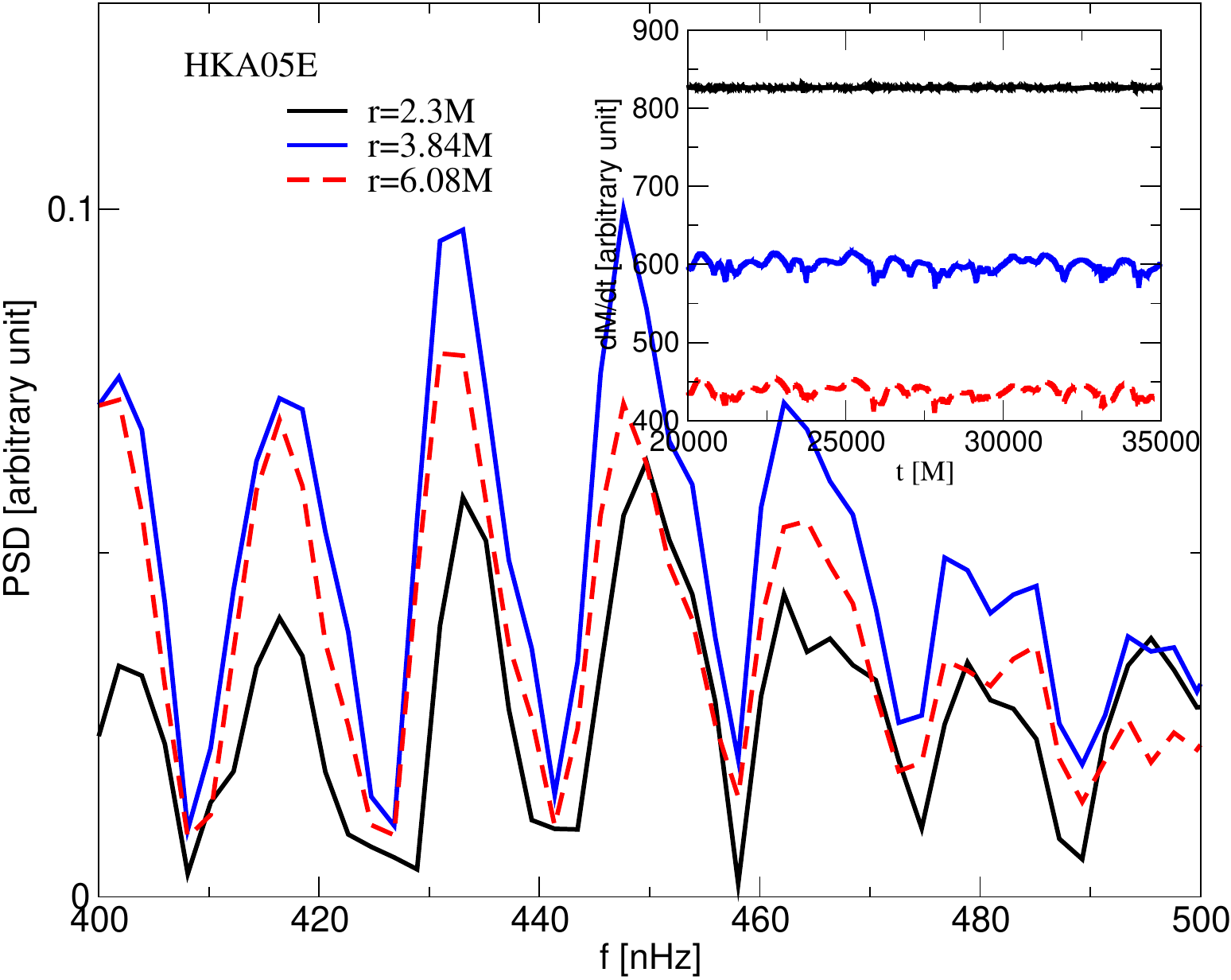,width=3.5cm, height=4.0cm}\hspace*{0.15cm}
     \psfig{file=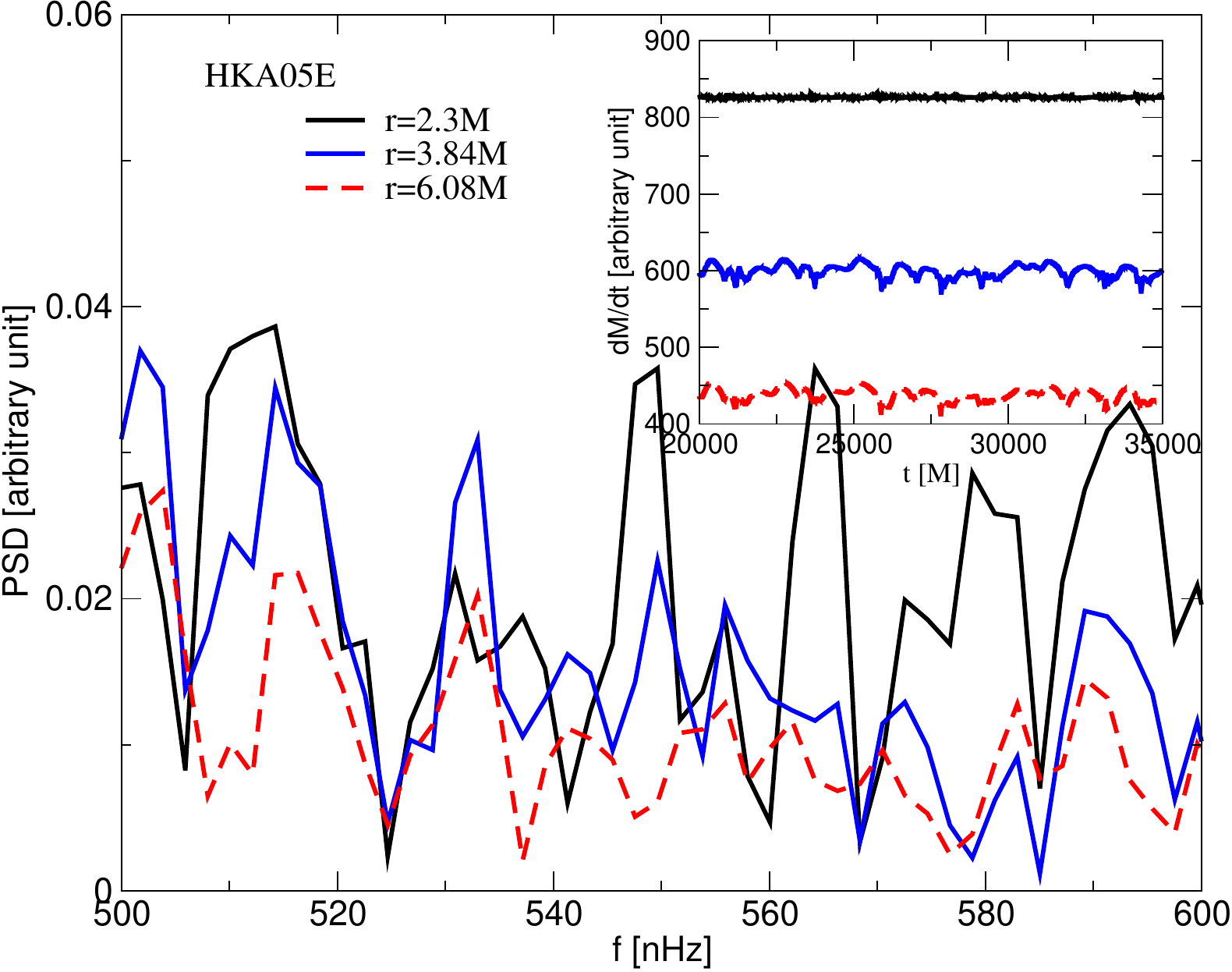,width=3.5cm, height=4.0cm}\hspace*{0.15cm}
     \psfig{file=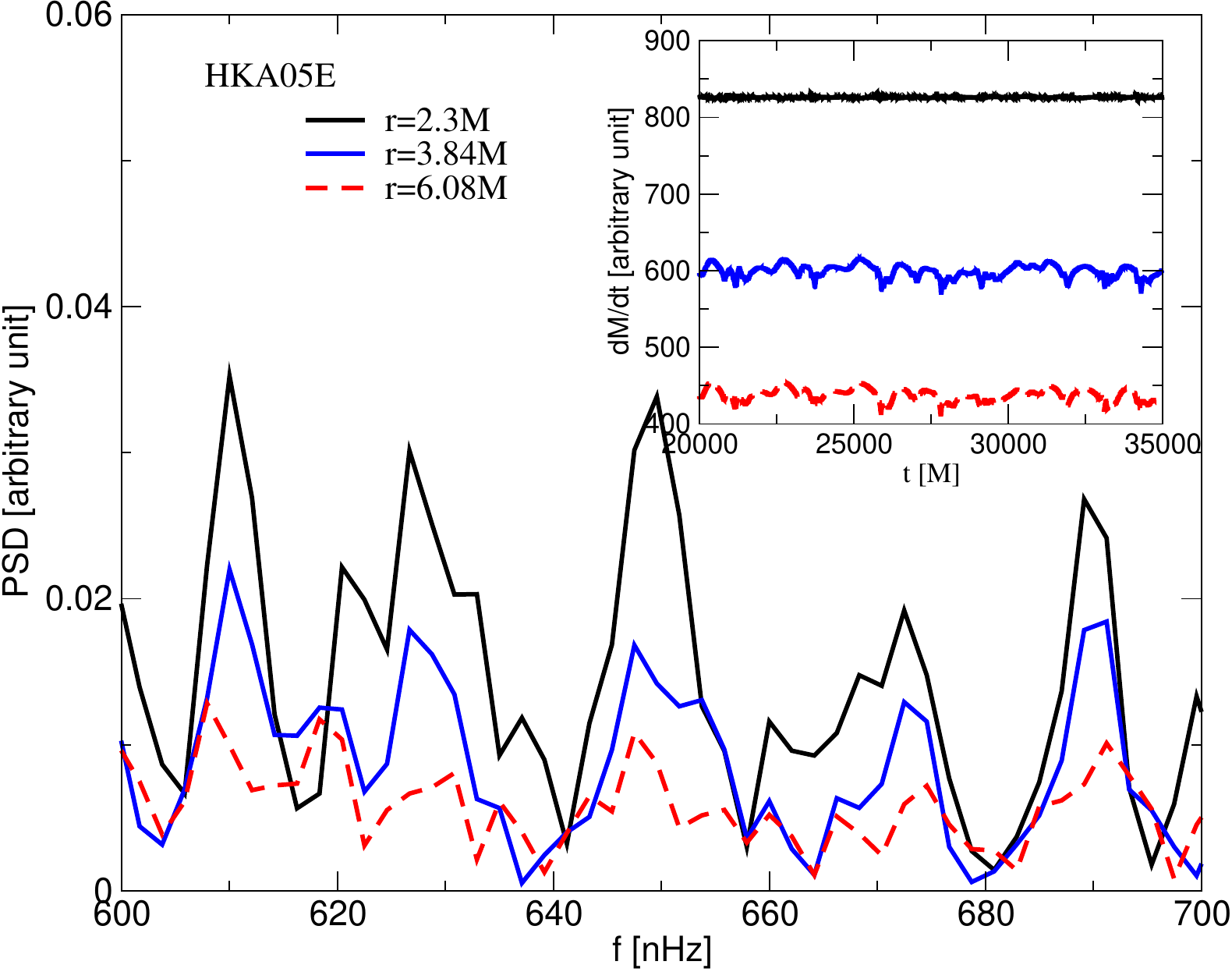,width=3.5cm, height=4.0cm}\hspace*{0.15cm}
     \psfig{file=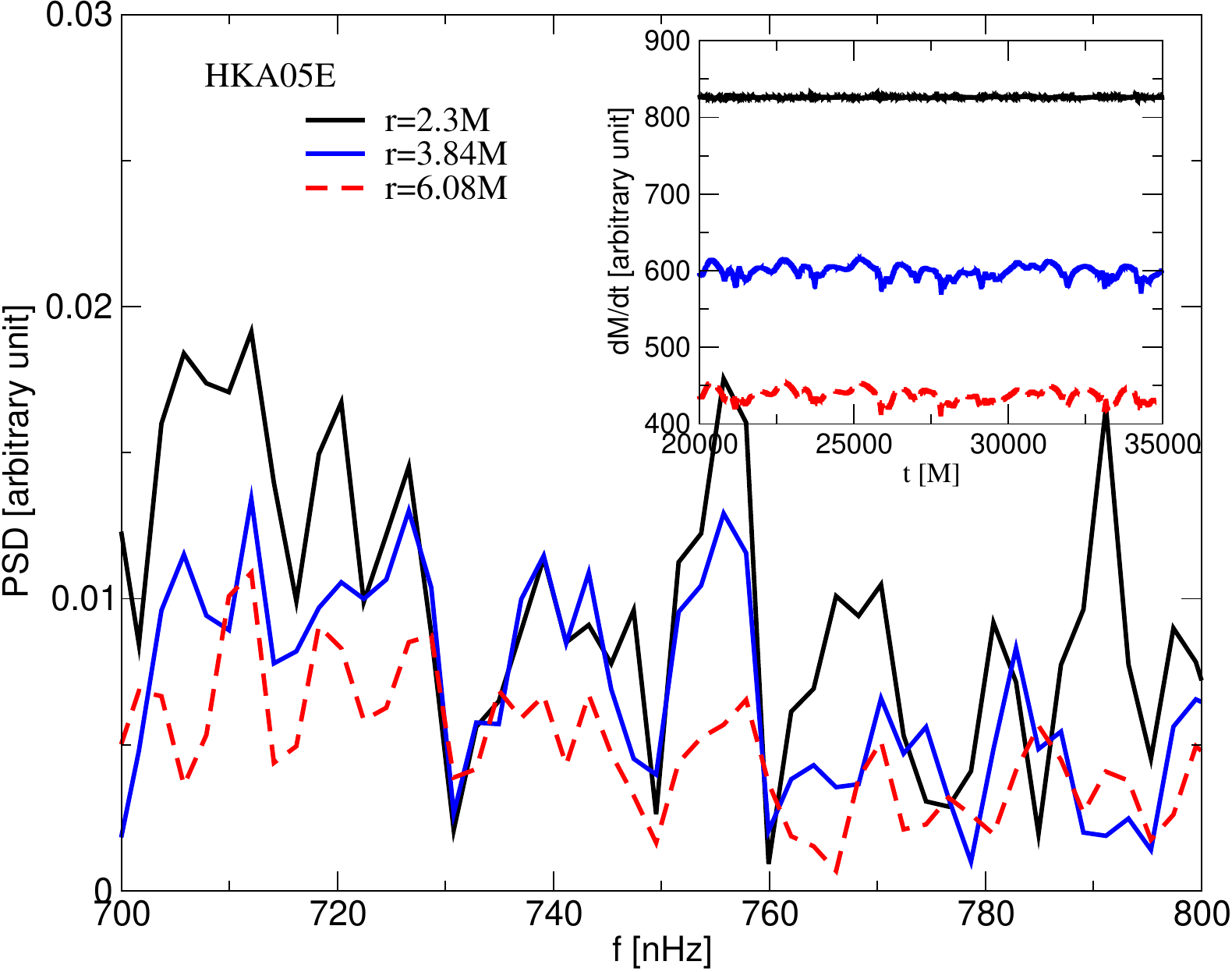,width=3.5cm, height=4.0cm}\\
     \vspace*{0.25cm}
     \psfig{file=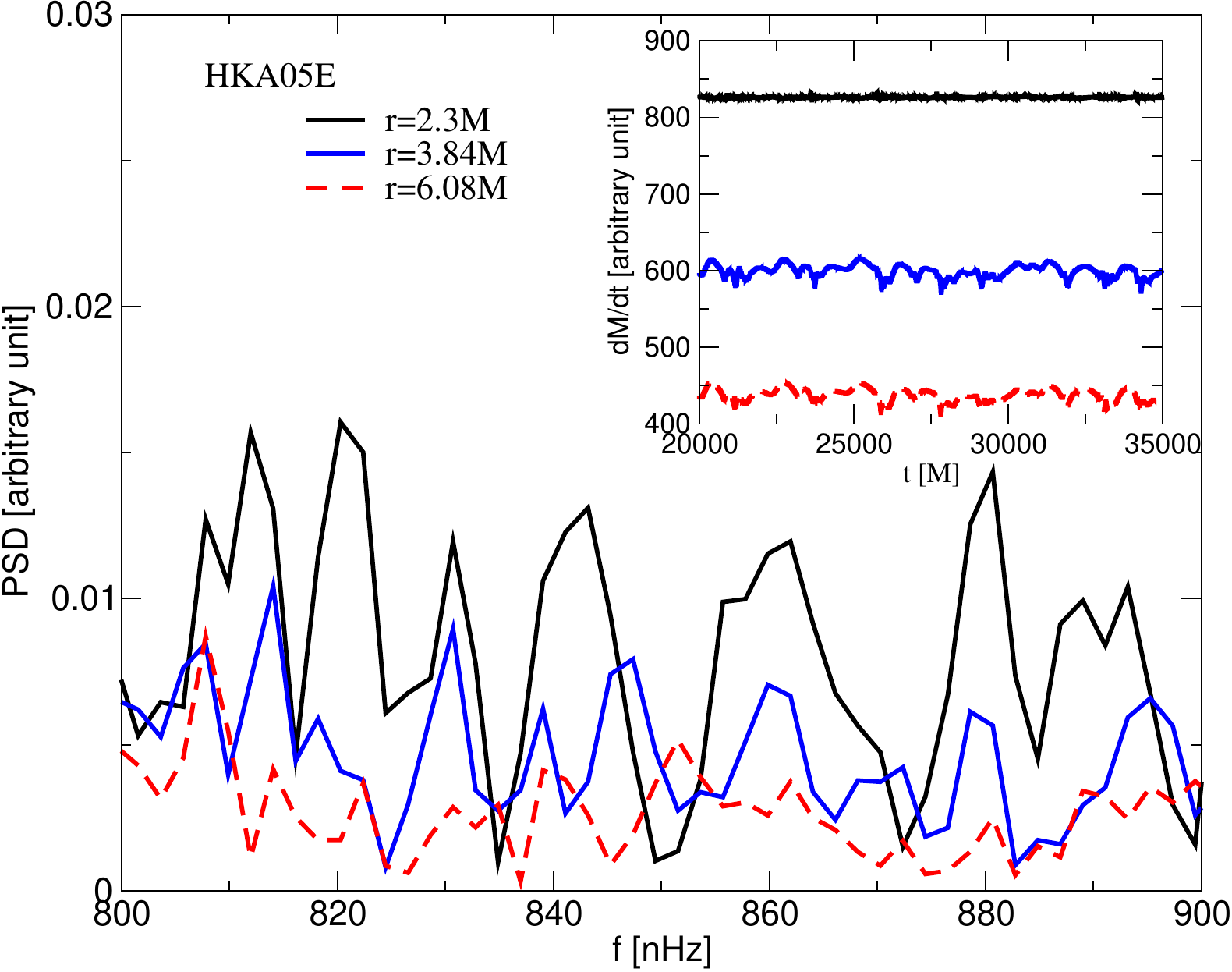,width=3.5cm, height=4.0cm}\hspace*{0.15cm}
     \psfig{file=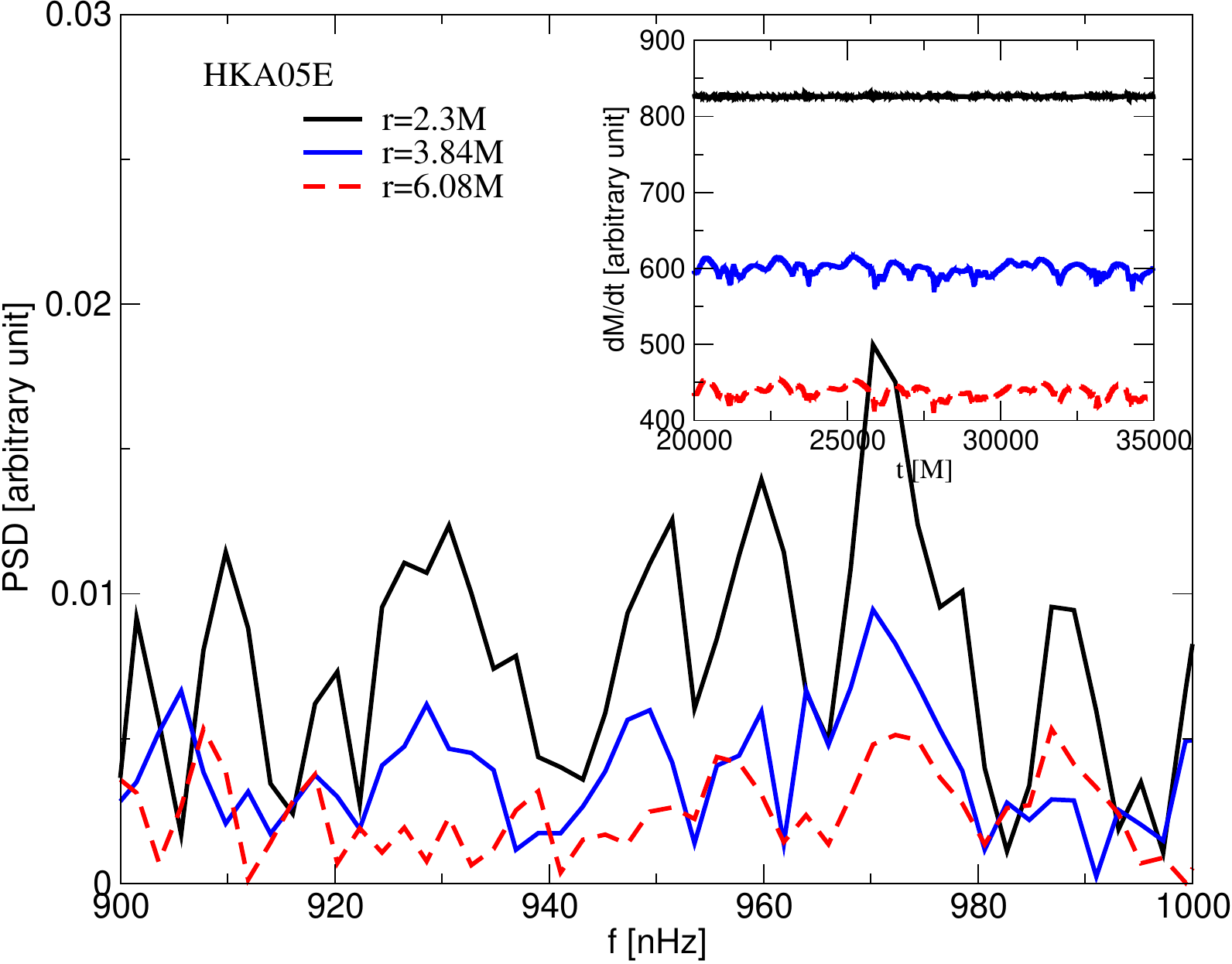,width=3.5cm, height=4.0cm}\hspace*{0.15cm}
     \psfig{file=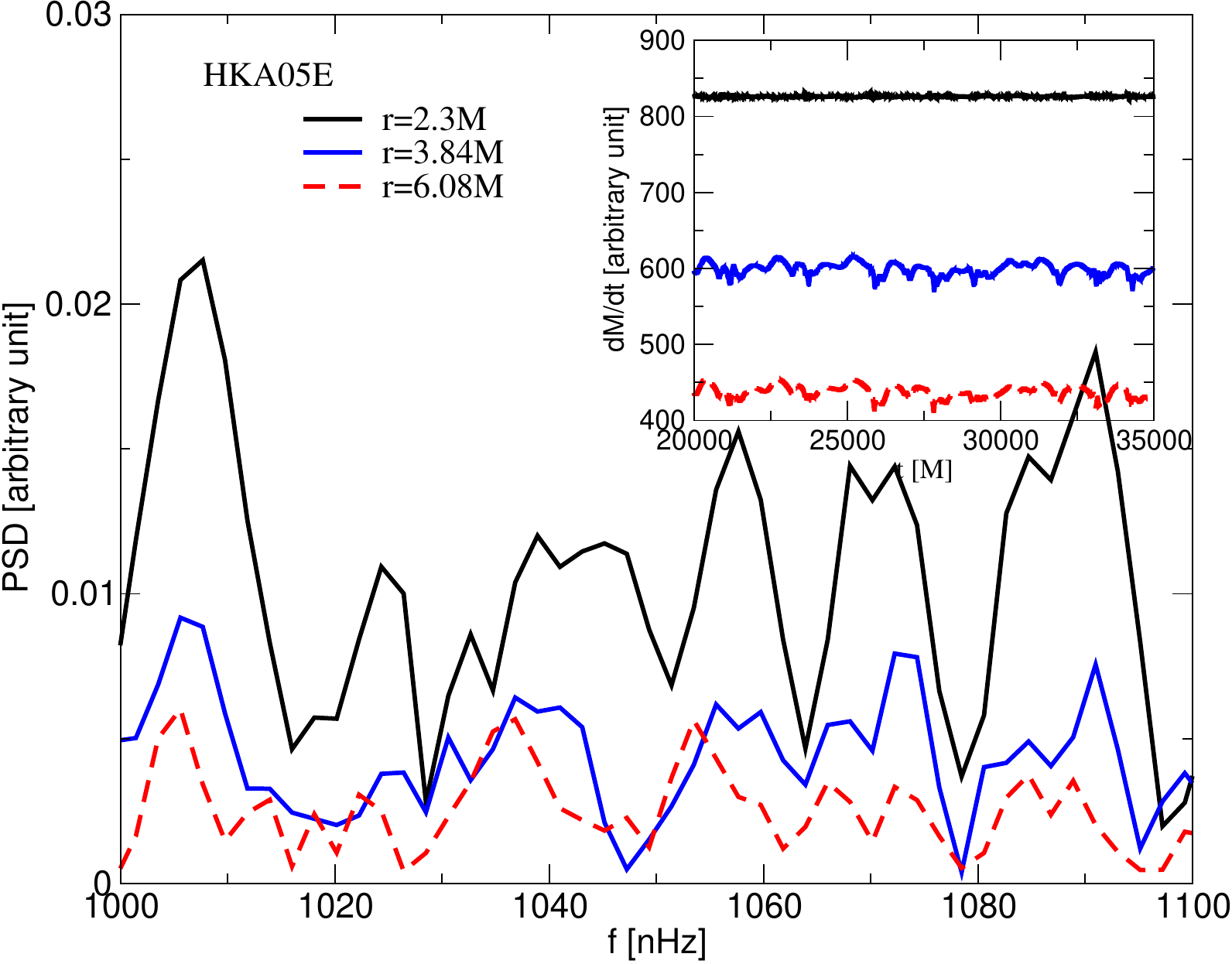,width=3.5cm, height=4.0cm}\hspace*{0.15cm}
     \psfig{file=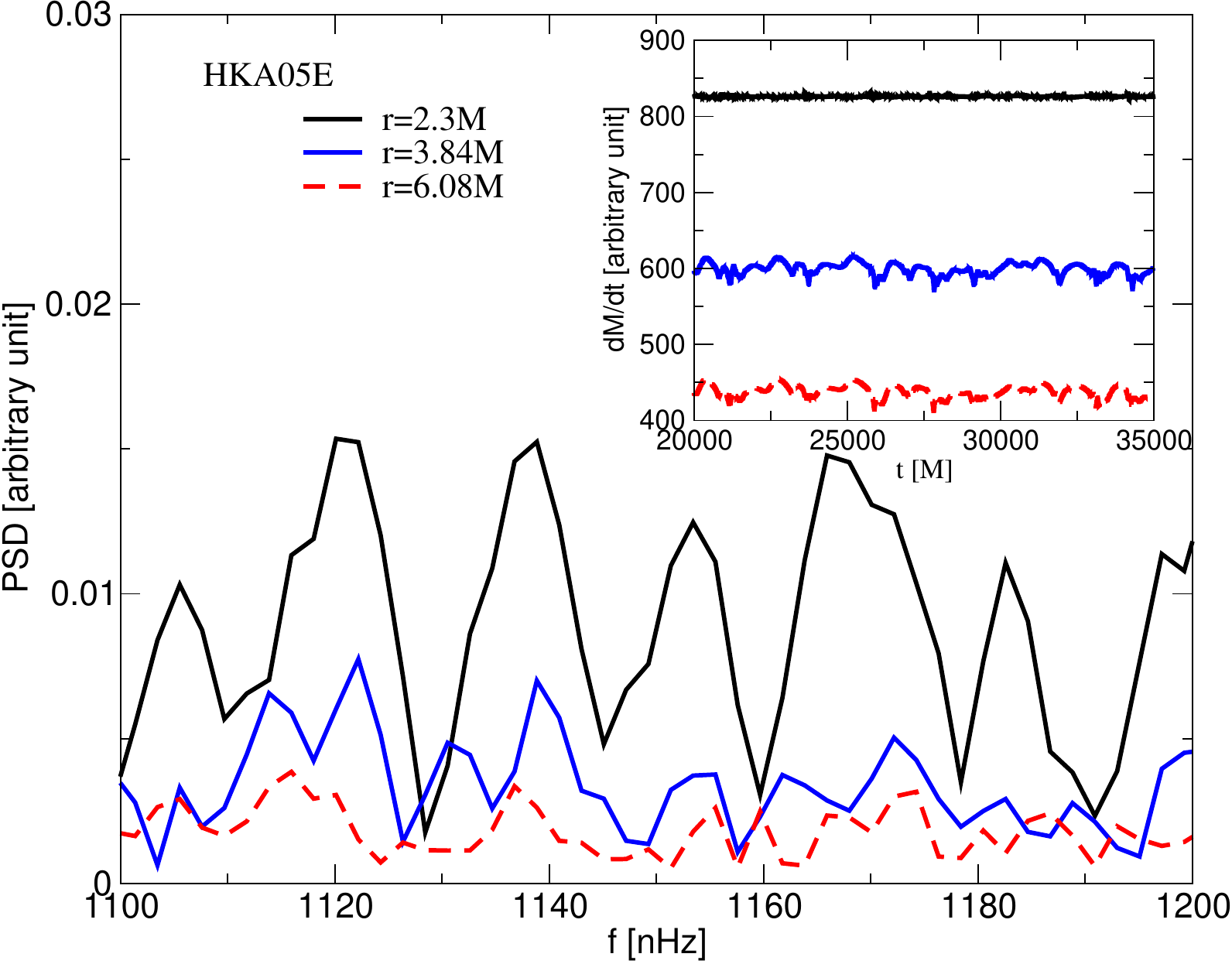,width=3.5cm, height=4.0cm}\\
     \vspace*{0.25cm}
     \psfig{file=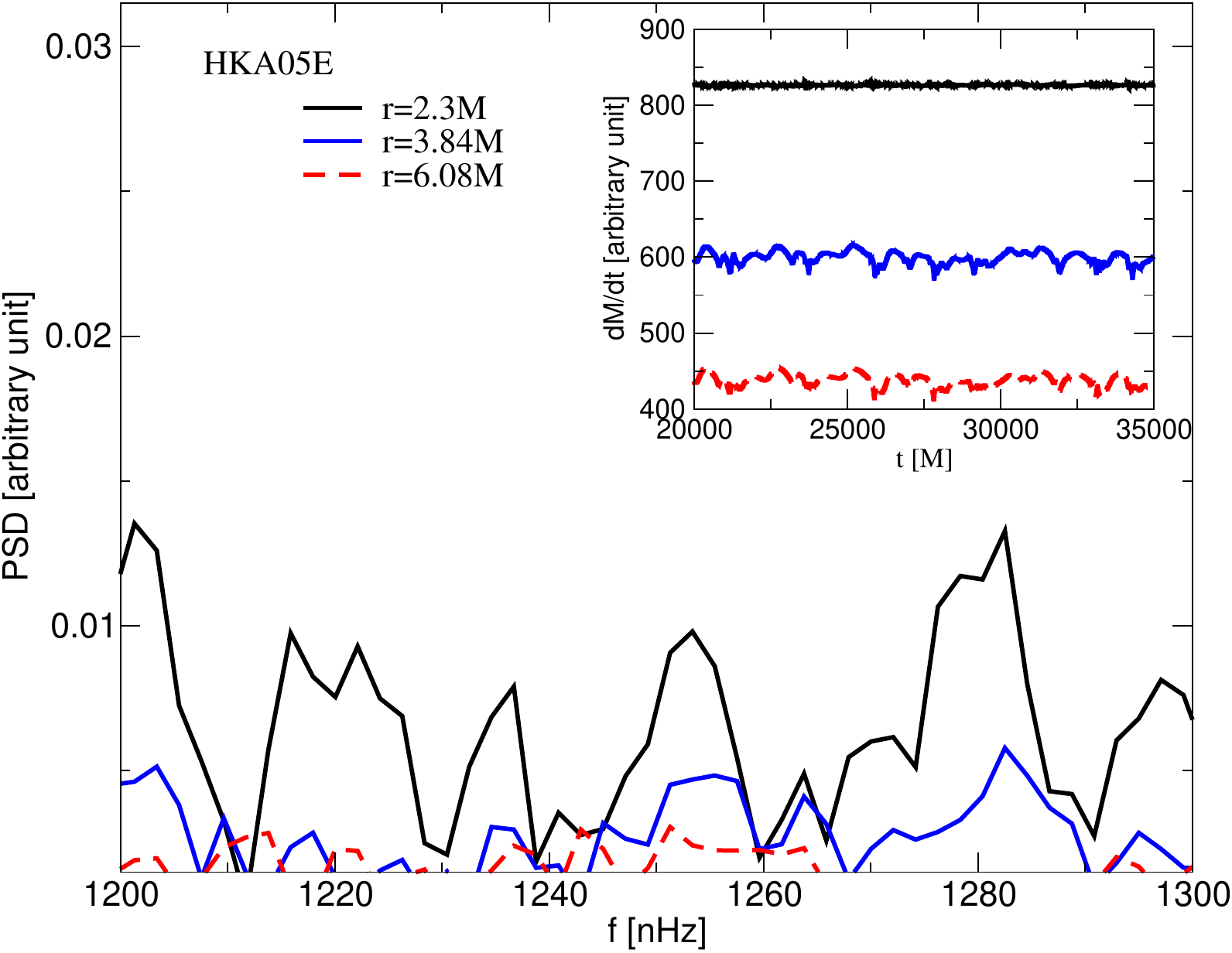,width=3.5cm, height=4.0cm}\hspace*{0.15cm}
     \psfig{file=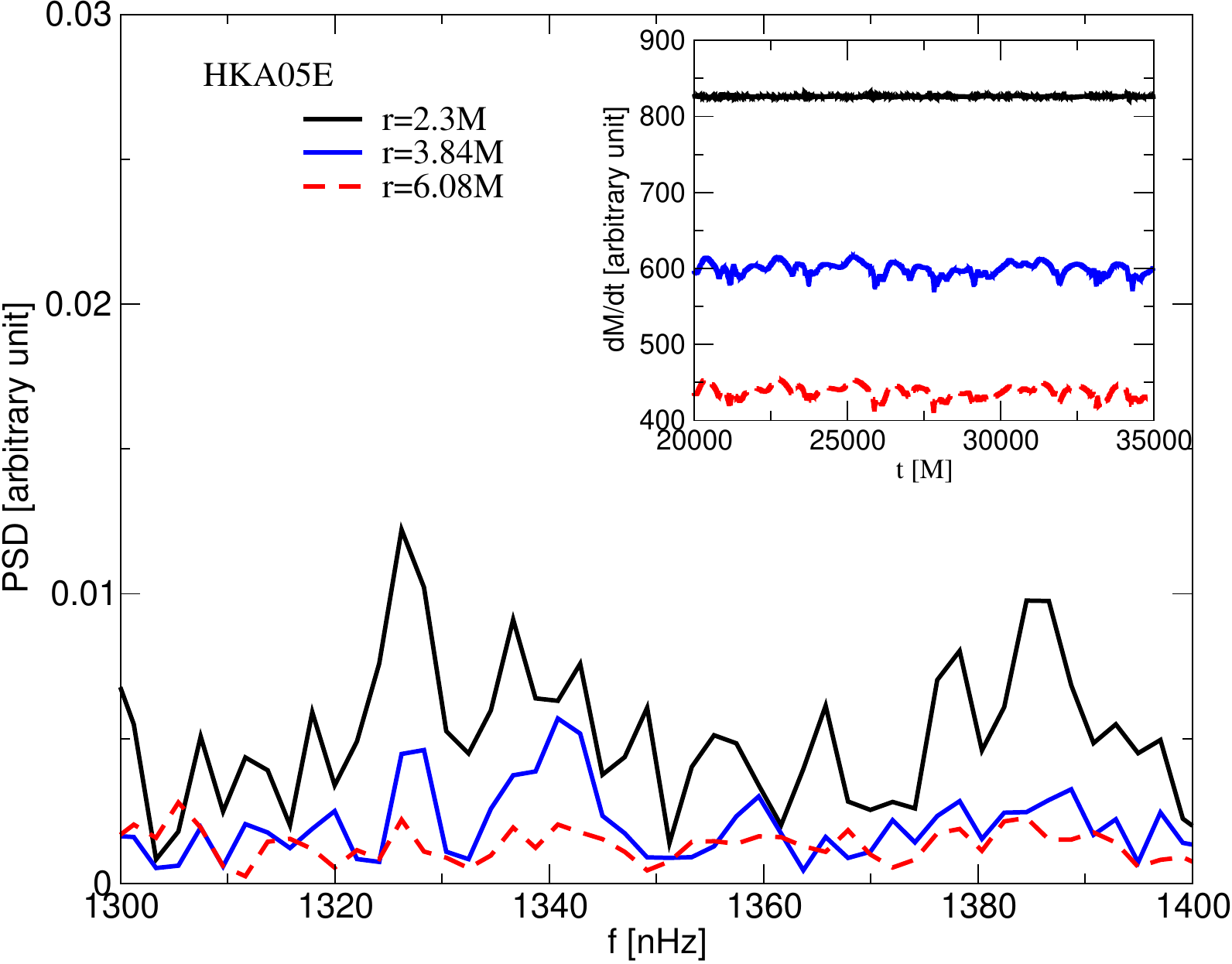,width=3.5cm, height=4.0cm}\hspace*{0.15cm}
     \psfig{file=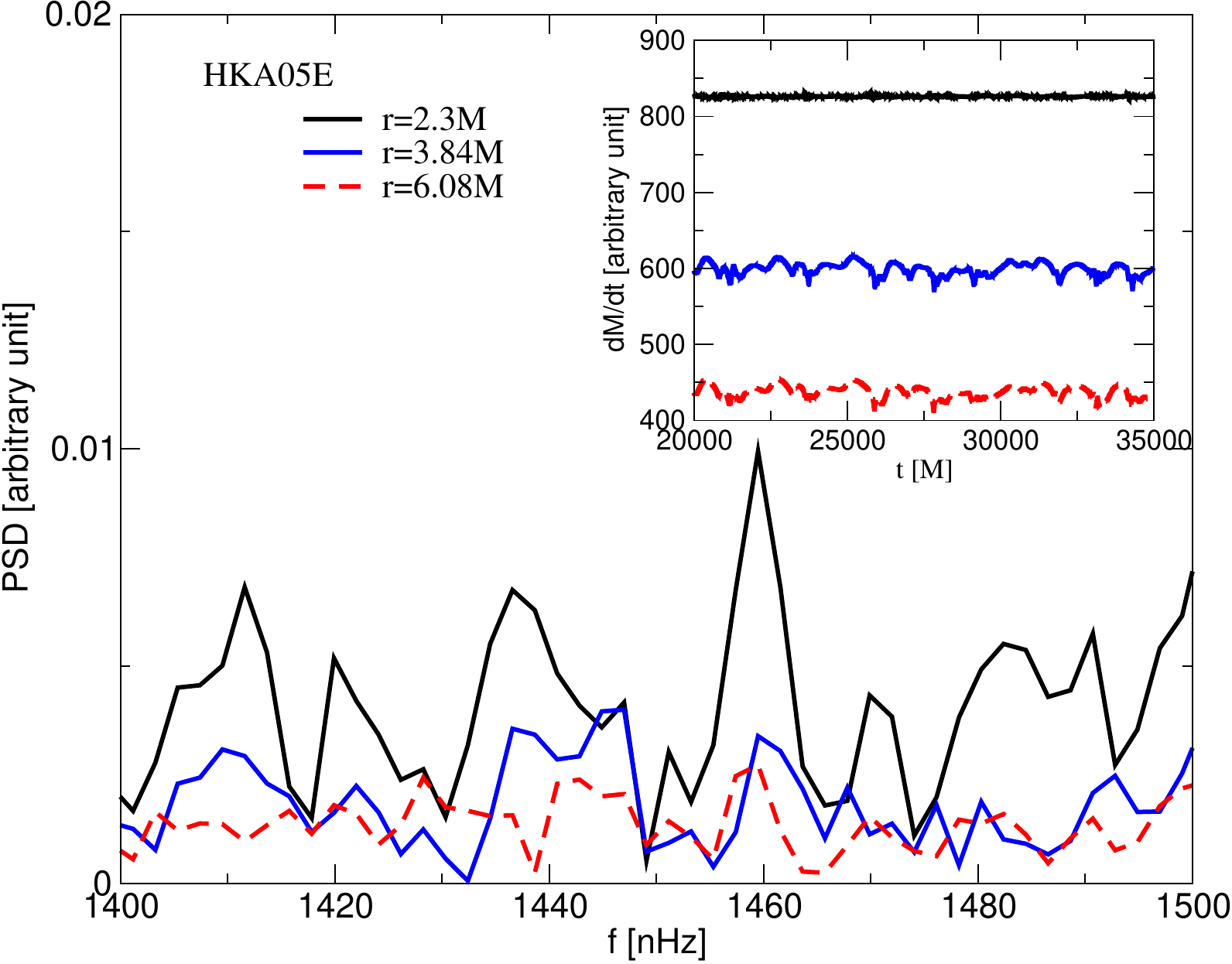,width=3.5cm, height=4.0cm}\hspace*{0.15cm}
     \psfig{file=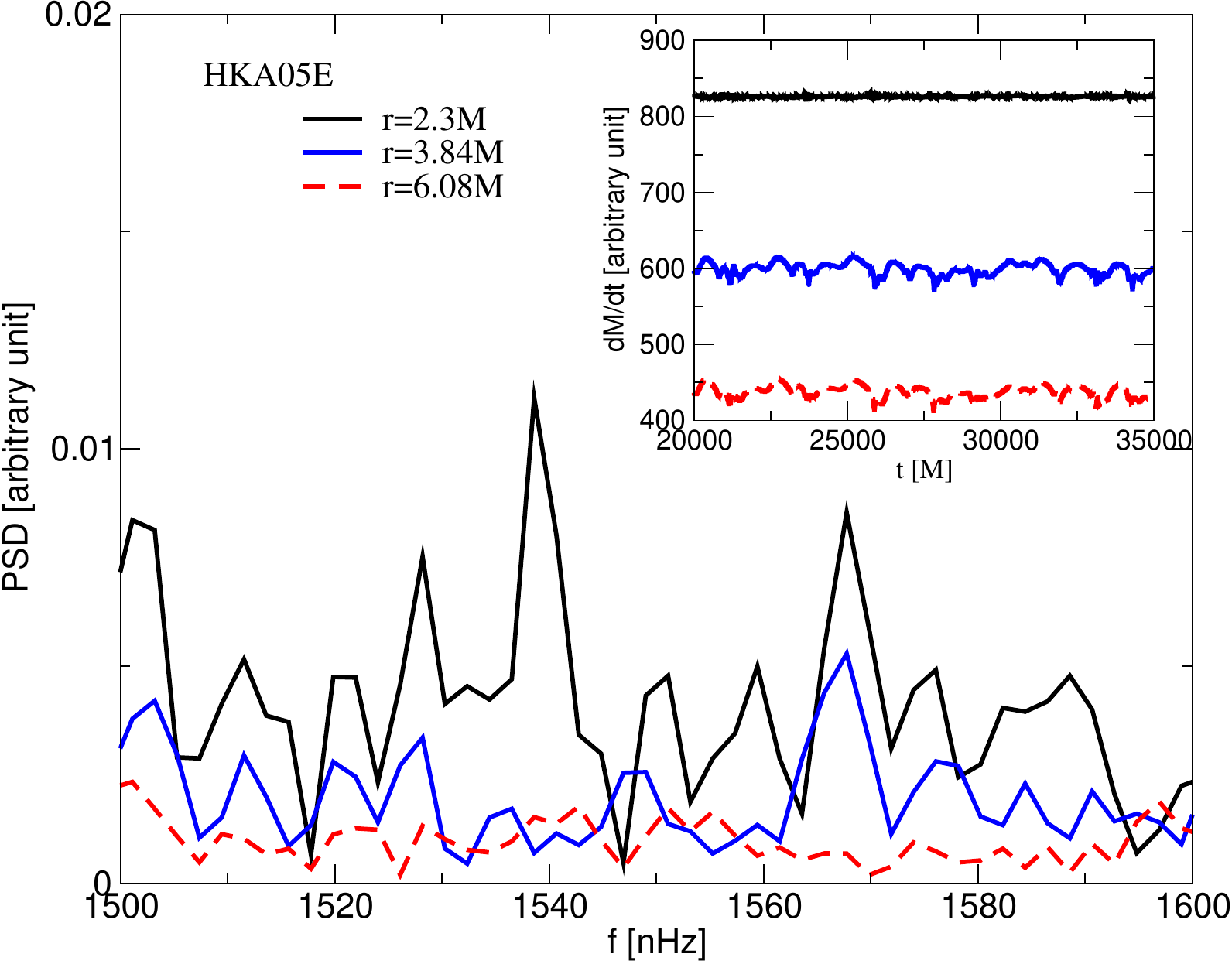,width=3.5cm, height=4.0cm}\\    
     \caption{Same as Fig.\ref{PSDHKA05_1}, but PSD analysis for model HKA05E at different frequency ranges.
    }
\vspace{1cm}
\label{PSDHKA05E_Diff_window}
\end{figure*}

In Fig.\ref{PSDHKA05_1}, the PSD analysis obtained from the accretion rate of the mass at different radial points of the shock cone is shown for various model conditions with $\eta=0.5$ as given in Table \ref{Inital_Con_2}. When these models, which have the same $\eta$ value, are compared with those in Fig.\ref{PSDHK05_1}, the spin parameter of the black hole increases from the first model to the last model in both figures. During this process, while the hair parameter used in the PSD analyses in Fig.\ref{PSDHK05_1} varies between $l_0/M=0.0608$ and $l_0/M=0.9113$, this value ranges from $l_0/M=0.7172$ to $l_0/M=0.9193$ for the models in Fig.\ref{PSDHKA05_1}. The PSD results in Fig.\ref{PSDHKA05_1}, when compared to those in Fig.\ref{PSDHK05_1}, show that the ones in Fig.\ref{PSDHK05_1} produce very complex and difficult-to-distinguish peaks. These peaks appear more clear and observable in Fig.\ref{PSDHKA05_1}. Thus, they produce QPO frequencies that are more clearly observable. In fact, such peaks are also seen in the HK05D model in Fig.\ref{PSDHKA05_1}. The hair parameter for this model is $l_0/M=0.9113$. Hence, it is clear that the hair parameter significantly affects the frequencies of the QPOs as well as their observability. On the other hand, all the explanations we made for the models in Fig.\ref{PSDHK05_1}  can also be made for the models in Fig.\ref{PSDHKA05_1}. That is, the frequencies formed here produce observable resonance modes like 1:2:3:4. Again, the fundamental QPO frequencies formed here result from the Lens-Thirring effect caused by the warping of spacetime around the black hole, as well as the excitation of pressure modes trapped within the disk. Other observed frequencies also result from nonlinear couplings, as previously explained. Similarly, as in Fig.\ref{PSDHK05_1}, the low frequencies seen in Fig.\ref{PSDHKA05_1} can be said to result from the resonance state of the  fundamental modes trapped within the cone, occurring due to the slow or moderate rotation of the black hole ($a/M<0.7$). However, as will be explained in Fig.\ref{PSDHKA05E_Diff_window}, the high frequencies result from the Lens-Thirring effect at $r=2.3M$, where the warping of spacetime due to the rapidly rotation of the black hole is most pronounced. At this point, the frequencies resulting from the Lens-Thirring effect suppress all other frequencies within the cone, so the observation of low-frequency QPOs at $r=2.3M$, as seen clearly in the HKA05E model in Fig.\ref{PSDHKA05_1}, is much lower compared to other points.
Thus, in this situation, which we have presented throughout the paper and is believed to be consistent with the theory which is the Lens-Thirring effect generates high-frequency QPOs around the black hole. The theoretical explanation of how high-frequency QPOs are generated due to the Lense-Thirring effect is provided in Section \ref{Theory_prediction}.

In Fig.\ref{PSDHKA05E_Diff_window}, for the HKA05E model given in Table \ref{Inital_Con_2}, which represents a rapidly rotating black hole with a high hair parameter, PSD analysis is provided for different frequency bands. Thus, the behavior of peaks in different frequency bands is revealed through PSD analyses conducted at various points. As seen in the figure, several frequencies are observed between 0-100 nHz. These are entirely the result of two fundamental modes trapped inside the cone and their nonlinear couplings. When Figures \ref{PSDHKA05_1} and \ref{PSDHKA05E_Diff_window} are examined together, it is seen that the fundamental modes occur at 16.6 nHz and 41 nHz. In contrast, $31$ nHz $\approx 2\times16.6$ nHz, and different combinations of 16.6 and 41 nHz produce QPO frequencies of 69, 75, 98, and 106 nHz.

On the other hand, as seen in Figures \ref{PSDHKA05_1} and \ref{PSDHKA05E_Diff_window} for the 0-100 nHz band range, the amplitudes of the peaks at $r=2.3M$ at low frequencies are very small. This is because, within this band range, the QPO frequencies excited inside the shock cone are suppressed by the Lense-Thirring effect, as we have previously mentioned many times. However, when looking at different band ranges, it is observed that the amplitudes of these modes increase and begin to have similar amplitudes to the QPO frequencies occurring at other r values. When the band range is 400-500 nHz, the same frequencies occur at almost the same amplitude at all calculated positions in the PSD analysis. This increases the likelihood of observing these frequencies in this situation. When looking at further band ranges, the frequencies resulting from the Lense-Thirring effect become dominant, while the diskoesmic pressure modes formed within the disk lose their dominance. Therefore, as in every rapidly rotating black hole model, it is predicted that high-frequency QPOs can be observed around the black hole due to the Lense-Thirring effect. When all band ranges are carefully examined, we can say that the physical cause of high-frequency QPOs is the QPO behaviors resulting from the warping of spacetime.

\begin{figure*}
  \vspace{1cm}
  \center
  \psfig{file=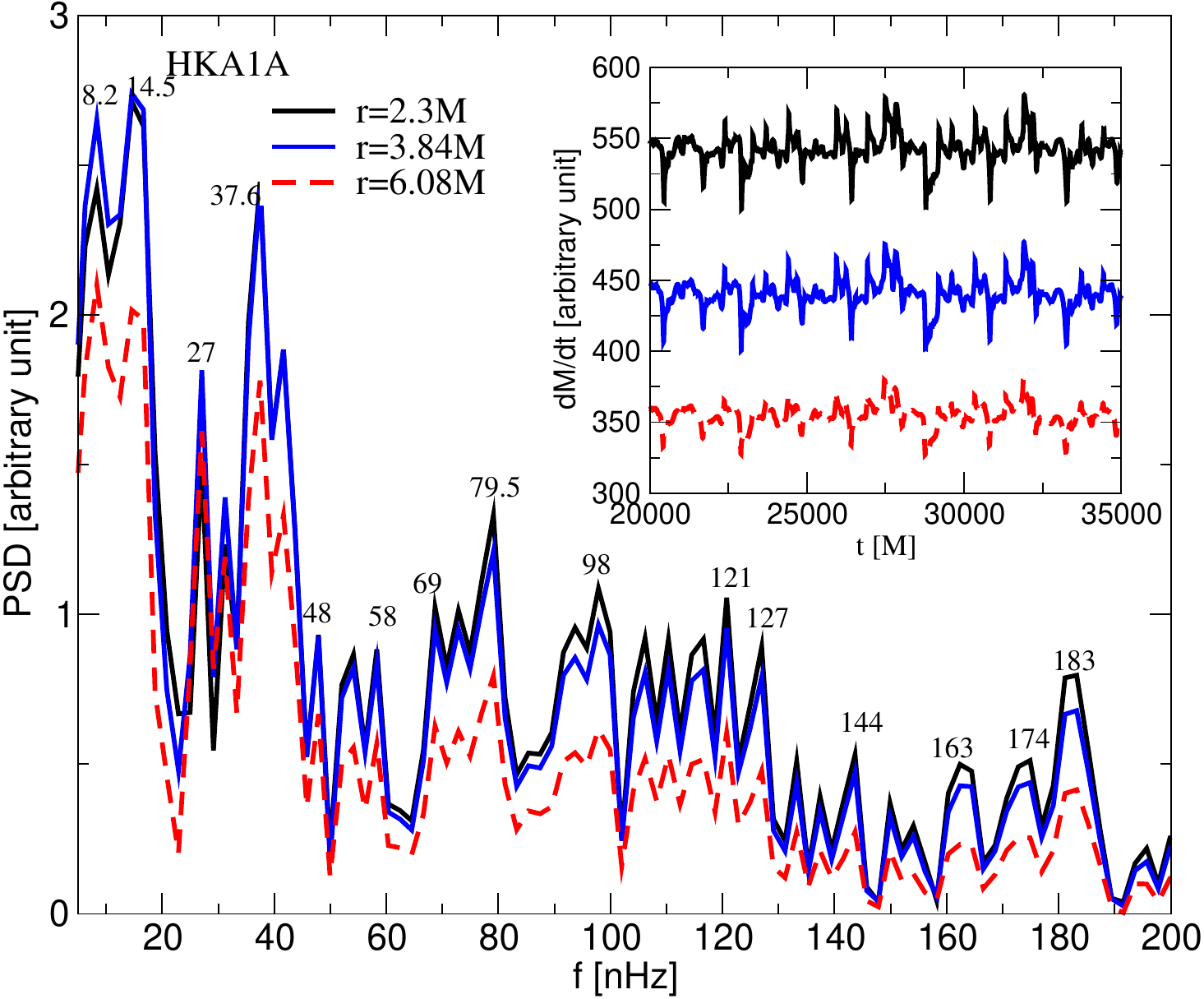,width=7.5cm, height=7.0cm}\hspace*{0.15cm}
  \psfig{file=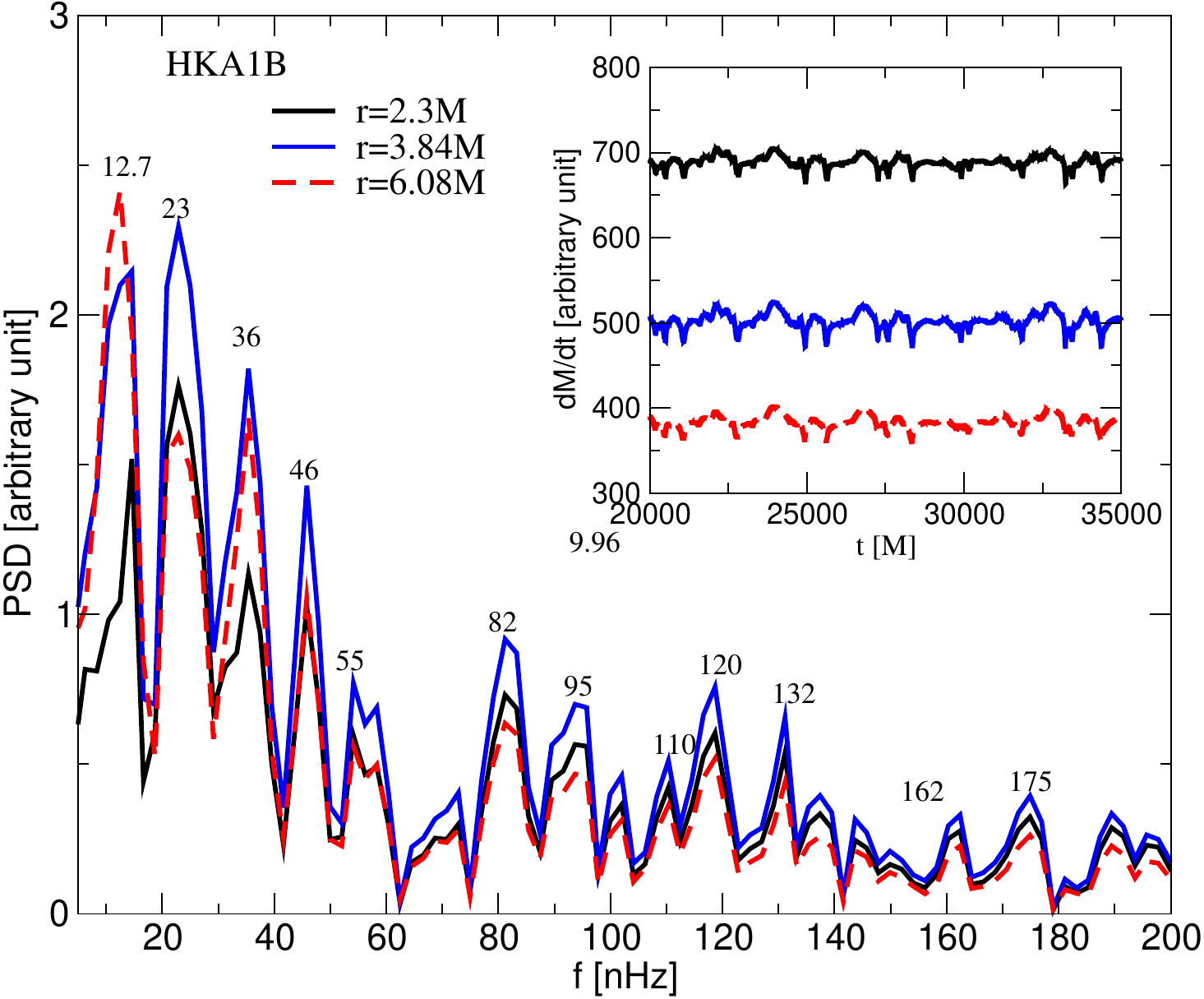,width=7.5cm, height=7.0cm}\\
  \vspace*{0.3cm}
  \psfig{file=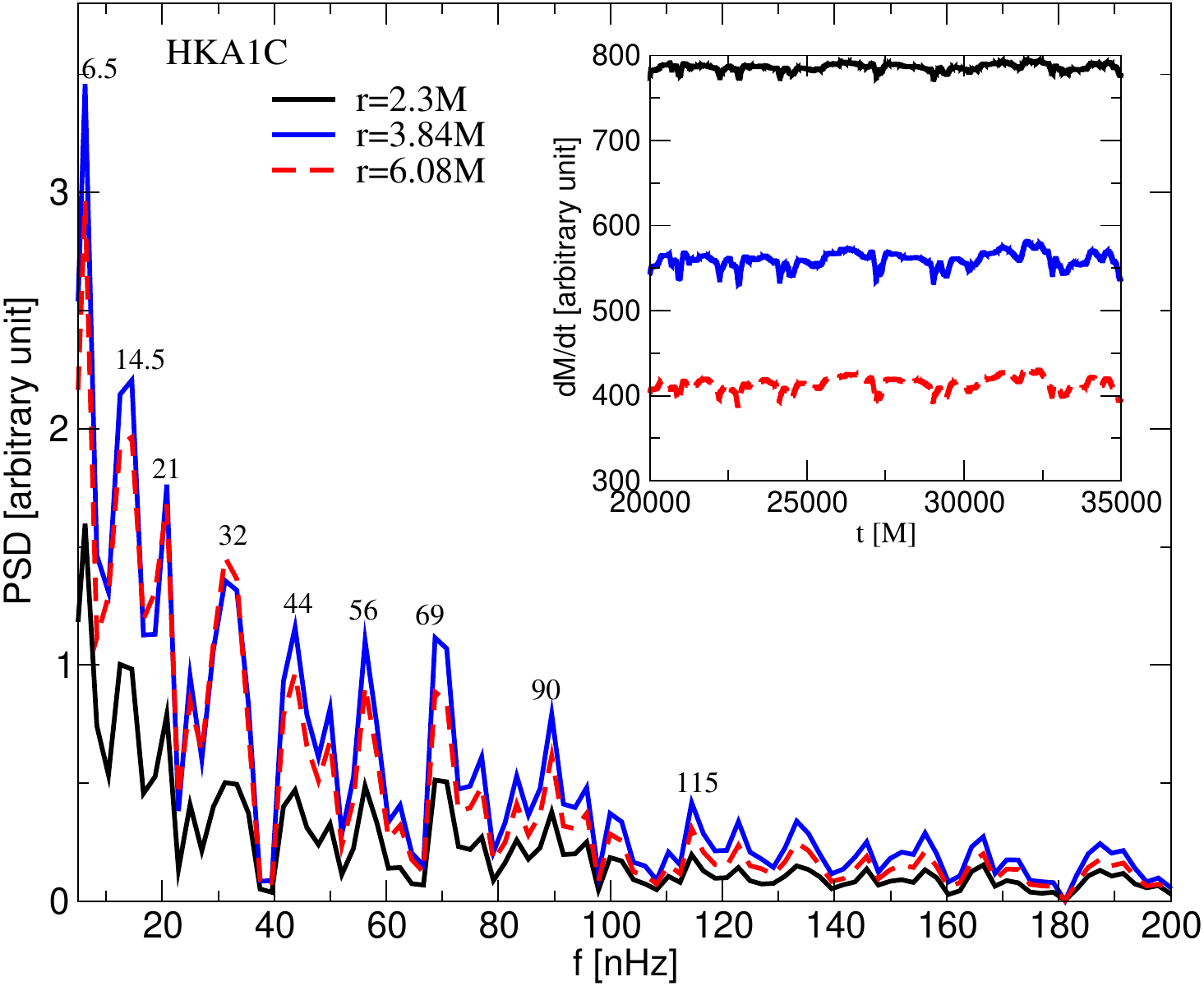,width=7.5cm, height=7.0cm}\hspace*{0.15cm}
  \psfig{file=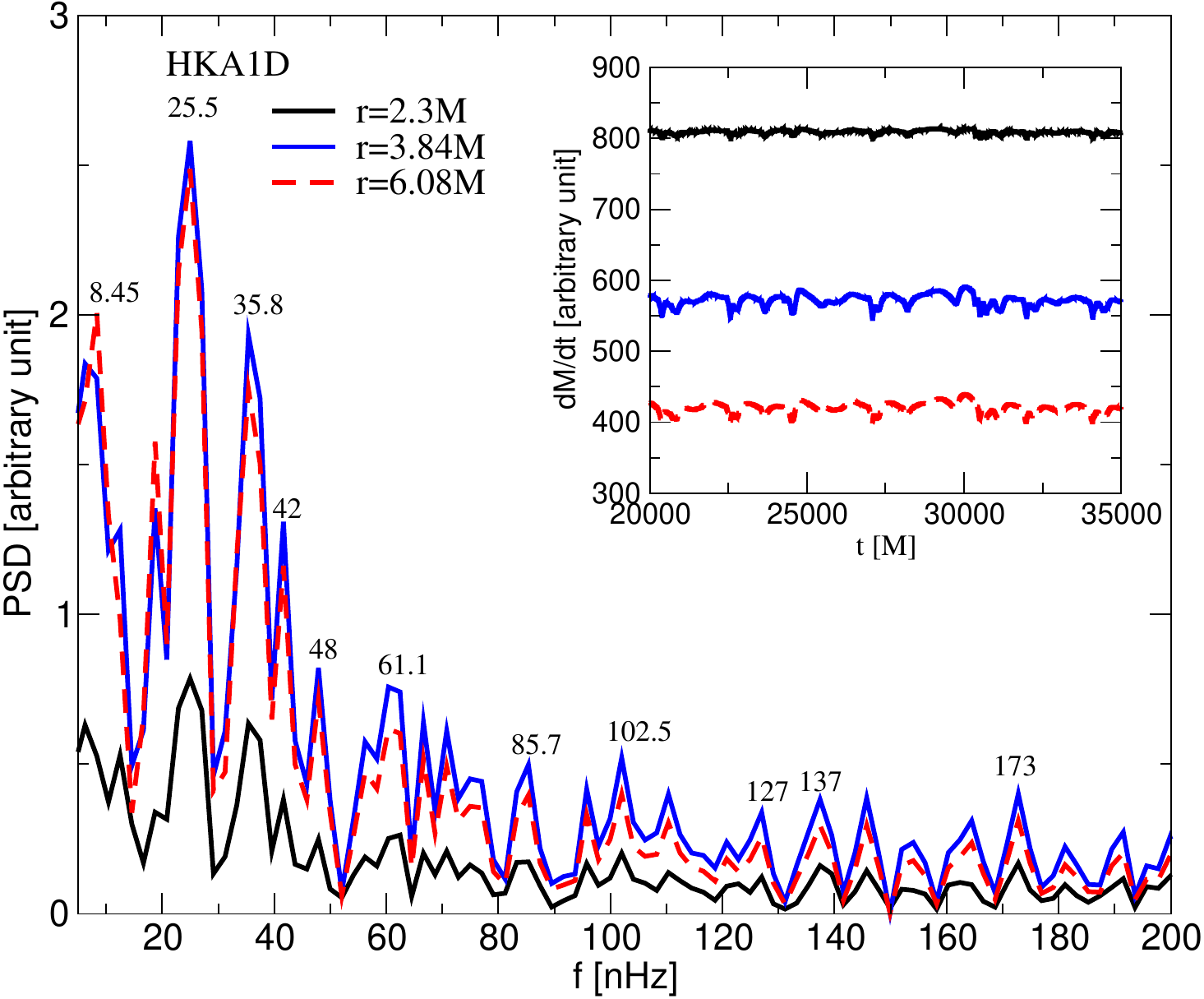,width=7.5cm, height=7.0cm}\\
  \vspace*{0.3cm}
  \psfig{file=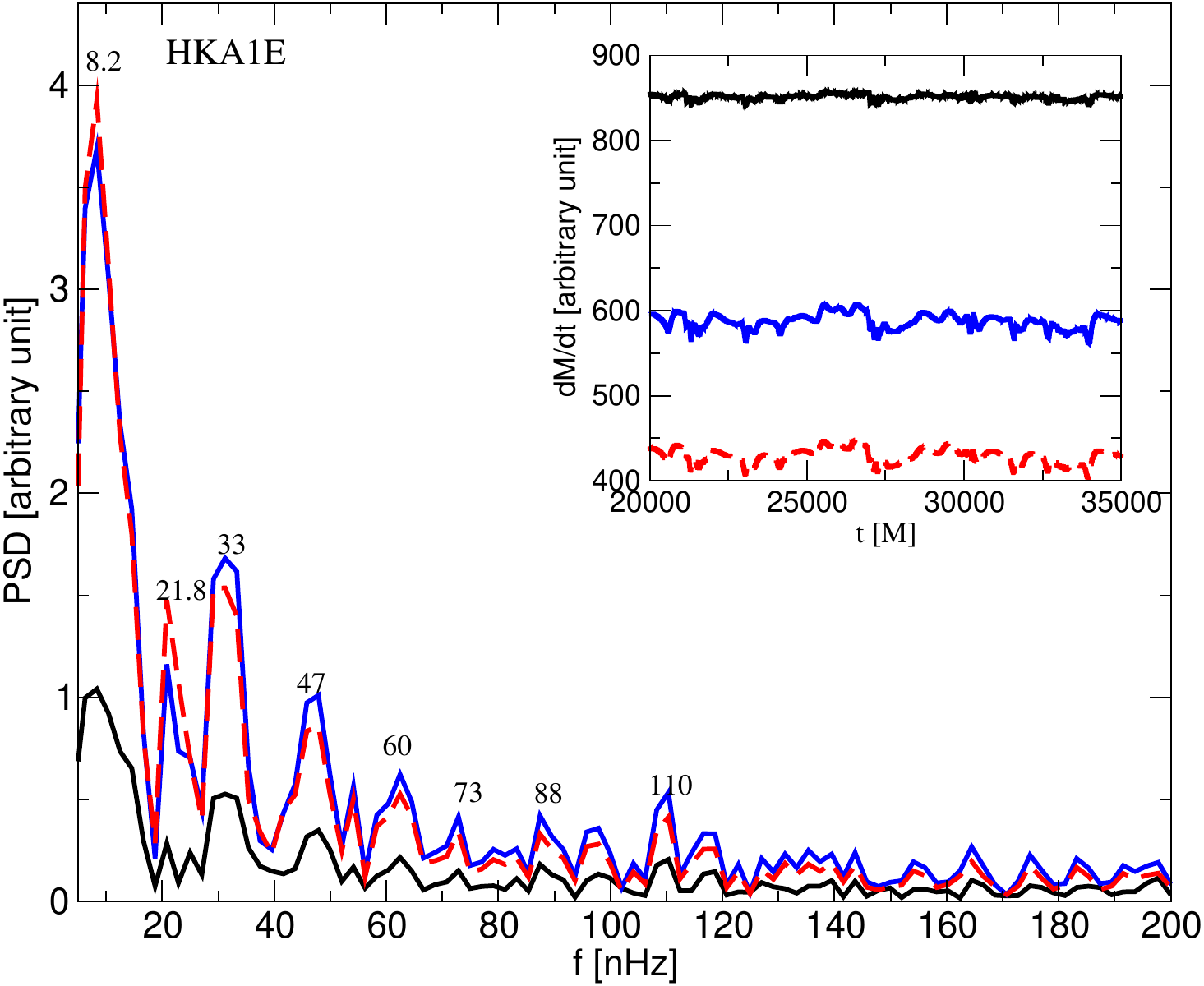,width=7.5cm, height=7.0cm}\\    
     \caption{Same as Fig.\ref{PSDHKA05_1} but for $\eta=1$.
    }
\vspace{1cm}
\label{PSDHKA1_1}
\end{figure*}

As in Figures \ref{PSDHKA05_1} and \ref{PSDHKA05E_Diff_window}, Fig.\ref{PSDHKA1_1} also presents a PSD analysis based on the initial conditions in Table \ref{Inital_Con_2}, for the case where the deviation parameter from Kerr is $\eta=1$, consistent with the EHT observational results. Everything we mentioned earlier regarding the QPO frequencies around rapidly rotating black holes applies to this situation as well. Additionally, the first four models in Fig.\ref{PSDHKA1_1} show the QPO frequencies for increasing spin parameters and almost the same hair parameter, while Model HKA1E demonstrates the behavior of peaks formed with the largest spin parameter and low hair parameter among these models. As seen in Fig.\ref{PSDHKA1_1}, the hair parameter significantly increases the number of peaks formed. This implies that the hair parameter of the black hole leads to the formation of new peaks compared to a Kerr black hole due to creating more complex situations around it. The increase in the value of the hair parameter confirms that it intensifies the instability around the black hole. This can also be verified by comparing Model HK1E in Fig.\ref{PSDHK1_1} with Model HKA1 in Fig.\ref{PSDHKA1_1}. The only distinguishing difference between HK1E and HKA1E is the hair parameter. When comparing these two models, it is observed that the hair parameter significantly increases the number of peaks resulting from the PSD analyses. Considering these situations, if the PSD analyses of the observed X-ray data exhibit more complex QPO behaviors, it strengthens the possibility that the central black hole has a hair parameter. These results are consistent with theoretical expectations \citep{Ingram2019,Rayimbaev2023Galaxy}. As the intensity of the hair parameter of the black hole increases, the spacetime is modified, leading to the formation of new instabilities. Consequently, new QPO frequencies emerge, differing from Kerr gravity.


\section{The effect of the hair parameter on the observed frequencies}
\label{effect_hair}

The hairy Kerr gravity, an extended alternative theory of general relativity, modifies spacetime due to the hair parameter and therefore affects the physical phenomena occurring around the black hole in the strong gravitational region. Identifying the impact of the hair parameter can help us understand whether the observed data result from Kerr or alternative gravity theories. In this context, we discuss the effect of the hair parameter on the QPO frequencies found in our numerical calculations in this section.

As previously discussed, the results of the hairy Kerr black hole clearly show the effect of the hair parameter on the resulting QPO frequencies when compared to Kerr. By comparing Fig.\ref{Kerr09} with Figures \ref{PSDHK05_1}, \ref{PSDHK1_1}, \ref{PSDHKA05_1}, and \ref{PSDHKA1_1}, the number of QPO peaks formed in the strong gravitational field around the Kerr black hole is fewer compared to the hairy Kerr black holes. This indicates that the hair parameter leads to more chaotic oscillations after modifying gravity, resulting in more complex and rich QPO behavior due to the formation of fundamental modes and their nonlinear couplings. In this case, it can be said that the observed QPO data can be used to distinguish whether they result from Kerr or alternative gravities.

We compare the HK1E model shown in Fig.\ref{PSDHK1_1} with the HKA1E models shown in Fig.\ref{PSDHKA1_1}. For both models, the deviation parameter from the Kerr black hole is $\eta = 1$. For HK1E, $a/M = 0.8056$ and $l_o/M = 0.9387$, while in the HKA1E case, these parameters are $a/M = 0.8291$ and $l_o/M = 0.2678$. In our discussions and numerical calculations in the previous sections, it is found that the amplitude of the QPO oscillation frequencies formed at $r = 2.3M$ around a rapidly rotating black hole is suppressed by the spin parameter. Considering the HK1E and HKA1E models, even though we expect the frequency for the HKA1E model to be more suppressed due to the larger spin parameter, this does not happen. We think this is due to the hair parameter. In the HK1E model, where the hair parameter is large and the black hole's spin parameter is slightly smaller, the Lense-Thirring effect does not occur as discussed above. This shows that the hair parameter also affects these QPO frequencies. Therefore, in the HKA1 model, the reason for the more suppressed amplitude at $r = 2.3M$ is the hair parameter. In fact, this behavior has been theoretically proven in detail in Ref.\citep{Wu2023PhRvD}. They demonstrated that the precession frequencies calculated due to the hair parameter are suppressed compared to Kerr. Thus, we can say that the numerical results we found here are consistent with the theoretical results.

On the other hand, two of the best models demonstrating the impact of the hair parameter on potential QPO frequencies around the $M87^*$ black hole  are the models HKA1D and HKA1E, given in Table \ref{Inital_Con_2}. Although the $a/M$ parameter in the HKA1D model is 10\% smaller than in HKA1E, the small hair parameter for the rapidly rotating black hole model causes the horizon radius of the black hole to be $R_{BH}=0.990M$. Therefore, the QPO frequencies calculated at $r=2.3M$ have similar peak values in both models. While the black hole in HKA1E rotates slightly faster than in HKA1D, the hair parameter in HKA1D is almost three times that of HKA1E. This causes significant changes in the shadow, affecting the QPOs formed at the same radial point. Consequently, we can say that the change in the hair parameter has a significant impact on the dynamics of the resulting QPO frequencies and their observability.


\section{Understanding the QPO behavior around the $M87^*$ black hole from  the relativistic precession model}
\label{Theory_prediction}

The mass accretion rate is an important parameter in revealing the oscillation properties of the matter around the black hole. It illustrates the relationship between the oscillations on the accretion disk and the physical mechanism. By examining the mass accretion rate, QPO oscillations occurring around black holes can be identified \citep{Donmez6, Donmez2024arXiv240216707D, Donmez2024Submitted}. Especially in strong gravitational fields, by calculating the mass accretion rate at different points, the changes and observability of QPO frequencies can be determined. If QPOs are identified numerically, their agreement with theoretical predictions must also be demonstrated.

Theoretically, resonance frequencies around black holes have been derived \citep{Stella1999PhRvL,Katopasj2001}. These frequencies include the orbital frequency known as the Kepler frequency at a given radius, the Lense-Thirring precession frequency is a frequency resulting from the curvature of spacetime due to the rotation of the black hole, the radial epicyclic frequency describing oscillations in the radial direction, and the vertical epicyclic frequency describing oscillations in the vertical direction. In this paper, since mass accretion on the equatorial plane is considered, there is no vertical oscillation, and therefore we do not discuss the vertical epicyclic frequency here. The theoretically derived frequencies based on the mass and spin parameter of the black hole are the Kepler frequency, $\nu_K= \frac{1}{2 \pi}\left(\frac{M}{r^3}\right)^{1/2}$, the Lense-Thirring precession frequency $\nu_{LT}= \left(\frac{2M^2}{r^3}\right)^{1/2}$, and the radial epicyclic frequency $\nu_r= \nu_K \left(1- \frac{6M}{r} + \frac{8aM^2}{r^{3/2}} \right)^{1/2}$.

Using these frequencies, a relationship can be established between the frequencies obtained from numerical calculations and the theoretical frequencies, thereby verifying the accuracy of the numerical frequencies. In our numerical analyses, we calculated the mass accretion rate at $r=2.3M$, $r=3.84M$, and $r=6.08M$. Using these data, we performed the power spectrum density analyses and identified the QPO behavior around the black hole for each model case. By comparing the QPO behavior observed in the numerical results with the theoretical results, we can predict the potential QPOs around the $M87^*$ black hole.

Since these radii define the region of strong gravitational field, we can test Einstein's general relativity theory with the observed QPOs. At the same time, we can determine whether the observed shadows of $M87^*$ can be explained by a hairy black hole. Due to the strong gravitational effects, gravitational redshift and frame dragging significantly influences the formation of QPOs. Consequently, high-frequency QPOs would be observed at $r=2.3M$  due to the strong gravitational field \citep{Strohmayer_APJ_2001}. Moderate frequency QPOs would be observed at $r=3.84M$ due to moderate effects \citep{Miller_APJ_2001}, and lower frequency QPOs would be observed at $r=6.08M$  compared to $r=2.3M$ due to the lesser relativistic effects \citep{Remillard2006ARA&A}. These results are consistent with observations.

As detailed in Section \ref{NumRes}, our numerical PSD analyses have shown that, despite small variations, the lowest frequency QPOs in all models occurred around $15$ nHz. Both in Kerr and Hairy Kerr black hole models, numerous QPOs are observed within the given frequency range. The reason for these QPOs is that they are resonance frequencies caused by different physical mechanisms, and these frequencies have combined through nonlinear coupling to produce new frequencies. As discussed in Section \ref{NumRes}, the fundamental mode frequencies are emphasized as being caused by the radial or angular oscillations of modes trapped within the shock cone. These oscillation frequencies are also influenced by the strong gravitational field resulting from the curvature of spacetime due to the black hole's spin parameter.

The curvature of spacetime due to the strong gravitational field near the black hole's horizon affects the location of shock cone, as clearly seen in Figures \ref{DensHK05} and \ref{DensHKA05}. This change leads to the formation of the Lense-Thirring precession frequency, which, as theoretical studies show, manifests as high-frequency QPOs \citep{Stella1999PhRvL,Katopasj2001}. When comparing results in Figures \ref{Kerr09_Diff_window}, \ref{PSDHKA1A_Diff_window}, and \ref{PSDHKA05E_Diff_window}, it's evident that for the spin parameter around $\sim 0.9$, low-frequency QPOs at $r=2.3M$ have much lower amplitudes compared to those at $r=3.84M$ and $r=6.08M$, while high-frequency QPOs at $r=2.3M$ are more prominent and observable. This indicates that high-frequency QPOs are primarily due to Lense-Thirring precession.

On the other hand, as shown in Fig.\ref{PSDHKA1A_Diff_window}, when the black hole's spin parameter is nearly zero, the frequencies observed at any position maintain similar amplitudes, indicating that in the absence of a strong Lense-Thirring effect, other modes are significant in the formation of QPOs around the black hole. In summary, the nHz-range QPO frequencies potentially observed around the rapidly rotating $M87^*$ black hole result from the excitation of modes trapped within the shock cone, while high frequencies up to mHz range in maximum are due to Lense-Thirring precession. Conversely, for slowly or moderately rotating black holes ($a/M< 0.7$) , as seen in Fig.\ref{PSDHKA1A_Diff_window}, QPOs result entirely from the excitation of modes within the cone. Possible QPOs that can be excited within the cone are discussed in detail in Ref.\citep{CruzOsorio2023JCAP, Donmez2024Submitted, Donmez2024arXiv240216707D}.

\section{Expected QPO frequencies for the $M87^*$ and comparison with numerical results}
\label{Expected_frequency}

Observing QPOs in the strong gravitational field around the M87 black hole not only facilitates understanding the physical properties of the black hole but also reveals the dynamic structure of the disk in the region close to the event horizon. However, as discussed below, detecting these frequencies, which range from nHz to mHz around such a massive black hole like $M87^*$, is currently not feasible due to the limitations of existing technology and the distance and characteristics of the source. Additionally, the timescale of QPOs formed around the M87 black hole is much larger compared to stellar black holes. As a result, continuous observations spanning years are needed to accurately reveal the formation of QPOs around this black hole. Therefore, the QPOs we have identified numerically in this paper make a significant contribution to the literature and pave the way for future observations.

To validate the accuracy of the QPOs found numerically in the strong gravitational field, this section discusses the potential QPO frequency range around  $M87^*$, based on observed QPOs around other massive black holes. As known, the literature contains various QPO observations around different massive black holes. One of these massive black holes is RE J1034+396. RE J1034+396 is a narrow-line Seyfert 1 galaxy. The QPO frequency of this black hole, which has a mass of $4 \times 10^6 M_{\odot}$, has been observed to be $2.7$ mHz \citep{Jinmnras2020}. Based on this, the expected QPO frequencies of the massive black hole $M87^*$ can be estimated.

Since QPOs occur near the event horizon of the black hole in a strong gravitational field, we generally look for QPOs in regions smaller than $r<10M$. On the other hand, considering that particles can move at maximum light speed in this strong field around the black hole, the mass of the black hole changes the distance of the point where QPOs form from the singularity of the black hole, increasing the orbital period of the particles around the black hole and thus decreasing the frequency. Taking these situations into account, we can predict the range of QPO frequencies that can be observed in the $M87^*$ black hole.

The mass of the $M87^*$ black hole is $6.5 \times 10^9 M_{\odot}$. Considering the inverse scaling relationship between frequency and mass, the QPO frequency that can be observed around $M87^*$ can vary from nHz to mHz. This is found as follows: considering the source RE J1034+396, the frequency of the black hole with a mass on the order of $10^6 M_{\odot}$ is around $1$ mHz. Applying this to $M87^*$, which has a mass of $6.5 \times 10^9 M_{\odot}$, the possible QPO frequency observed for $M87^*$ would be around $14$ nHz or greater. The oscillation period of a $140$ nHz frequency is approximately $82.5$ days. If sufficient sensitivity and monitoring are conducted, the QPOs of $M87^*$ can be revealed.

Based on the QPOs observed from the massive black holes sources, the QPOs predicted to form around the $M87^*$ black hole in a strong gravitational field are consistent with those found in numerical calculations. Although observing this type of QPO seems unlikely in the short term due to both technical challenges and the nature of the source such as the low inclination of the observer,  the small magnitude  and long period of the variations\citep{Medeiros2022ApJ, Cui2023Natur}, we believe it is important as it would contribute to the literature.

\section{Discussion and Conclusion}
\label{Concl}

Observing the shadows of the $M87^*$ and $SgrA^*$ black holes and revealing their physical properties is crucial for testing the strong gravitational field around black holes \citep{Akiyama1,Akiyama3, Vagnozzi2}. Unveiling the characteristics of the strong gravitational field helps understand the QPOs that may result from black hole-disk interactions in this region. On the other hand, the characteristics of the scalar field, which is thought to define dark matter in the universe's mysterious objects, can also be revealed \citep{Matos2000PhRvD, Cunha2015PhRvL, Banerjee2017PhRvD, Guo2023EPJC}. Using only the well-known Kerr black hole defined in General Relativity to uncover the characteristics of the strong gravitational field may be insufficient to explain certain phenomena. In other words, using alternative gravities is scientifically necessary to impose different constraints from Kerr gravity that align with observations for both black hole shadows and gravitational wave detections \citep{Ayzenberg2018CQGra, Glampedakis2021PhRvD}. At this point, by using alternative gravity theories, more detailed or different solutions can be proposed for observed phenomena that have been attempted to be explained using the Kerr black hole \citep{Khodadi1}. In this paper, we reveal the effects of deviations from the Kerr metric and the hair parameters on the QPO frequencies that may form around the $M87^*$ black hole using the hairy Kerr black hole.

The modification of spacetime not only changes the physical properties of the black hole but also alters the interactions occurring around it. Numerical simulations have shown that changes in the deviation parameter of Hairy Kerr gravity from Kerr gravity affect the QPOs formed around the black hole. This change is observed not only in the value of QPO frequencies but also in the amplitude of the resulting peaks. As this parameter's value increases, the amplitude of the QPO frequencies also increases. Considering that the QPO frequencies around the $M87^*$ black hole are in the nHz-mHz range, having larger peak amplitudes would enhance their observability. Therefore, the parameter defining the deviation of gravities from each other contributes to the observability of the resulting QPOs.

It has been revealed from the numerical simulations that the increase in the spin parameter of the black hole significantly affects the amplitudes of the resulting QPO frequencies. In the case of an increased spin parameter which is $a/M>0.7$, it has been shown that the amplitude of the low-frequency QPOs calculated at $r=2.3M$ is much lower compared to other points at $r=3.84M$ and $r=6.08M$. This demonstrates the complex effects of the black hole's spin parameter on the oscillation modes of the surrounding matter. When the black hole spins rapidly, it warps the surrounding space more, and this warping suppresses the pressure mode oscillations trapped within the shock cone, leading to lower amplitude oscillation frequencies at $r=2.3M$. However, as the distance from the black hole increases, the effect of the spin parameter in warping the shock cone diminishes, thus not preventing the formation of low-frequency QPOs in these regions. This situation indicates that the interaction between the black hole and the disk in the strong gravitational field around rapidly spinning black holes leads to highly chaotic behaviors. Additionally, since the deviation parameter from the Kerr black hole, $\eta$, and the hair parameter, $h/M$, significantly affect the radius of the resulting shadow, the horizons of black holes with the same spin parameter differ at different $\eta$ and $h/M$ values. Consequently, the resulting QPO frequencies and their observability are seriously affected by the changes in these parameters.

The complex effects of the interaction between the black hole and the disk, resulting from the warping of spacetime around a rapidly spinning black hole, can be used to explain the continuously varying and QPO oscillations observed in well-known sources like GRS 1915+105. The frequencies of the GRS 1915+105 black hole observed over the past decades highlight the interaction between the calculated spin parameter of the black hole and the disk. As a result of this interaction, the observed QPO frequencies exhibit continuous variability, ranging from a few Hertz to hundreds of Hertz \citep{Strohmayer2001ApJ, Belloni2013MNRAS, Rawat2019ApJ, Motta2023, Chauhan2024MNRAS, Donmez2024Submitted}. Therefore, the impact of the black hole's spin parameter on the resulting QPO frequencies, as revealed in this paper, explains the physical reasons for the different frequencies observed from this source. In other words, the interaction between a rapidly spinning black hole and its disk leads to a variety of oscillatory behaviors.

The numerical demonstration of the effect of the black hole's spin parameter on the formation of low and high-frequency QPOs around the black hole shows the significant impact of the strong gravitational field on the oscillation modes of the disk. In other words, it demonstrates that the pressure modes trapped within the cone are damped by the influence of the strong gravitational field. As the distance from the black hole horizon increases, the decreasing effect of spacetime warping allows  for the observation of low frequencies in this region. Therefore, our study has revealed that determining the region of the disk where the observed X-ray data are produced plays an important role in identifying the observed QPO frequencies.

As theoretically expected, the warping of spacetime around rapidly rotating black holes significantly affects the physical phenomena occurring on the accretion disk around the black hole. Our numerical results here have shown that to understand the Lense-Thirring effect in physical phenomena like QPOs caused by the warping of spacetime, the observational data must be collected as close as possible to the black hole's horizon. Otherwise, at slightly further points like $r=3.8M$ from the black hole's horizon, the physical phenomena within the disk or the shock cone could suppress this warping effect, making it less discernible. This result has been clearly observed in numerical simulations.

Finally, the numerical results have shown that QPO frequencies can form around the $M87^*$ black hole using both Kerr and Hairy Kerr gravities. In both cases, oscillation peaks that are harmonics of each other have been observed. The spin parameter of the black hole significantly affects the formation, excitation, or suppression of QPOs. For example, QPO frequencies have been observed in all models, including Kerr gravity. In scenarios where the black hole is spinning slowly, there are numerous low-frequency peaks, whereas in rapidly spinning black hole models, the number of these peaks significantly decreases. This confirms the Lens-Thirring effect, which suppresses or even completely eliminates many peaks at points closest to the black hole's horizon, as continuously emphasized in this paper. On the other hand, It is shown that in all models of slowly rotating black holes, a large number of low-frequency QPOs have occurred inside the shock cone due to the excitation of pressure modes in the weak frame-dragging region. The scenarios related to the formation of QPO frequencies around both slowly and rapidly rotating black holes show perfect agreement with the literature \citep{Stella1998ApJ, Remillard2006ARA&A, Ingram2009MNRAS}.


\section*{Acknowledgments}
All simulations were performed using the Phoenix  High
Performance Computing facility at the American University of the Middle East
(AUM), Kuwait.\\

\newpage

\bibliographystyle{JHEP} 
\bibliography{paper.bib}

\end{document}